\documentclass[preprint]{aastex62}

\DeclareRobustCommand{\ion}[2]{\textup{#1\,\textsc{\lowercase{#2}}}}
\newcommand{\HI}{\ion{H}{i}}

\newcommand{\NHIthin}{N_{\scriptsize \ion{H}{i}}^\mathrm{thin}}
\newcommand{\HII}{\ion{H}{ii}}
\newcommand{\NHII}{N_{\scriptsize \ion{H}{ii}}}
\newcommand{\NH}{N_\mathrm{H}}
\newcommand{\NHi}{N_\mathrm{H, i}}
\newcommand{\NHmod}{N_\mathrm{H, mod}}
\newcommand{\NHbk}{N_\mathrm{H, bk}}

\newcommand{\NHtau}{N_\mathrm{H, \tau_{353}}}
\newcommand{\Htwo}{\mathrm{H}_{2}}

\newcommand{\WCO}{W_\mathrm{CO}}
\newcommand{\WHI}{W_{\scriptsize \ion{H}{i}}}
\newcommand{\Dem}{D_\mathrm{em}}
\newcommand{\HIs}{\mbox{{\scriptsize H}{\tiny\,I}}}

\newcommand{\THI}{\tau_{\HIs}}

\received{January 1, 2018}
\revised{January 7, 2018}
\accepted{\today}
\submitjournal{ApJ}

\shorttitle{CRs and ISM in Local HI Clouds}
\shortauthors{Mizuno et al.}

\begin{document}

\title{Study of the Cosmic Rays and Interstellar Medium in Local $\HI$ Clouds using \textit{Fermi}-LAT Gamma-Ray Observations}

\correspondingauthor{T. Mizuno}
\email{mizuno@astro.hiroshima-u.ac.jp}

\author[0000-0001-7263-0296]{T. Mizuno}
\affiliation{Hiroshima Astrophysical Science Center, Hiroshima University, 1-3-1 Kagamiyama, Higashi-Hiroshima, Hiroshima, 739-8526, Japan
}

\author{S. Abdollahi}
\affiliation{Department of Physical Sciences, Hiroshima University, Higashi-Hiroshima, Hiroshima 739-8526, Japan}

\author{Y. Fukui}
\affiliation{Departiment of Physics and Astrophysics, Nagoya University, Nagoya, Aichi 464-8602, Japan}

\author{K. Hayashi}
\affiliation{Departiment of Physics and Astrophysics, Nagoya University, Nagoya, Aichi 464-8602, Japan}

\author{T. Koyama}
\affiliation{Department of Physical Sciences, Hiroshima University, Higashi-Hiroshima, Hiroshima 739-8526, Japan}

\author{A. Okumura}
\affiliation{Institute for Space-Earth Environmental Research, Nagoya University, Nagoya, Aichi 464-8601, Japan}

\author{H. Tajima}
\affiliation{Institute for Space-Earth Environmental Research, Nagoya University, Nagoya, Aichi 464-8601, Japan}
\affiliation{W.~W.~Hansen Experimental Physics Laboratory, Kavli Institute for Particle Astrophysics and Cosmology, 
Department of Physics and SLAC National Accelerator Laboratory, Stanford University, Stanford, CA 94305, USA}

\author{H. Yamamoto}
\affiliation{Departiment of Physics and Astrophysics, Nagoya University, Nagoya, Aichi 464-8602, Japan}







\begin{abstract}
An accurate estimate of the interstellar gas density distribution is crucial 
to understanding the interstellar medium (ISM) and Galactic cosmic rays (CRs).
To comprehend the ISM and CRs in a local environment, 
a study of the diffuse $\gamma$-ray emission 
in a mid-latitude region of the third quadrant 
was performed.
The $\gamma$-ray data in the 0.1--25.6~GeV energy range 
of the \textit{Fermi} Large Area Telescope (LAT) and other interstellar gas tracers
such as the HI4PI survey data and the \textit{Planck} dust thermal emission model were used,
and the northern and southern regions were analyzed separately.
The variation of the dust emission $\Dem$ with the total neutral gas column density $\NH$ was studied in high dust-temperature areas,
and the $\NH$/$\Dem$ ratio was calibrated using $\gamma$-ray data
under the assumption of a uniform CR intensity in the studied regions.
The measured integrated $\gamma$-ray emissivities above 100~MeV are
$(1.58\pm0.04)\times10^{-26}~\mathrm{photons~s^{-1}~sr^{-1}~H\mbox{-}atom^{-1}}$
and 
$(1.59\pm0.02)\times10^{-26}~\mathrm{photons~s^{-1}~sr^{-1}~H\mbox{-}atom^{-1}}$
in the northern and southern regions, respectively,
supporting the existence of a uniform CR intensity in the vicinity of the solar system.
While most of the gas can be interpreted to be $\HI$ with a spin temperature
of $T_\mathrm{S} = 125~\mathrm{K}$ or higher, an area dominated by optically thick $\HI$
with $T_\mathrm{S} \sim 40~\mathrm{K}$ was identified.
\end{abstract}



\section{Introduction}

High-energy $\gamma$-rays (energy $E \gtrsim 100~\mathrm{MeV}$) are
produced by
interactions of cosmic-ray (CR) particles with the gas 
and the radiation fields 
in the interstellar medium (ISM).
Because the ISM is essentially transparent to those $\gamma$-rays
\citep[e.g.,][]{Moskalenko2006}, observations of
high-energy $\gamma$-rays
are a useful probe of CRs and other components of the ISM.
Assuming an electron--to--proton ratio
in the interstellar space to be ${\sim}100$ at 10~GeV (the value measured at the Earth), 
$\gamma$-rays produced via nucleon--nucleon interactions 
is dominant compared to those by electron bremsstrahlung.
Because the $\gamma$-ray production cross section is independent of the chemical or thermodynamic state 
of the ISM gas, 
if the gas column density is well-established at medium to high Galactic latitudes,
then the CR proton intensity can be inferred under the assumption of a
uniform intensity and a known contribution of heavier elements
(and an uniform electron--to--proton ratio).

Usually, the distribution of neutral atomic hydrogen ($\HI$) is measured via 21-cm line surveys \citep[e.g.,][]{Dickey1990}
by assuming the optically thin approximation or a uniform spin temperature ($T_\mathrm{S}$),
and the distribution of molecular hydrogen ($\Htwo$) is 
indirectly estimated from the carbon monoxide (CO) line-emission surveys \citep[e.g.,][]{Dame2001}
assuming a linear conversion factor (usually called $X_\mathrm{CO}$).
Although the volume fraction of ionized gas is large, its column density
is usually small \citep[e.g.,][]{Ferriere2001} 
and can be neglected compared to the neutral gas.
The total neutral gas column density can also be indirectly estimated from dust using 
the extinction, reddening, or emission
\citep[e.g.,][]{Bohlin1978}.
A significant amount of gas not traced properly via $\HI$ and CO surveys 
in the solar neighborhood 
has been revealed
by combining the EGRET $\gamma$-ray data, $\HI$, CO, and dust extinction maps
and has been referred to as ``dark gas'' \citep{Grenier2005}.
Its column density is comparable to that of $\Htwo$ gas traced by CO.
Subsequently,
the work by \citet{Grenier2005} has
been confirmed and improved in terms of significance and accuracy
by recent observations of Galactic diffuse $\gamma$-rays by \textit{Fermi}
Large Area Telescope \citep[LAT;][]{Atwood2009}, as exemplified by 
\citet{Fermi2ndQ} and \citet{Fermi3rdQ,FermiCham}.
In addition, the \textit{Planck} mission has provided an
all-sky dust thermal emission model \citep{Planck2011,Planck2014a}
that is useful for the study of the ISM gas distribution 
because of its precision and high angular resolution. 

However, both $\gamma$-ray emission and dust emission $\Dem$
(or extinction/reddening) studies have limitations, and therefore the ISM gas distribution 
(and CR intensity in the interstellar space) remains uncertain even
in the solar neighborhood. On the $\gamma$-ray side, measurements suffer from low photon statistics
to trace the ISM gas distribution at high angular resolution,
contamination from point sources,
and background due to inverse Compton (IC) emission and isotropic background signal at high Galactic latitude. 
On the dust side, a procedure to convert $\Dem$
into the total neutral gas column density ($\NH$) has not yet been established.
For example, \citet{Fukui2015} compared the integrated $\HI$ 21-cm line intensity ($\WHI$) and
the \textit{Planck} dust optical depth at 353~GHz ($\tau_{353}$) 
in their all-sky data analysis (with a low latitude region ($|b| \le 15\arcdeg$) and
several other areas masked)
and 
interpreted the strong dust temperature ($T_\mathrm{d}$) dependence of the
$\WHI$ to $\tau_{353}$ ratio as a significant amount of optically thick $\HI$.
Meanwhile,
the \citet{Planck2014a} found that the dust radiance $R$ (integrated intensity)
correlated well with $\WHI$ over a wide range of $T_\mathrm{d}$ in the diffuse ISM
and proposed that it would be a better tracer of the dust (and the total gas) column 
density.
We
also note that the dust-to-gas conversion 
may be affected by dust and gas properties that can vary over the region.
In particular the uncertainty on the $\HI$ gas $T_\mathrm{S}$ have not been fully accounted for in previous studies
of $\gamma$-ray data \citep[e.g.,][]{Fermi2ndQ,Fermi3rdQ,FermiCham}.
This contributes to the uncertainties on the total ISM gas column density and 
then on the CR intensity distribution estimates \citep[see, e.g.,][]{Grenier2015}.

In this paper, we describe a detailed analysis of the \textit{Fermi}-LAT data for 
a mid-latitude region of the third quadrant 
(see Section~2.1 and Appendix~A for details of the region definition).
Two regions of interest (ROIs), spanning northern and southern Galactic latitude ranges ($22\arcdeg \le |b| \le 60\arcdeg$), do not contain any known large molecular clouds.
Most of the atomic hydrogen is expected to be within 1~kpc of the solar system and therefore 
the dust--to--gas ratio and CR intensity are expected to be uniform
due to the ROIs covering medium--to--high Galactic latitudes.
They were analyzed in an early publication of \textit{Fermi}-LAT analysis using six months of data
\citep{FermiHI} to study the CR intensity in the vicinity of the solar system.
We now aim to better
constrain the ISM gas distribution and the CR intensity/spectrum using eight years of \textit{Fermi}-LAT data
and newly available gas data such as the HI4PI survey data \citep{HI4PI} and the \textit{Planck} dust emission models.
In the light of studies by \citet{Fukui2015} and \citet{Planck2014a}, we consider both $\tau_{353}$ and $R$.

This paper is organized as follows. 
We describe the properties of the 
ISM tracers ($\WHI$ for the HI4PI survey and $R$ or $\tau_{353}$ for \textit{Planck}) in the studied regions in Section ~2
and the $\gamma$-ray observations, data selection, and modeling in Section~3. 
To model the $\gamma$-day data we take into account the neutral gas component ($\WHI$, $R$, or $\tau_{353}$),
IC emission, isotropic background, emission from the Sun and Moon, and $\gamma$-ray point sources. We also include ionized gas contribution as a fixed component.
The results of the data analysis are presented in Section~4,
in which we use the \textit{Fermi}-LAT $\gamma$-ray data as a 
robust tracer of $\NH$.
We discuss the ISM and CR properties of the studied region in Section~5. 
A summary of this study and future prospects are presented in Section~6.

\clearpage

\section{ISM Gas Tracer Properties and Map Preparation}

\subsection{Properties of the ISM Gas in the ROI}
Prior to preparing templates of the ISM gas for the $\gamma$-ray data analysis,
we investigated the properties of their tracers. As neutral gas tracers, we
prepared dust maps and a $\WHI$ map
stored in a HEALPix \citep{Gorski2005} equal-area sky map of order 9
\footnote{This corresponds to the total number of pixels of $12 \times (2^{9})^{2} = 3145728.$ (9 comes from the resolution index.)}
(with a mean spacing of $6\farcm9$ that is commensurate with the ${\sim}5'$ resolution of the Planck dust maps.)
We used the \textit{Planck} dust maps 
(of $R$, $\tau_{353}$, and $T_\mathrm{d}$) 
from the public data release 1 (version R1.20)\footnote{\url{http://irsa.ipac.caltech.edu/data/Planck/release_1/all-sky-maps/}}
described by \citet{Planck2014a}.\footnote{We note that a specific choice of the data release version is not crucial to constrain
the $\NH$ distribution and CR intensity because we use $\gamma$-ray data as a robust tracer of the ISM gas
as described in Section~4.
We also confirmed that the \textit{Planck} public data release 2 gives similar $\WHI$--$\Dem$ relationships
as those in Section~2.1 (stronger $T_\mathrm{d}$ dependence and non-linearity for the case of $\tau_{353}$
in the northern and southern regions, respectively).}
Assuming a uniform dust temperature along the line of sight,
they constructed maps by modeling the dust thermal emission with a single modified black body
\citep[for details of their procedure, see][]{Planck2014a}.
As described in \citet{Planck2014a},
the dust optical depth is the product of the dust opacity
and the total neutral gas column density $\NH$. 
Therefore, if the dust--to--gas ratio and dust cross section at 353~GHz are spatially constant,
$\tau_{353}$ is proportional to $\NH$.
The dust radiance $R$ is also expected to trace the total gas column density 
because it is proportional to $\NH$ under the assumption that
the dust--to--gas ratio, the optical and UV absorption cross section of the dust, 
and the interstellar radiation field are spatially uniform.
These hypotheses have to be tested because dust evolution models predict that
the dust emission properties change across the ISM 
\citep[e.g.,][]{Roy2013,Ysard2015}; which severely discourages the use of a single
conversion factor without verification.
For this reason we investigate the
change in the 
$\NH$-$\Dem$ relationship with dust properties using $\gamma$-ray data (Section~4).
We analyzed 
the northern ROI ($200\arcdeg \le l \le 260\arcdeg$ and $22\arcdeg \le b \le 60\arcdeg$)
and
the southern ROI ($210\arcdeg \le l \le 270\arcdeg$ and $-60\arcdeg \le b \le -22\arcdeg$).
The northern ROI is identical to that adopted by \citet{FermiHI}
but the southern ROI is shifted by $10\arcdeg$ (see below).
To construct the $\WHI$ map, we used the HI4PI survey data \citep{HI4PI}
integrated over the full velocity range of the survey (from $-600$ to $600~\mathrm{km~s^{-1}}$).
In the $\WHI$ map, we identified several bright radio sources and
intermediate velocity clouds \citep[e.g.,][]{Wakker2001}.
We removed these radio sources from the $\WHI$ map by filling them with the average of the peripheral pixels.
We also identified an area where the $\WHI$--$\Dem$ relation is affected by the
contamination of an intermediate velocity cloud, and masked the area
when studying the $\WHI$--$\Dem$ relation and $\gamma$-ray data analysis.
We also masked the Orion-Eridanus superbubble 
\citep[e.g.,][]{Ochsendorf2015}
because the CR and ISM properties inside the bubble
can be appreciably different from those in other areas (see Appendix~A for details).
To compensate the loss of photon statistics because of the mask,
we shifted the longitude of 
the southern ROI by $10\arcdeg$ 
toward the positive direction (i.e., away from the mask)
from that of \citet{FermiHI}.
In the \textit{Planck} dust maps, we identified several spots with high $R/\tau_{353}$ ratios and
high $R$. They are likely infrared sources and were removed from the dust maps
by filling them with the average of the
peripheral pixels (see Appendix~B). 
Finally, we examined the \textit{Planck} type~3 CO map 
that has the highest signal--to--noise ratio among three types of the map
\citep{Planck2014b} and confirmed that
there is no strong CO 2.6-mm emission in our ROI (see Appendix~C).
We identified a weak spot at ($l$, $b$) $\sim$ ($221\fdg4$, $45\fdg1$).
This spot can also be seen in the $R$ and $\tau_{353}$ maps and is therefore likely to be an infrared source.
Therefore, we removed the source from the dust maps
by filling the source area with the average of the peripheral pixels,
and do not consider CO-bright $\Htwo$ in constructing gas models hereafter.

The correlations between $\WHI$ and $R$, and those between $\WHI$ and $\tau_{353}$,
are shown in Figures~1 and 2, respectively, in which different colors represent different $T_\mathrm{d}$.
To match the resolution of the $\WHI$ map,
we smoothed the dust maps using a Gaussian kernel with a full width at half maximum (FWHM)
of $15\farcm4$.
In the northern region, we can confirm two trends of the dust--gas relation found by previous studies 
of the high-latitude sky described in Section~1:
(1) we observe in Figure~1(a) a rather good correlation between $\WHI$ and $R$
over a wide range of $T_\mathrm{d}$,
as shown by \citet{Planck2014a} that proposed $R$ as a good neutral gas tracer,\footnote{
They reported a good correlation up to column densities of (at least)
$5 \times 10^{20}~\mathrm{cm^{-2}}$
(Figure~20 of the reference), 
which corresponds to a $\WHI$ of $\mathrm{{\sim}280~K~km~s^{-1}}$.
}
and
(2) we observe in Figure~1(b) a stronger $T_\mathrm{d}$ dependence of the $\WHI$--$\tau_{353}$ relationship,
which \citet{Fukui2015} interpreted as being primarily due to optically thick $\HI$ in low-$T_\mathrm{d}$ 
areas.
In the southern region, while both $R$ and $\tau_{353}$ show a small $T_\mathrm{d}$ dependence
in the relation with $\WHI$,
a possible non-linear relationship between $\WHI$ and $\Dem$
(a break in the $\WHI$/$\Dem$ ratio) can be identified, particularly for the case of $\tau_{353}$.

\begin{figure}[ht!]
\figurenum{1}
\gridline{
\fig{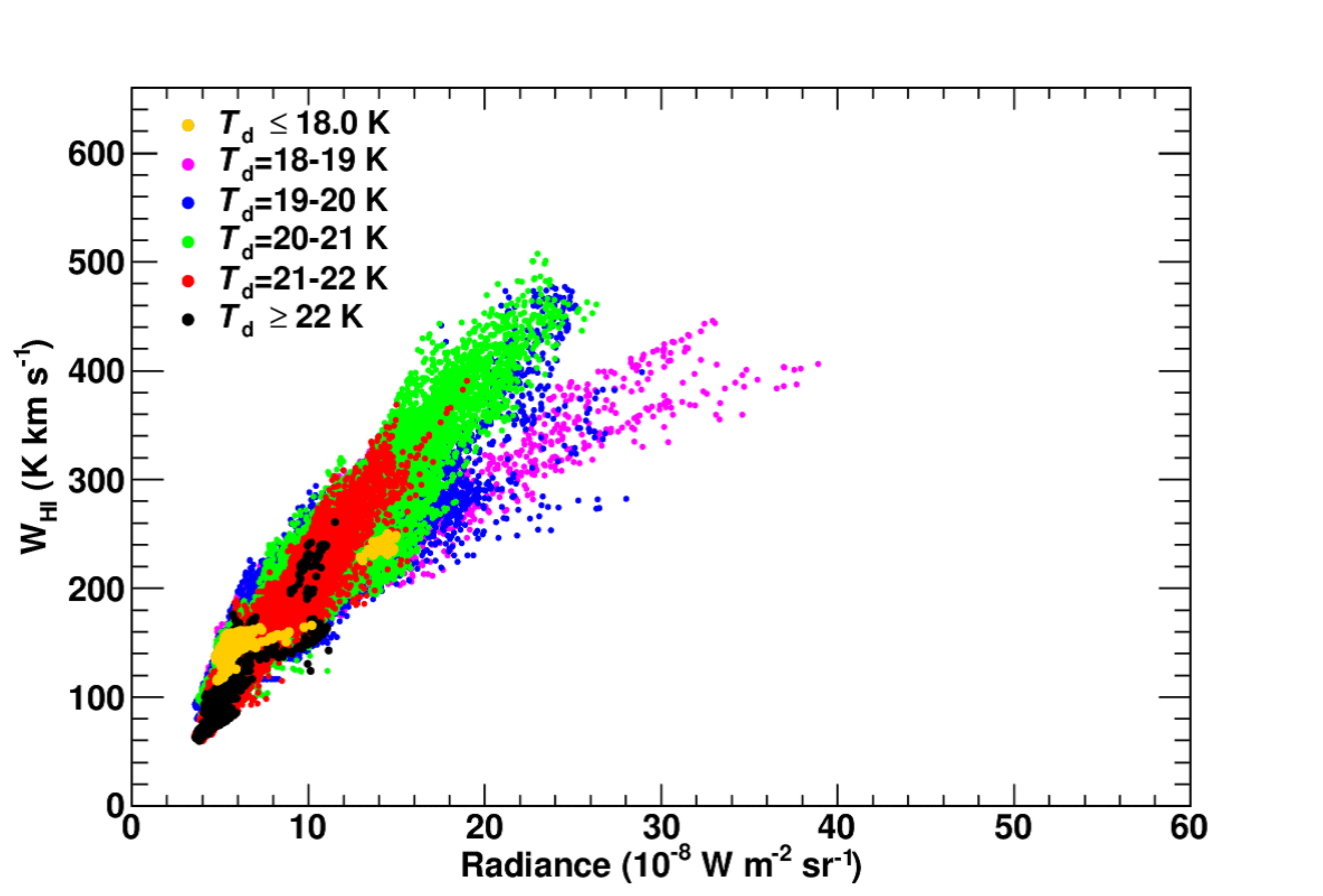}
{0.5\textwidth}{(a)}
\fig{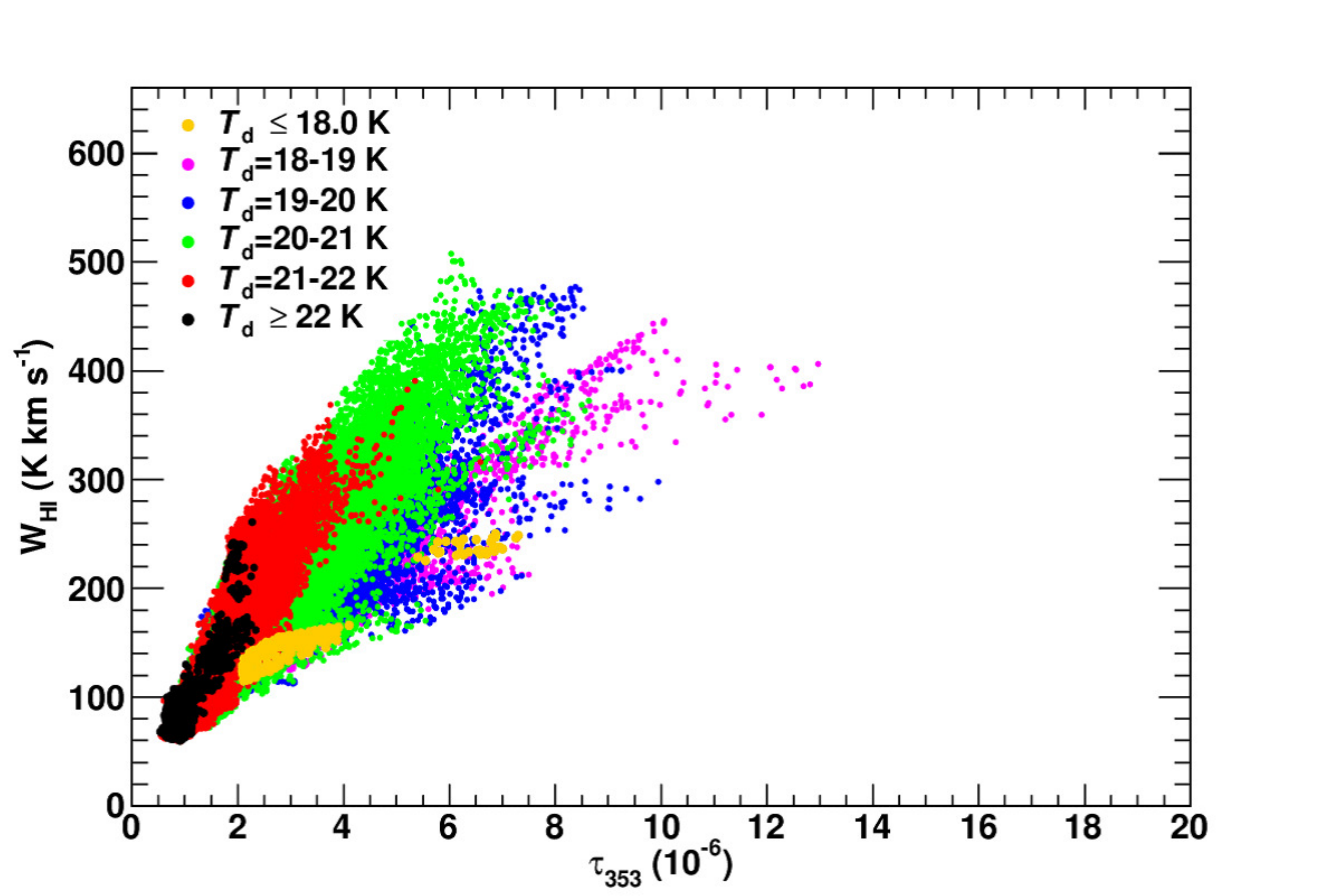}
{0.5\textwidth}{(b)}
}
\caption{
Correlations between $\WHI$ and the dust tracers in the northern region.
(a) Scatter plot of $\WHI$ versus $R$ and (b)
scatter plot of $\WHI$ versus $\tau_{353}$. 
In constructing these plots, the $\Dem$ ($R$ and $\tau_{353}$) maps were smoothed
using a Gaussian kernel with the FWHM of $15\farcm4$.
Each point represents one pixel in the underlying HEALPix map
(order 9; with a mean spacing of $6\farcm9$).
The $\NH \propto \Dem$ relations calibrated using data in the high $T_\mathrm{d}$ area
will be used to construct the initial $\NH$ template maps in
the $\gamma$-ray data analysis (see Section~2.2).
\label{fig:f1}
}
\end{figure}

\begin{figure}[ht!]
\figurenum{2}
\gridline{
\fig{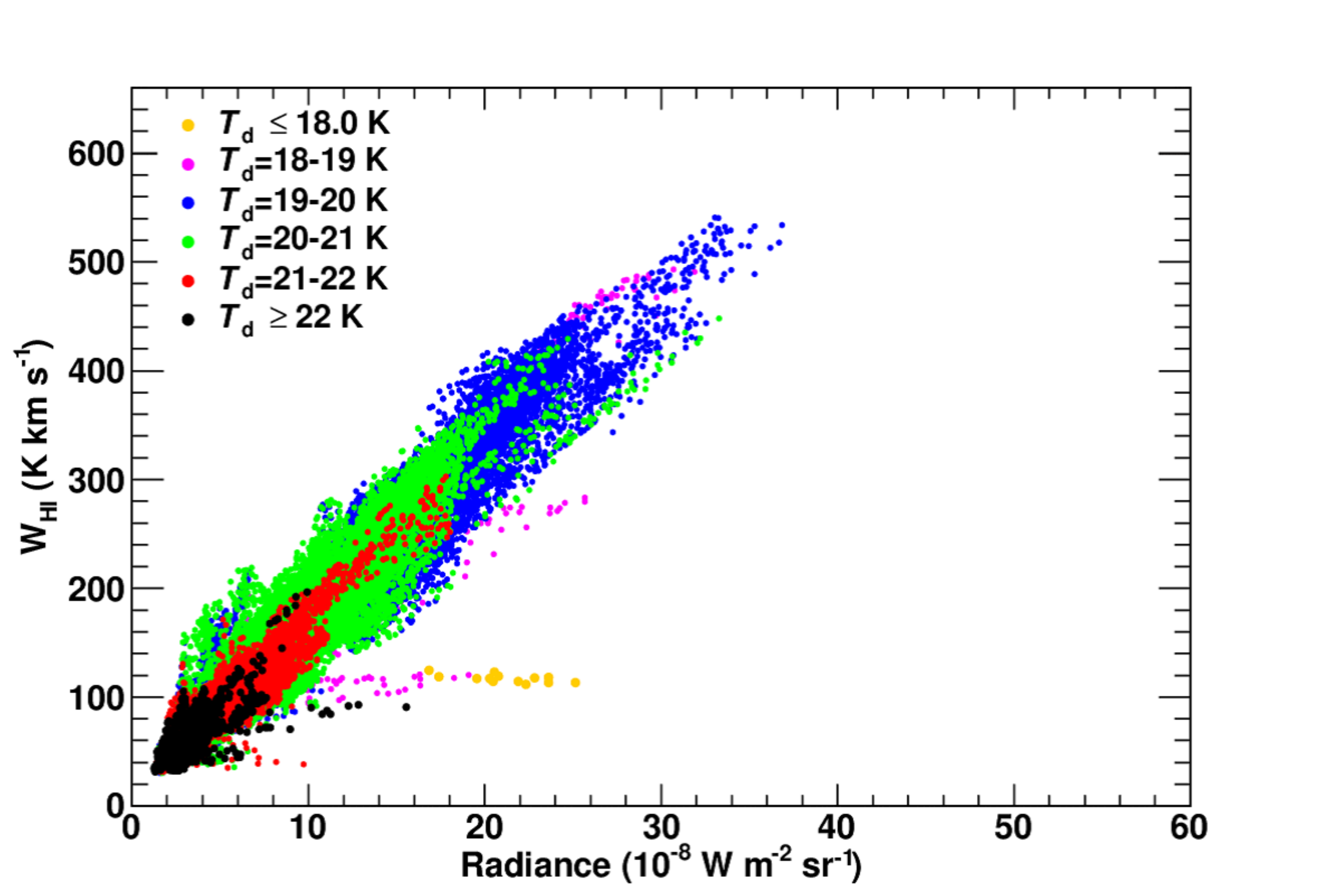}
{0.5\textwidth}{(a)}
\fig{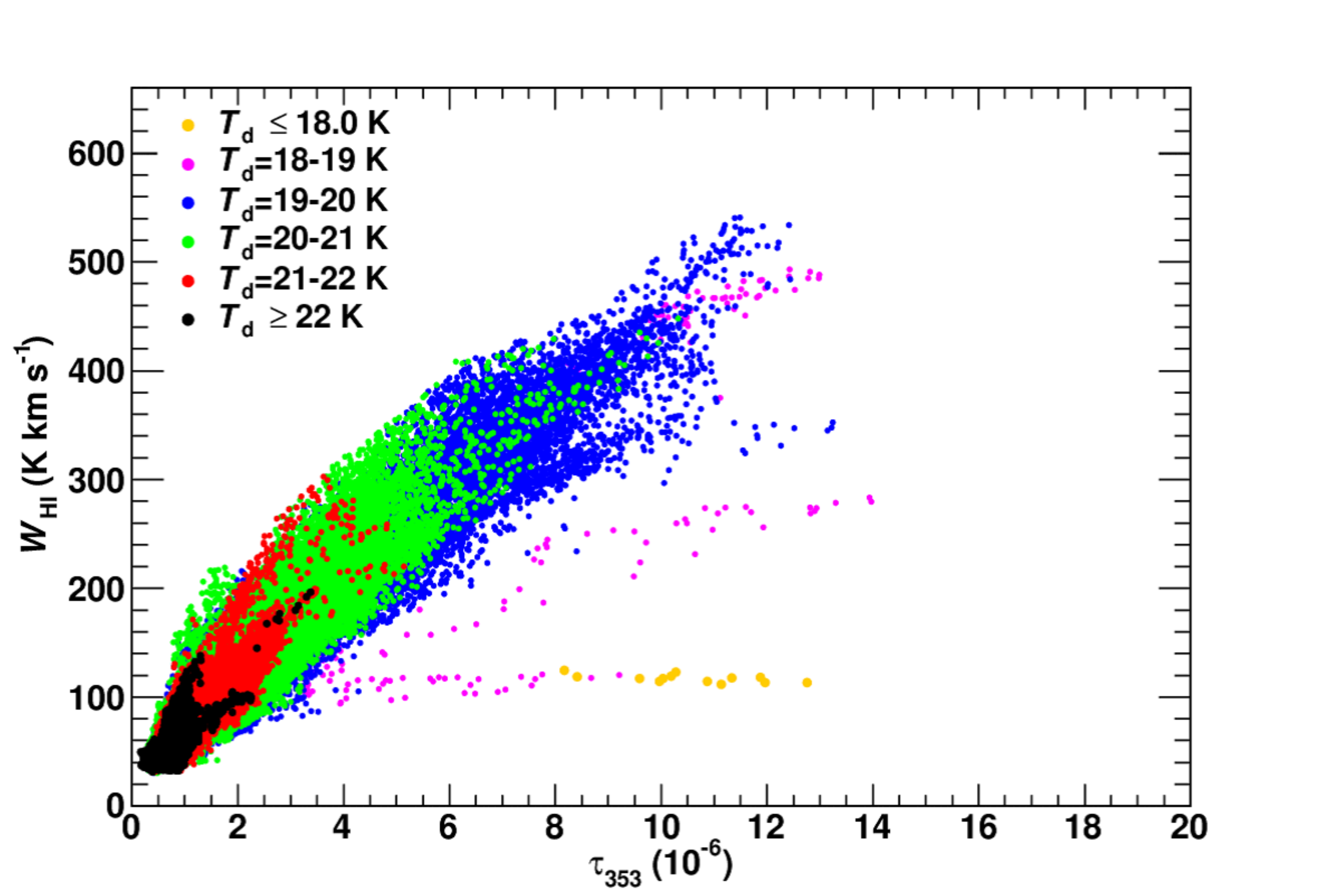}
{0.5\textwidth}{(b)}
}
\caption{
The same as Figure~1 but for the southern region.
$\NH \propto \Dem$ relations calibrated using data in the high $T_\mathrm{d}$ and low $\WHI$ area
will be used to construct the initial $\NH$ template maps in
the $\gamma$-ray data analysis (see Section~2.2).
\label{fig:f2}
}
\end{figure}

The correlation
between the dust tracers and $\WHI$ alone cannot distinguish
which variable ($R$ or $\tau_{353}$) is the better tracer of the total dust and total gas column densities,
and the correlation with the $\gamma$-ray intensity is crucial to reveal the true $\NH$ distribution.
Therefore, we prepared two types of $\NH$ model maps based on $R$ and $\tau_{353}$,
in addition to the previously described $\NH$ map based on $\WHI$,
and tested them against the \textit{Fermi}-LAT $\gamma$-ray data,
in which the northern and southern regions were analyzed separately.

\clearpage

\subsection{Construction of Gas Templates}

We converted the $\WHI$ map into  
an atomic hydrogen column density map
by assuming the optically thin approximation
($\NHIthin(\mathrm{cm^{-2}}) = 1.82 \times 10^{18} \cdot \WHI(\mathrm{K~km~s^{-1}})$).
The obtained $\NHIthin$ map
and the $T_\mathrm{d}$ map in our ROI are shown in panels (a) and (b), respectively,
in Figures~3 and 4.
Under the assumption that all neutral gas is atomic and $\HI$ is optically thin,
these $\NHIthin$ maps will be used as $\NH$ template maps
in the $\gamma$-ray data analysis (Section~3).

\begin{figure}[ht!]
\figurenum{3}
\gridline{
\fig{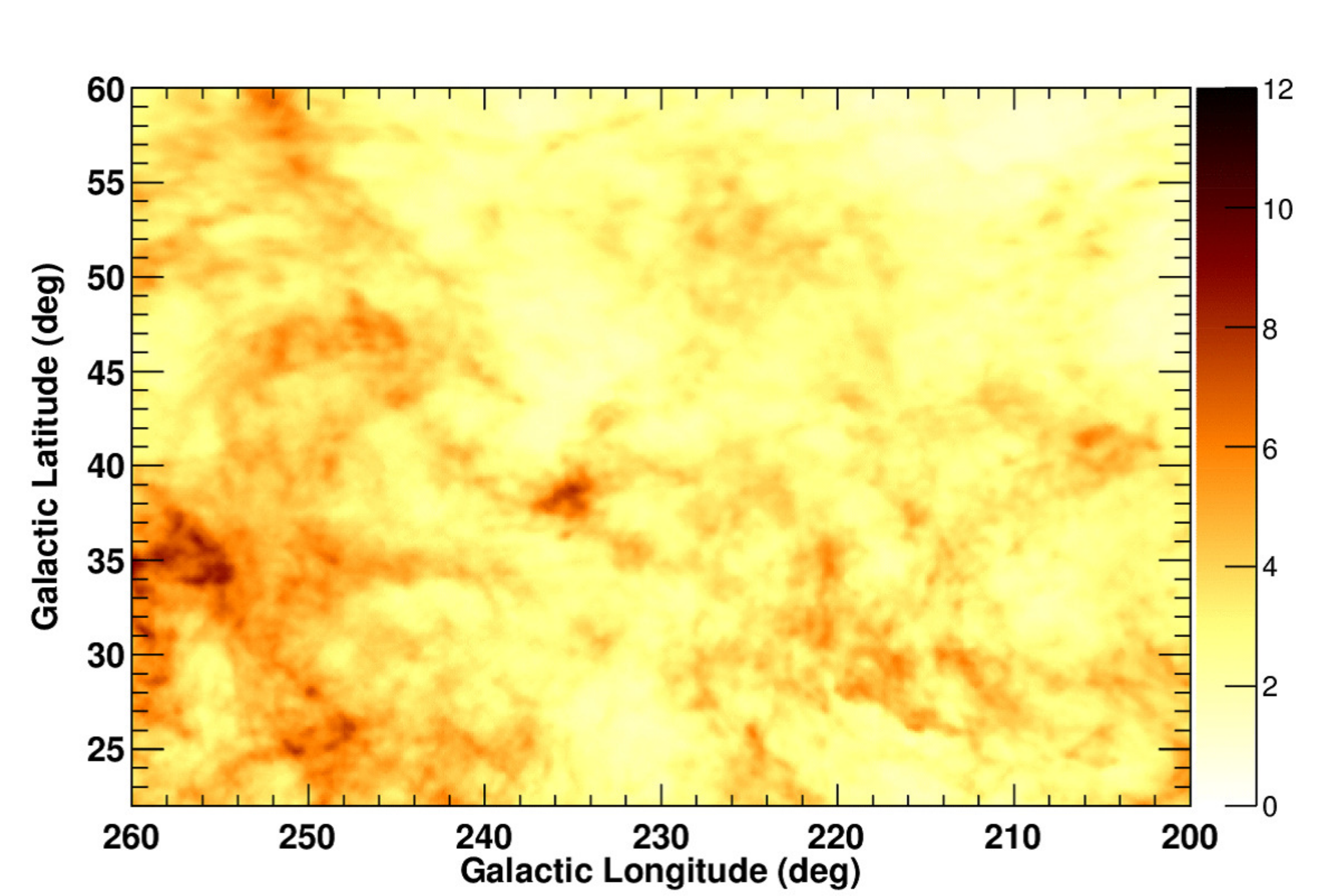}
{0.5\textwidth}{(a)}
\fig{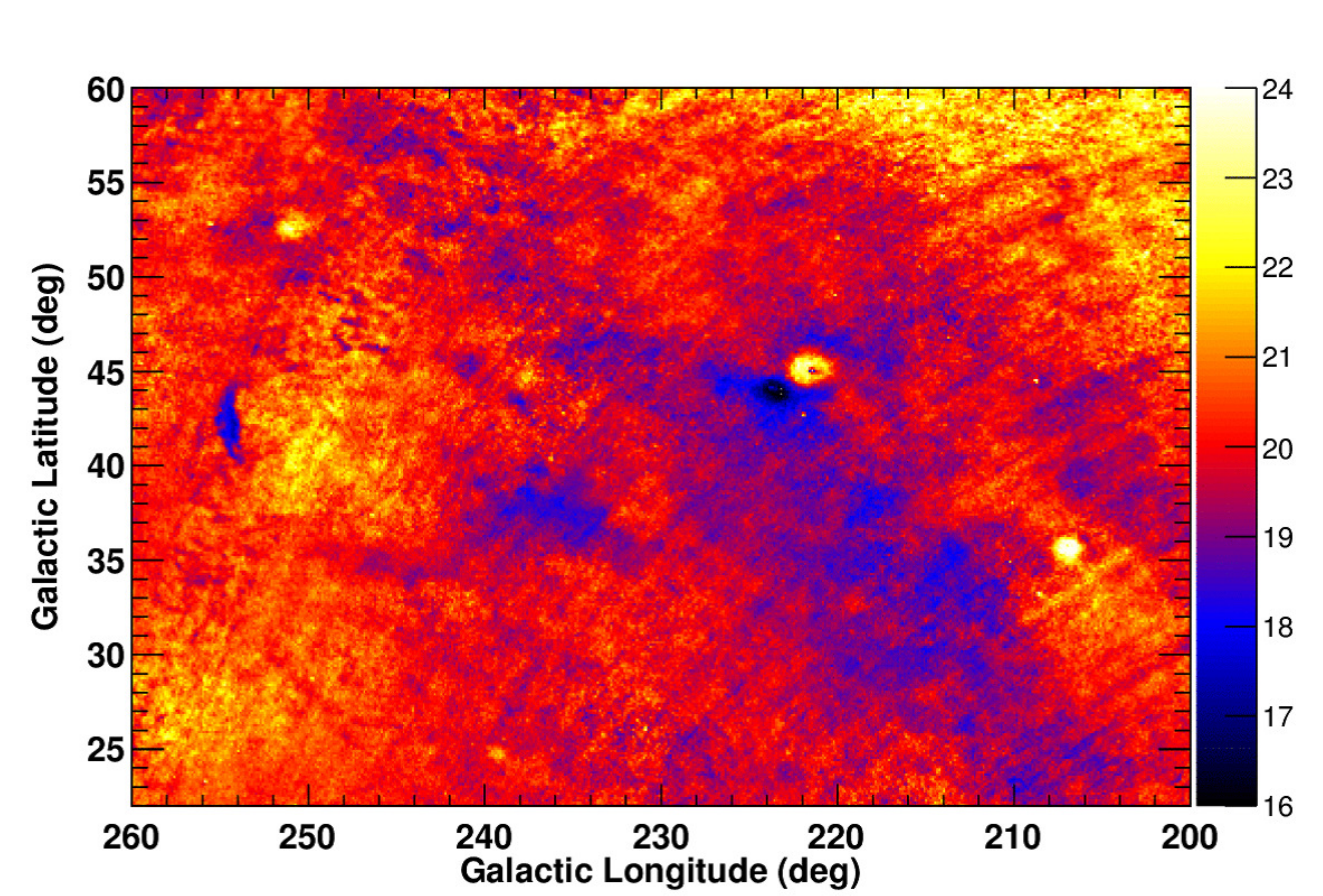}
{0.5\textwidth}{(b)}
}
\gridline{
\fig{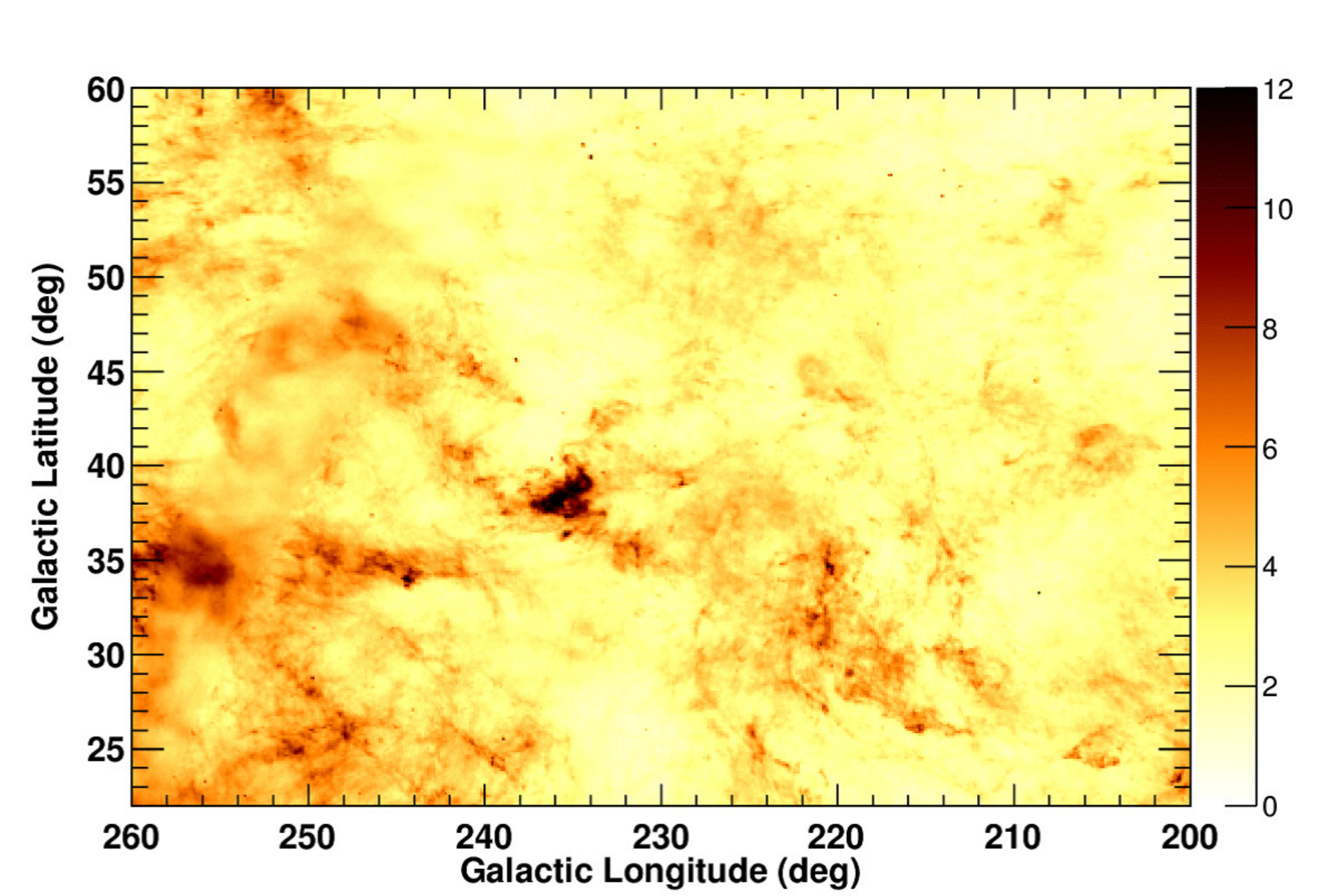}
{0.5\textwidth}{(c)}
\fig{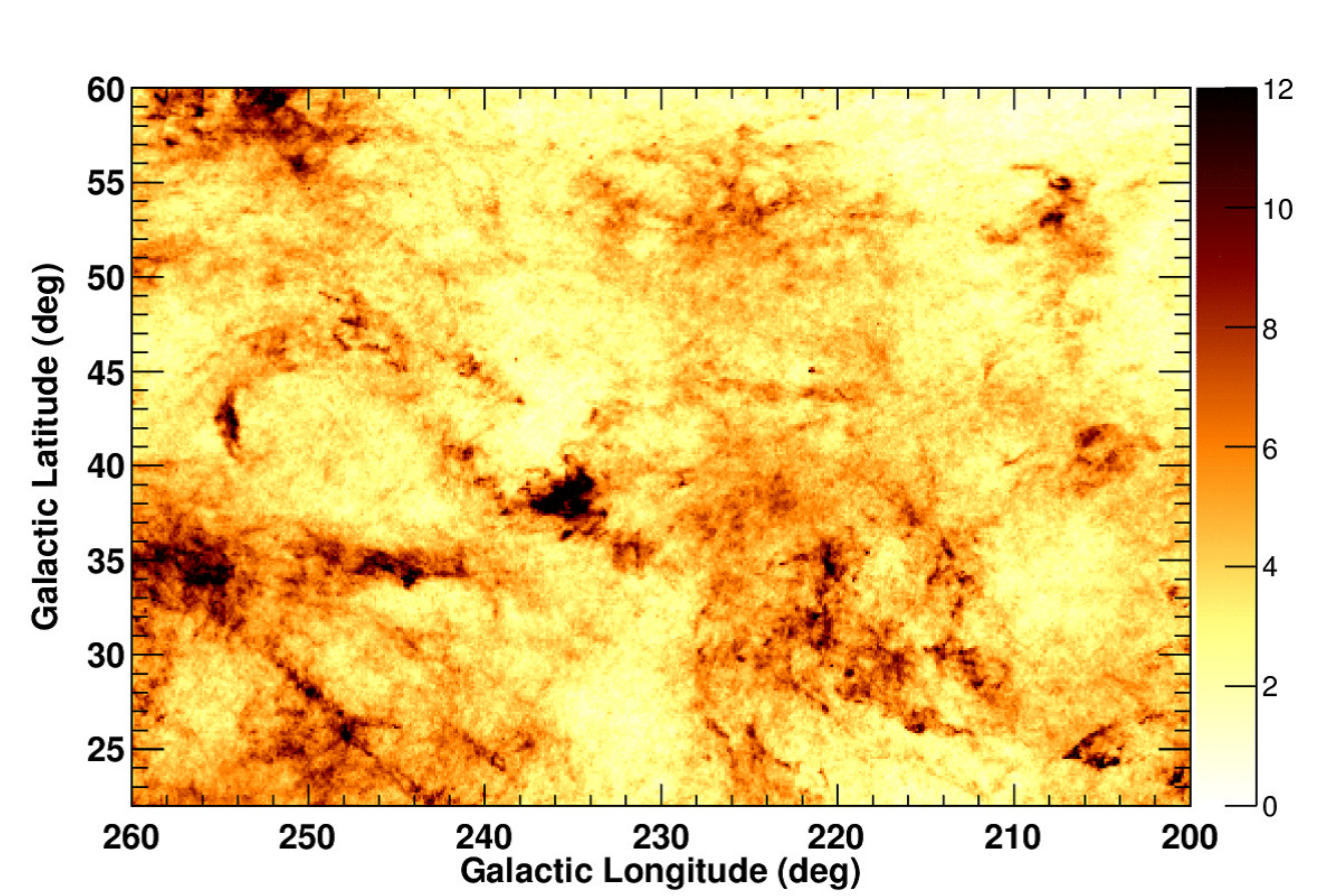}
{0.5\textwidth}{(d)}
}
\caption{
(a) The $\WHI$ map converted to $\NH$ 
under the assumption that all neutral gas is atomic and $\HI$ is optically thin
($\NH(\mathrm{cm^{-2}}) = 1.82 \times 10^{18} \cdot \WHI(\mathrm{K~km~s^{-1}})$);
(b) the $T_\mathrm{d}$ map (K);
(c) the $\NH$ template map proportional to $R$
($\NH(\mathrm{cm^{-2}}) = 38.4 \times 10^{26} \cdot R(\mathrm{W~m^{-2}~sr^{-1}})$);
and (d) the $\NH$ template map proportional to $\tau_{353}$
($\NH(\mathrm{cm^{-2}}) = 159 \times 10^{24} \cdot \tau_{353}$)
for the northern region.
The maps in panels (a), (c), and (d) are 
shown in units of $10^{20}~\mathrm{cm^{-2}}$.
\label{fig:f3}
}
\end{figure}

\clearpage

\begin{figure}[ht!]
\figurenum{4}
\gridline{
\fig{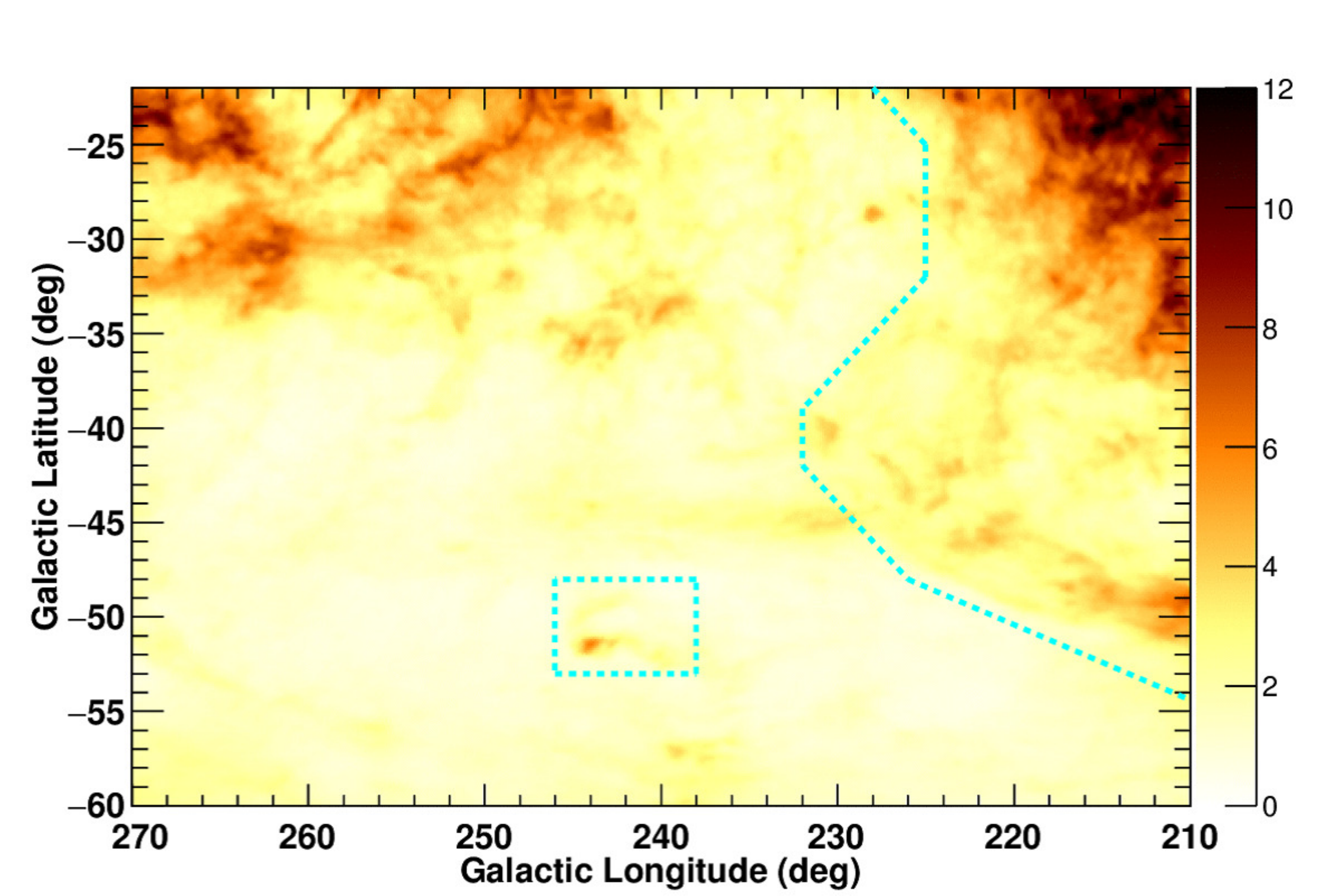}
{0.5\textwidth}{(a)}
\fig{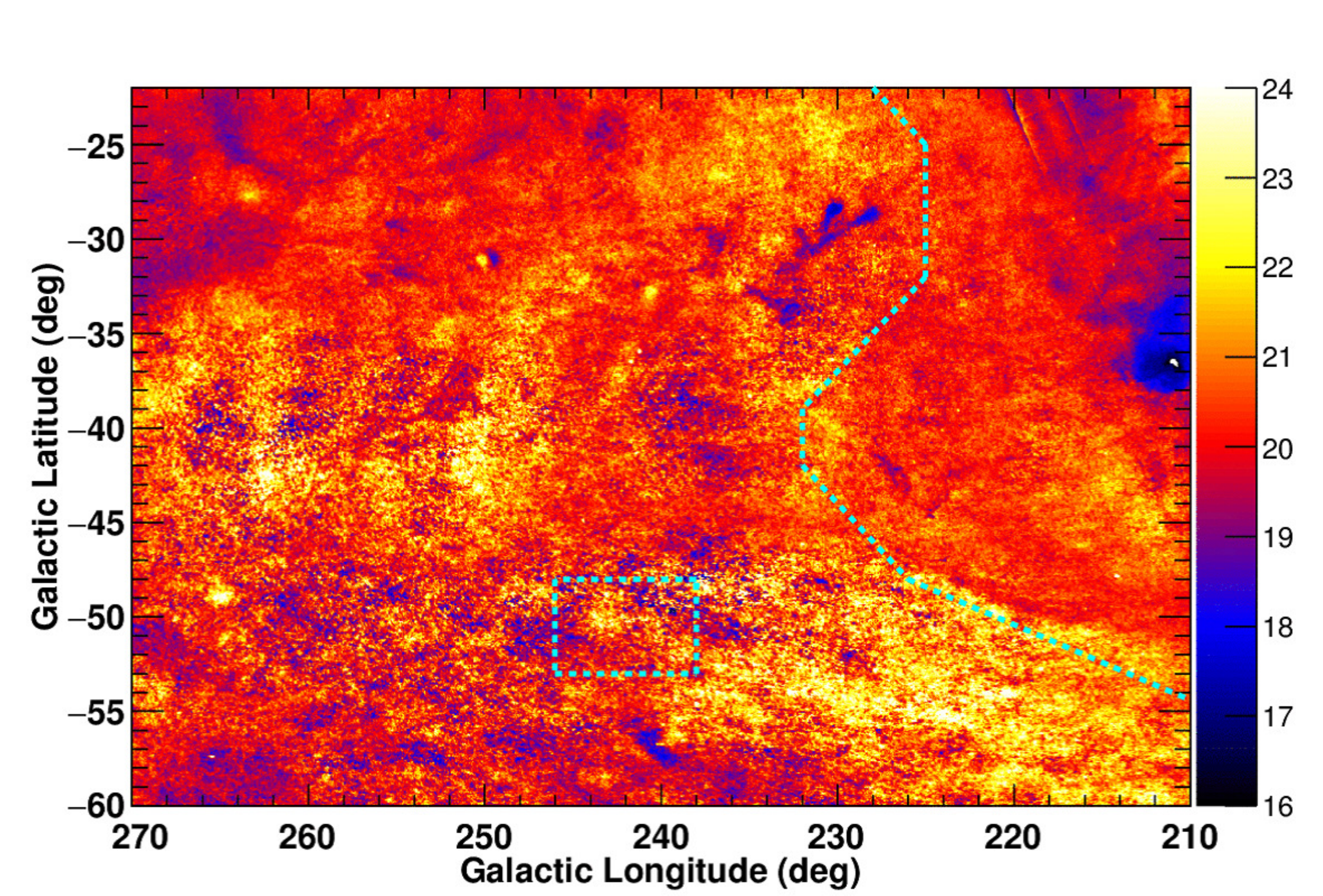}
{0.5\textwidth}{(b)}
}
\gridline{
\fig{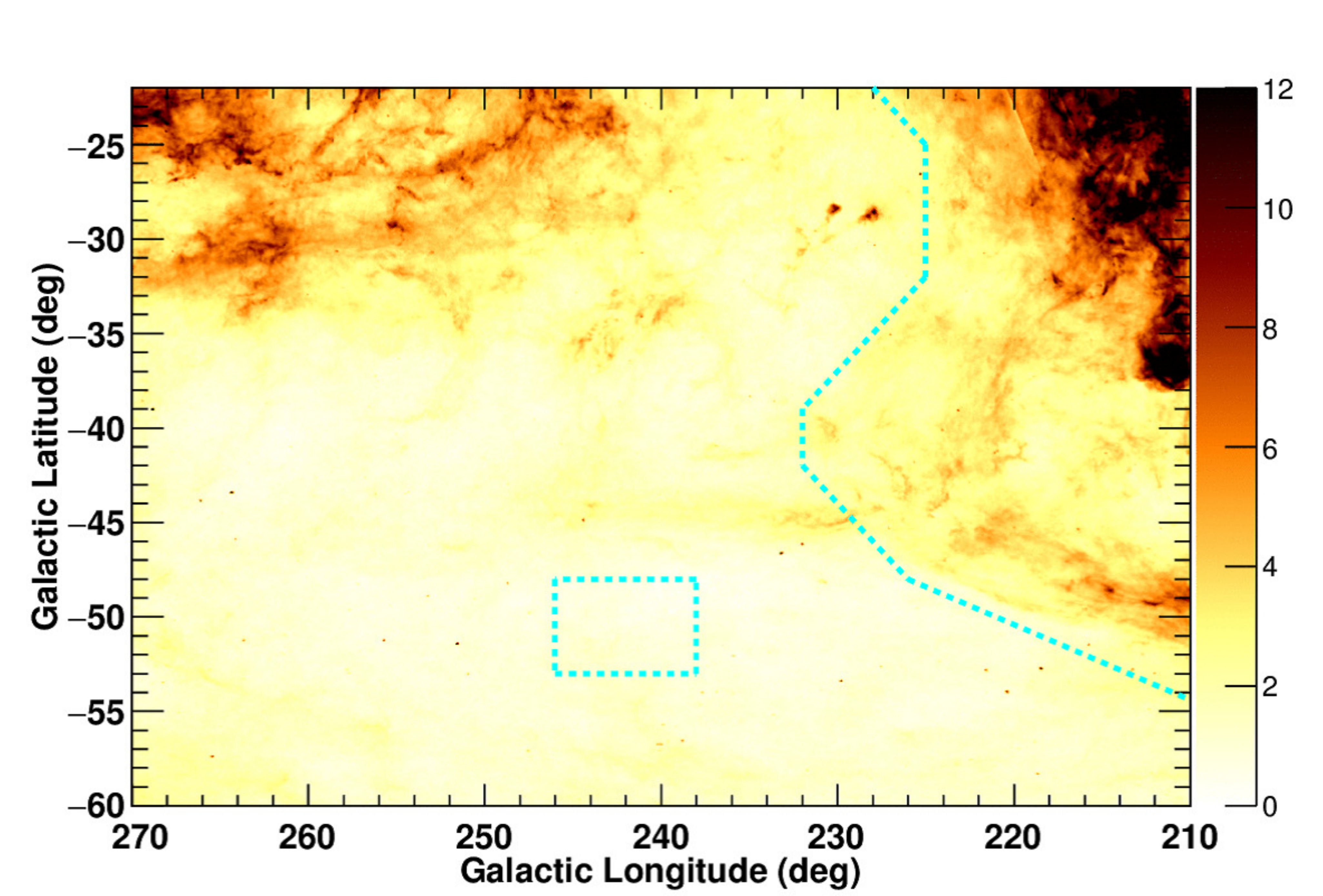}
{0.5\textwidth}{(c)}
\fig{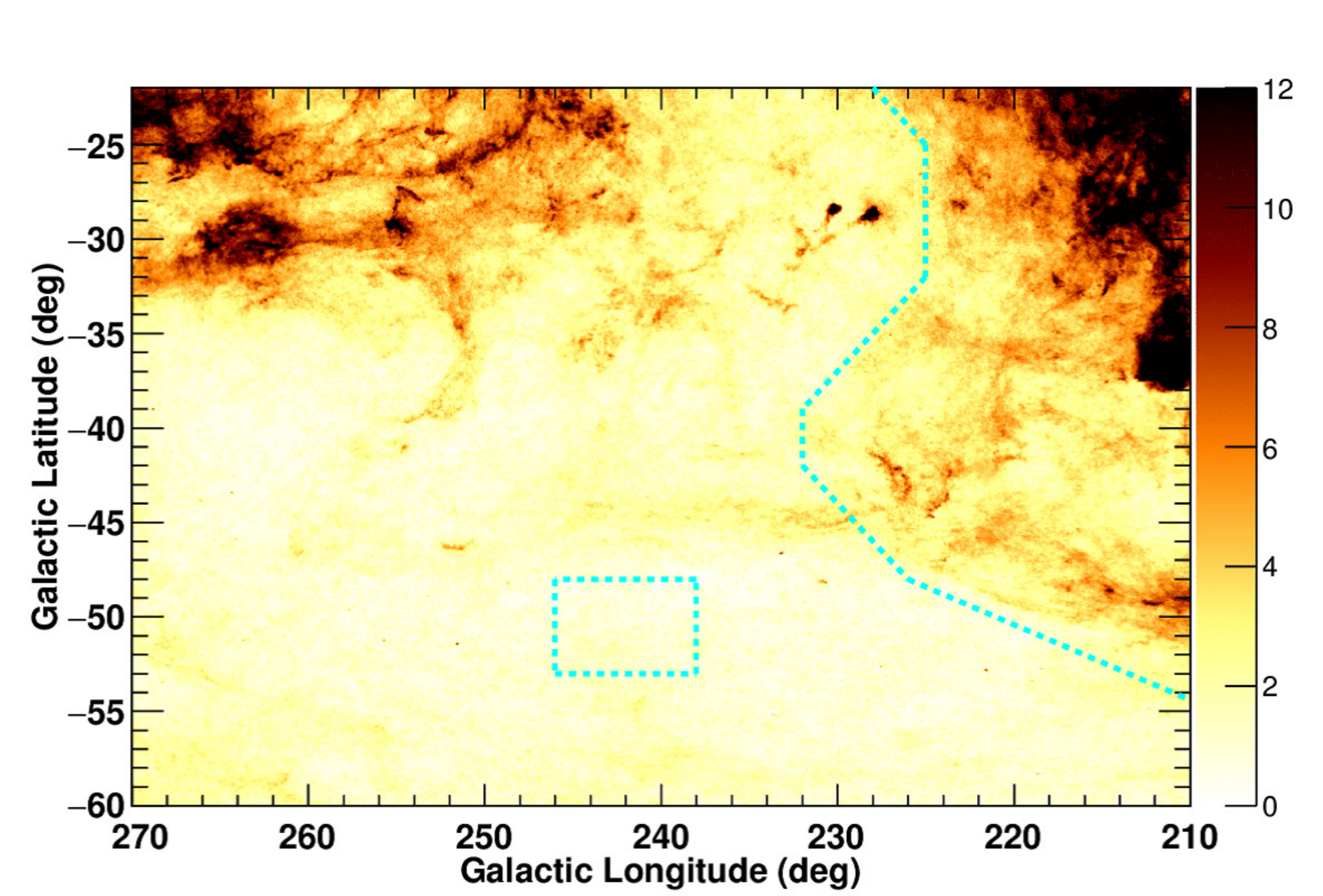}
{0.5\textwidth}{(d)}
}
\caption{
The same as Figure~3 but for the southern region. The region boundaries used for masking areas in the study of 
the $\WHI$--$\Dem$ relation and $\gamma$-ray data analysis are overlaid.
$\NH(\mathrm{cm^{-2}}) = 32.0 \times 10^{26} \cdot R(\mathrm{W~m^{-2}~sr^{-1}})$
and
$\NH(\mathrm{cm^{-2}}) = 122 \times 10^{24} \cdot \tau_{353}$
are used in panels (c) and (d), respectively.
One can see that areas of dense gas are away from the boundaries and therefore the spillover outside the masks is not severe.
\label{fig:f4}
}
\end{figure}

To construct the $\NH$ model maps based on the \textit{Planck} dust models for the northern region,
we assumed that $\NH$ and $\Dem$ ($R$ or $\tau_{353}$) are proportional
and that $\HI$ is optically thin, at least for the areas with high dust temperature ($T_\mathrm{d} \ge 21.0~\mathrm{K}$).
This assumption is based on the fact that high-$T_\mathrm{d}$ areas 
show a $\tau_{353}/\WHI$ ratio smaller than that in low-$T_\mathrm{d}$ areas 
and are expected to have relatively high gas temperatures, and are therefore
likely to be dominated by optically thin $\HI$.
The uniformity of the $\NH$/$\Dem$ ratio over the whole ROI will be tested in the 
$\gamma$-ray data analysis (Sections~4.1.2 and 4.2.2).
First, we made a least-squares fit to the 
dust--$\WHI$ relationship for $T_\mathrm{d} \ge 21.0~\mathrm{K}$
in Figure~1 using a linear function
with an intercept fixed at 0 because the zero-level of the dust emission
is already subtracted in the \textit{Planck} dust models \citep{Planck2014a}.
We obtained coefficients of 
$21.1 \times 10^{8}~\mathrm{K~km~s^{-1}~(W~m^{-2}~sr^{-1})^{-1}}$
and 
$87.2 \times 10^{6}~\mathrm{K~km~s^{-1}}$
for $R$ and $\tau_{353}$, respectively.
In these fits, $\WHI$ is treated as an independent variable because the signal-to-noise ratio is
much greater than that of $\Dem$.
Then, we converted $R$ (or $\tau_{353}$) into $\NH$ maps
using the obtained coefficients and multiplied by $1.82 \times 10^{18}~\mathrm{cm^{-2}~(K~km~s^{-1})^{-1}}$.
This gives a conversion factor for $R$ of $38.4 \times 10^{26}~\mathrm{cm^{-2}~(W~m^{-2}~sr^{-1})^{-1}}$,
and for $\tau_{353}$ of $159 \times 10^{24}~\mathrm{cm^{-2}}$.
The obtained $\NH$ template maps are shown in Figure~3 (panels (c) and (d), respectively).
By comparing these $\NH$ model maps
we can see that the map based on $\WHI$ shows the weakest contrast.
The $\tau_{353}$-based map predicts the strongest contrast
and approximately a factor of 2 higher $\NH$ in the dense clouds 
compared to the $\WHI$-based map, while the $R$-based map shows a moderate contrast.
Thus, they will give different contrasts in the predicted $\gamma$-ray map and can be tested
by the fit quality to the $\gamma$-ray data (Section~4).

For the southern region, we also assumed that
$\WHI$ and $\Dem$ are proportional 
and followed the same procedure as that for the northern region.
The obtained coefficients for the dust--$\WHI$ relationship are
$17.6 \times 10^{8}~\mathrm{K~km~s^{-1}~(W~m^{-2}~sr^{-1})^{-1}}$
and 
$66.9 \times 10^{6}~\mathrm{K~km~s^{-1}}$,
and the conversion factors to construct $\NH$ model are
$32.0 \times 10^{26}~\mathrm{cm^{-2}~(W~m^{-2}~sr^{-1})^{-1}}$ and
$122 \times 10^{24}~\mathrm{cm^{-2}}$
for $R$ and $\tau_{353}$, respectively.
The obtained $\NH$ template maps are shown in Figure~4;
the contrast of the gas density is weakest for the map based on
$\WHI$,
while that based on $\tau_{353}$ is the strongest.

Another possible source of diffuse $\gamma$-ray emissions is CR interactions with ionized gas.
To estimate this contribution, we referred to
\citet{FermiHI2} and used the free-free intensity map at a frequency of 22.7~GHz
extracted from nine years of \textit{WMAP} observations, namely 
wmap\_K\_mem\_freefree\_9yr\_v5.fits \citep{Bennett2013},
as a template for the $\gamma$-ray emission
correlated with the ionized hydrogen ($\HII$).
We used the scaling factor adopted by \citet{FermiHI2},
$1.3 \times 10^{20}~\mathrm{cm^{-2}~mK^{-1}}$, and constructed 
an $\HII$ column density ($\NHII$) model map. We confirmed that the average column density is only a few \% of the neutral gas column density
estimated by $\WHI$ shown in Figure~3(a) and 4(a).
Therefore the contribution of the ionized gas to $\gamma$-ray emission is small,
and the $\NHII$-related term will be fixed in the $\gamma$-ray data analysis (Section~3).

\clearpage

\section{Gamma-ray Data and Modeling}


\subsection{Gamma-ray Observations and Data Selection}

The LAT on board the \textit{Fermi Gamma-ray Space Telescope}, launched in 
June 2008,
is a pair-tracking $\gamma$-ray telescope, detecting photons in the energy range from {$\sim$}20~MeV to more than
300~GeV. Details of the LAT instrument and the pre-launch performance expectations
can be found in \citet{Atwood2009}, and the on-orbit calibration is described in \citet{Abdo2009}.
Past studies of the Galactic diffuse emissions by \textit{Fermi}-LAT
can be found in, e.g., \citet{FermiPaper2}, and \citet{FermiHI2}.

Routine science operations with the LAT started on 
August 4, 2008.
We used a data taking interval
from August 4, 2008 to August 1, 2016
(i.e., eight years), to study diffuse $\gamma$-rays in our ROI
($200\arcdeg \le l \le 260\arcdeg$ and $210\arcdeg \le l \le 270\arcdeg$ for the
northern and southern regions, respectively, with $22\arcdeg \le |b| \le 60\arcdeg$ for both). 
During most of this time,
the LAT was operated in sky survey mode, obtaining complete sky coverage every
two orbits (which corresponds to ${\sim}3$~h) and a relatively uniform exposure.
Because the diffuse $\gamma$-ray emission from the ISM is greatly extended and the intensity is rather weak
at high latitudes, having a clean (low background) sample of photons is crucial. Therefore, we used the
latest release of the Pass~8 data (P8R3) recently made public by the LAT collaboration.
The previous data set (P8R2)
is known to suffer from a residual
background with a peak along the ecliptic plane which is inside our ROI in the northern region.
We used the standard LAT analysis software, \textit{Fermi} Science Tools
\footnote{\url{http://fermi.gsfc.nasa.gov/ssc/data/analysis/software/}}
version v10r00p05
to select events above 100~MeV
\footnote{In Section~4.2.1, we also used data below 100~MeV as a preparatory stage of the analysis in order to dissolve a coupling 
among fit components.} 
because good angular resolution is essential for examining the correlation between
$\gamma$-rays and gas distribution, and selected
events of the SOURCE class.
The instrument response function (IRF) that matches our data set and event selection, P8R3\_SOURCE\_V2,
was used in the following analysis.
We also analyzed data using the cleanest ULTRACLEANVETO class and corresponding IRF and found that
a decrease of the isotropic component (see Section~3.2), which includes the residual background, was marginal (within $\le 10\%$). 
Therefore, we keep using the SOURCE class to maximize the photon statistics.
In addition, we required that the reconstructed zenith angles of the
arrival direction of the photons be less than $100\arcdeg$ and $90\arcdeg$ for energies above and below 200~MeV,
respectively, to reduce contamination by photons from Earth's atmosphere.
To accommodate the rather poor angular resolution at low energy, below 200~MeV,
we used events and the responses of point-spread function (PSF) event types 2 and 3;
meanwhile, above 200~MeV, we did not apply selections based on PSF event types and 
used events and the responses of Front + Back.
As described in Section~3.3, we carried out a bin-by-bin likelihood fitting, and data below and above 200 MeV are analyzed separately.
We stopped at 25.6~GeV because of poor photon statistics above that energy.
We used {\tt gtselect} command to apply the selections described above.

In addition, we excluded periods of time during which the LAT
detected bright $\gamma$-ray bursts or solar flares (by using {\tt gtmktime} command); 
the integrated period of time excluded in this procedure was negligible (less than 1\%) compared to the total observation time.
The data count maps in the northern and southern regions of our ROI
are given in Figure~5.

\begin{figure}[ht!]
\figurenum{5}
\gridline{
\fig{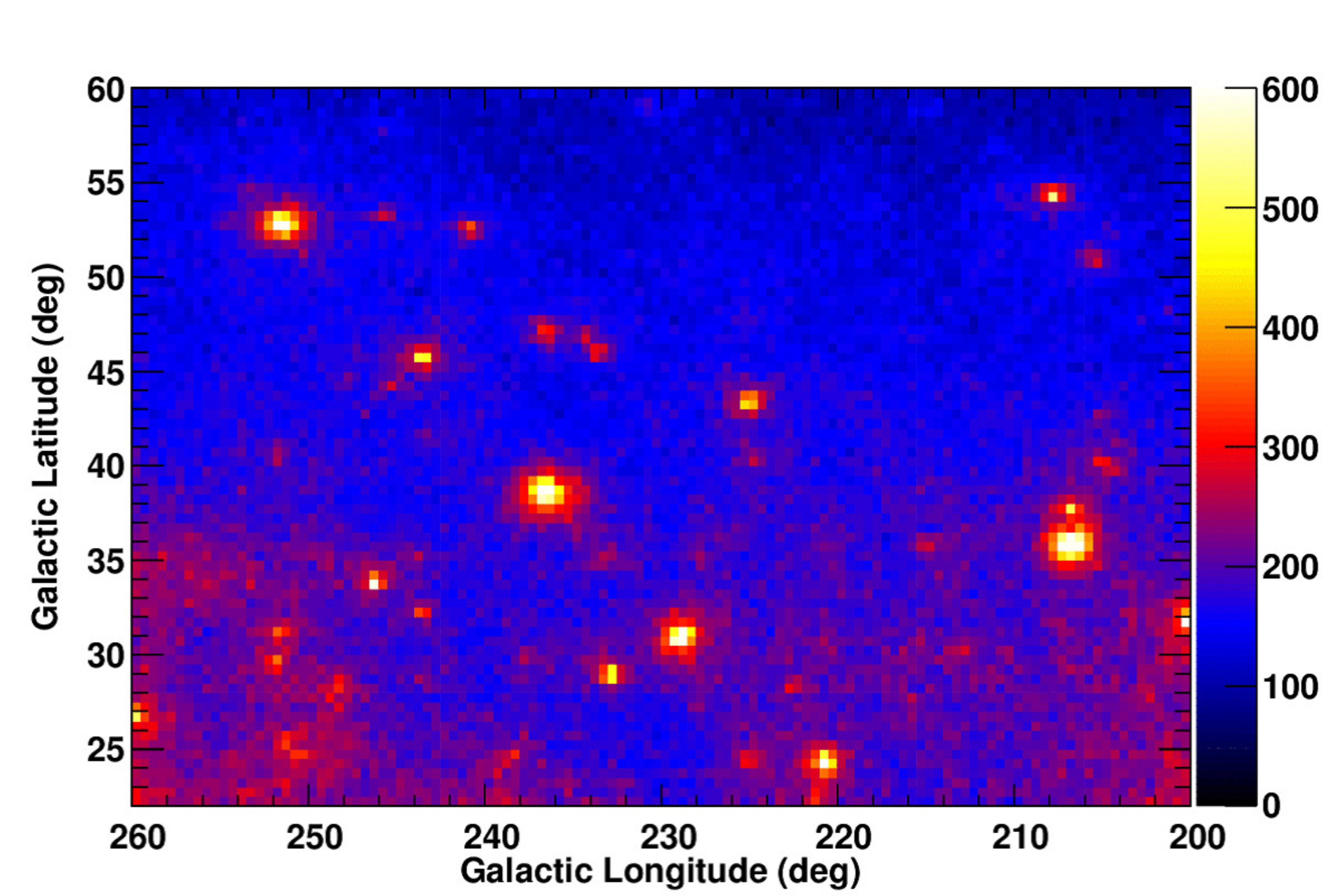}
{0.5\textwidth}{(a)}
\fig{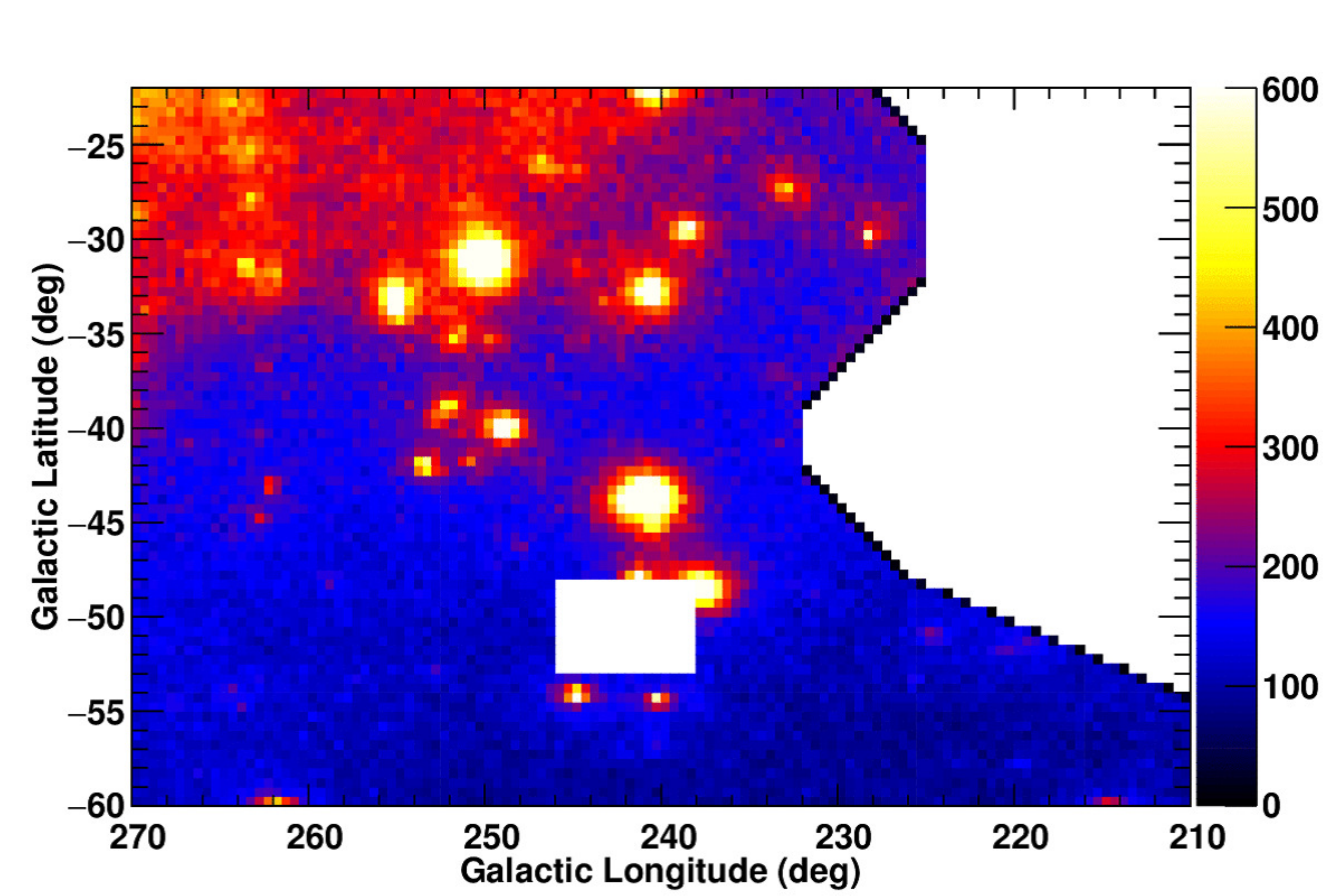}
{0.5\textwidth}{(b)}
}
\caption{
\textit{Fermi}-LAT $\gamma$-ray count maps ($E \ge \mathrm{100~MeV}$) of the regions we analyzed.
The maps are in a Cartesian projection, and the northern and southern regions are shown in 
panels (a) and (b), respectively.
Although the fit has been performed
in $0\fdg25 \times 0\fdg25$ bins
above 400~MeV, all the data
have been re-binned in $0\fdg5 \times 0\fdg5$ pixels for display.
\label{fig:f5}
}
\end{figure}

\clearpage

\subsection{Model to Represent the Gamma-ray Emission}

We modeled the $\gamma$-ray emission above 100~MeV observed by \textit{Fermi}-LAT
as a linear combination of the $\NH$ model map constructed from 
the HI4PI survey data ($\WHI$) or the \textit{Planck} dust map ($R$ or $\tau_{353}$) (see Section~2.2),
the template map for $\NHII$,
the IC emission, the isotropic component, the emission from the Sun and Moon \citep[e.g.][]{FermiSun,FermiMoon}, and the $\gamma$-ray point sources.
The use of the gas column density map as a template is based on the assumption that 
$\gamma$-rays are generated
via interactions between CRs and the ISM gas and that the CR intensity does not vary significantly
over the scale of the interstellar complexes in this study.
This assumption is simple
but very
plausible, particularly in high-latitude regions (where most of the gas is expected to be local),
such as the one studied here. 
We started with a single $\NH$ map (in Sections~4.1.1 and 4.2.1)
and then tested
multiple $\NH$ maps based on the \textit{Planck} dust model 
maps
(in Sections~4.1.2 and 4.2.2).
To model the 
$\gamma$-rays
produced via IC scattering, we used GALPROP 
\footnote{\url{http://galprop.stanford.edu}}
\citep[e.g.,][]{Galprop1,Galprop2},
a numerical code that solves the CR transport equation within the Galaxy and predicts
the $\gamma$-ray emission produced via the interactions of CRs with interstellar gas
and low-energy photons.
The IC emission is calculated from the distribution
of the propagated electrons and the interstellar radiation field developed by \citet{Porter2008}.
Here, we adopted the IC model map produced in the GALPROP
run 54\_Yusifov\_z4kpc\_R30kpc\_Ts150K\_EBV2mag,\footnote{The file is available at \url{https://galprop.stanford.edu/PaperIISuppMaterial/}}
which was developed by \citet{FermiPaper2} and 
has been used in other LAT collaboration publications\footnote{\url{https://www-glast.stanford.edu/cgi-bin/pubpub}}
such as \citet{Planck2015}, and \citet{Remy2017}. To model
individual $\gamma$-ray sources, we referred to the 
preliminary \textit{Fermi}-LAT eight-year point source list (FL8Y),\footnote{\url{https://fermi.gsfc.nasa.gov/ssc/data/access/lat/fl8y/}}
which is based on the first eight years of the science phase of the mission and 
includes more than 5000 sources detected at a significance of {$\ge$}4$\sigma$.
We note that the list is built using the same lowest energy threshold (0.1~GeV) as the present study.
For our analysis, we considered 151 and 142 FL8Y sources (detected at a significance of {$\ge$}4$\sigma$)
in the northern and southern regions, respectively, and 27 bright sources just outside the ROIs
({$\ge$}50$\sigma$ within $10\arcdeg$ and {$\ge$}20$\sigma$ within $1\arcdeg$)
to take into account possible contaminations from their emission
caused by overlap because of the breadth of the PSF.
We modeled a (quasi-) isotropic component
to represent the contributions to our ROI due to extragalactic diffuse emission
and the residual instrumental background from misclassified CR interactions
in the LAT detector by the uniform emission in our ROI.
A template of the ionized gas was constructed based on 
the \textit{WMAP} free-free intensity map as described in Section~2.2.
We included the templates for the $\gamma$-ray emission from the Sun and Moon 
that were used in the initial sky models for the fourth LAT source catalog (4FGL) \citep{Fermi4FGL}, 
which is based on the first eight years of LAT observations.
Specifically, we used template\_SunDiskMoonv3r2\_8years\_zmax105.fits and
template\_SUNICv2\_8years\_zmax105.fits.\footnote{These files will be made available at \url{http://www-glast.stanford.edu/pub_data}}

Then, the $\gamma$-ray intensity $I_{\gamma}(l, b, E)~\mathrm{(ph~s^{-1}~cm^{-2}~sr^{-1}~MeV^{-1})}$
can be modeled as

\begin{eqnarray}
I_{\gamma}(l, b, E) & = & q_{\gamma}(E) \left[c_{1}(E) \cdot \NH(l, b) + \NHII(l, b) \right] + c_{2}(E) \cdot I_\mathrm{IC}(l, b, E) \nonumber \\
& & + I_\mathrm{iso}(E) + c_{3}(E) \cdot I_\mathrm{SM}(l, b, E) +
\sum_{j} \mathrm{P}_{j}(l, b, E)~~,
\end{eqnarray}
where $\NH(l, b)$ is the total neutral gas column density model 
(based on $\WHI$, $R$, or $\tau_{353}$)
in $\mathrm{cm^{-2}}$,
$\NHII(l, b)$ is the gas column density model from ionized gas in $\mathrm{cm^{-2}}$,
$q_{\gamma}(E)$ ($\mathrm{ph~s^{-1}~sr^{-1}~MeV^{-1}}$) is the model of the differential $\gamma$-ray yield 
or $\gamma$-ray emissivity per H atom,
$I_\mathrm{IC}(l, b, E)$, $I_\mathrm{iso}(E)$, and $I_\mathrm{SM}(l, b, E)$ 
are the IC model, (quasi-)isotropic background intensities, and model of the emission from the Sun and Moon
(each in units of $\mathrm{ph~s^{-1}~cm^{-2}~sr^{-1}~MeV^{-1}}$), respectively, and
$\mathrm{P}_{j}(l, b, E)$ represents the point source contributions.
We used the $\gamma$-ray emissivity model $q_{\gamma}(E)$ of the local interstellar spectrum
(LIS) of the CRs (protons and electrons) adopted in \citet{FermiHI}, specifically the model curve for the
nuclear enhancement factor $\epsilon_\mathrm{M}$ 
(a scale factor to take into account the effect of heavy nuclei in both CRs and the target matter)
of 1.84 \citep{Mori2009}.
In other words, $q_{\gamma}(E)$ is decomposed as $1.84 \cdot q_{\gamma}(\mathrm{pp})+q_{\gamma}(\mathrm{brems})$
where $q_{\gamma}(\mathrm{pp})$ and $q_{\gamma}(\mathrm{brems})$ are the $\gamma$-ray emissivity models due to the
proton--proton interaction and electron bremsstrahlung, respectively.
To accommodate the uncertainties in the LIS and $\epsilon_\mathrm{M}$, 
we included 
a scale factor
($c_{1}(E)$ in Equation~(1)) as a
free parameter.
This parameter is 1.0 if the measured $\gamma$-ray emissivity agrees with the LIS and $\epsilon_\mathrm{M}$ we adopted.
Because the estimated ionized gas column density is small (at the maximum ${\sim}1 \times 10^{20}~\mathrm{cm^{-2}}$),
we fixed the scale factor at 1.0 for the $\NHII(l, b)$ template
to obtain a stable fitting.
The IC emission model is also uncertain, and we included another
scale factor ($c_{2}(E)$ in Equation~(1)) as a free parameter.
The extragalactic diffuse emission and the residual background are modeled by the
isotropic term, $I_\mathrm{iso}$, as a uniform template.
The intensity of this component could exhibit large-scale fluctuations due to the uncertainty of the LAT
response and/or possible non-uniformity of the background.
Therefore we include the isotropic term in each energy band rather than 
adopt the template provided by the LAT collaboration\footnote{\url{https://fermi.gsfc.nasa.gov/ssc/data/access/lat/BackgroundModels.html}}
nor determine it by the fit outside of our ROI.
We note that the isotropic term is a dominant component (see Section~4); therefore we should be able to constrain its contribution
using the data in our ROI.
The scale factor for $I_\mathrm{SM}(l, b, E)$, $c_{3}(E)$, was taken to be a free parameter in the northern region
where the ecliptic circle goes through our ROI, even though it was fixed to 1.0 in the southern region.
The point source contributions were also taken 
to be free parameters
as a function of energy. 
The positions of the sources were fixed to the values in FL8Y.
To model the contamination from outside the ROI,
we used the $\NH$, $I_\mathrm{IC}$, and $I_\mathrm{SM}$ maps
including peripheral regions.

Among the model components described above, the $\gamma$-ray emission model from Sun and Moon
was adopted from the work for the 4FGL while it was in the development phase. The model
will be outdated when the work to construct the 4FGL is complete.
However, we confirmed that the choice of the Sun and Moon template does not affect the fit for the neutral gas component
as described in Sections~4.1.1 and 4.1.2. Because the IC model has uncertainty,
we also examined the effect as described in Sections~4.1.2 and 4.2.2.

\subsection{Model Fitting Procedure}

We divided the $\gamma$-ray data from 0.1 to 25.6~GeV into eight energy ranges
using logarithmically equally spaced energy bands,
and each band was divided into four sub bins.
Then, we fit Equation~(1) to data in each energy band in
$0\fdg5 \times 0\fdg5$ bins below 400~MeV (where the PSF (68\% contaminant radius) is $\ge 2\arcdeg$) and
$0\fdg25 \times 0\fdg25$ bins above 400~MeV
using the binned maximum-likelihood method with Poisson statistics implemented in the Science Tools.
In each narrow energy band, $c_{1}$, $c_{2}$, and $c_{3}$ were modeled as energy-independent normalization factors.
Because $I_\mathrm{iso}(E)$ is the most dominant component over the entire energy range,
it was modeled via a power-law function 
with both photon index and normalization as free parameters in each energy band.
$\mathrm{P}_{j}(l, b, E)$ were modeled via separate power-law functions in each energy band
with only the normalization allowed to vary; the photon index was fixed at 2.2 as a representative value
of the high-latitude LAT sources \citep[FSRQ and BL Lac;][]{Fermi3LAC}.

When modeling the point sources, we iteratively included them 
in several groups at a time in order of decreasing significance.
First, we included and fitted the 
brightest sources detected in FL8Y at more than 35$\sigma$;
then, we added and fit a second group 
detected at 20--35$\sigma$,
freezing the already included source parameters.
In this way, we worked down to the sources detected at more than 4$\sigma$ in FL8Y.
For each step, the parameters of the diffuse emission model ($c_{1}$, $c_{2}$, $c_{3}$, and $I_\mathrm{iso}$)
were always left free to vary.
After we reached the least-significant sources (more than 4$\sigma$) we went back
to the brightest ones (more than 35$\sigma$ significance),
and let them and diffuse emission models free to vary while parameters of other sources are kept fixed
to those already determined.
We repeated the process until the increment of the log-likelihoods, $\ln{L}$
\footnote{
$L$ is conventionally calculated as $\ln{L}=\sum_{i}n_{i} \ln(\theta_{i})-\sum_{i}\theta_{i}$,
where $n_{i}$ and $\theta_{i}$ are the data and the model-predicted counts in each pixel (for each energy band)
denoted by the subscript, respectively \citep[see, e.g.,][]{Mattox1996}.
}, 
was less than 0.1 over one loop in each energy band.
To model the contamination from outside the ROI,
we took into account 
sources (with the model parameters fixed to those of FL8Y) 
detected above 50$\sigma$ located 
at a distance ${\le}10\arcdeg$ from the region boundaries, or
detected above 20$\sigma$ and located within ${\le}1\arcdeg$ from the boundaries. Other sources outside of our ROIs
are not considered.

\clearpage

\section{Gamma-ray Data Analysis}
\subsection{Northern Region}

\subsubsection{Initial Modeling with a Single Gas Map}

First, we analyzed the northern region and started our data analysis 
using individual $\NH$ model maps based on 
the HI4PI survey data or the \textit{Planck} dust model maps described in Section~2.2:
We used
$\NH(\mathrm{cm^{-2}}) = 1.82 \times 10^{18} \cdot \WHI(\mathrm{K~km~s^{-1}})$, or
$\NH(\mathrm{cm^{-2}}) = 38.4 \times 10^{26} \cdot R(\mathrm{W~m^{-2}~sr^{-1}})$, or
$\NH(\mathrm{cm^{-2}}) = 159 \times 10^{24} \cdot \tau_{353}$.

We fit the $\gamma$-ray data as described in Section~3.3; 
$c_{1}$, $c_{2}$, $c_{3}$, $I_\mathrm{iso}$, and $P_\mathrm{j}$ are free parameters in each energy band.
As described in Section~3.3. point sources are included iteratively.
We remind the reader that the dust-based templates are equivalent to
$\NH = \left( \overline{\NHIthin/\Dem} \right) \cdot \Dem$ where $\overline{\NHIthin/\Dem}$ is the average of the
$\NHIthin$/$\Dem$ ratio calibrated in the high-$T_\mathrm{d}$ regions.
Therefore, in this initial modeling, we implicitly assume that the $\NH$/$\Dem$ ratio is constant over
the whole $T_\mathrm{d}$ range.

The obtained values of $\ln{L}$, 
summed over the entire energy range
with the $R$-based and $\tau_{353}$-based $\NH$ maps are 62.2 and -455.5, respectively,
when compared to that of the $\WHI$-based $\NH$ map.
Therefore, the $R$-based $\NH$ map is preferred by the $\gamma$-ray data
and the $\tau_{353}$-based map is disfavored. 
This conclusion is unchanged even if we do not include weak sources 
(detected at less than 5$\sigma$),
or if we change the energy threshold higher (200 or 400~MeV) to reduce 
the coupling with sources.
By looking at the residual maps summarized in Figure~6, we can identify extended positive residuals
covering the range of $l$ in 245--260$\arcdeg$ and $b$ in 30--55$\arcdeg$ in the $\tau_{353}$-based map
which makes the $\ln{L}$ significantly worse.
Although the difference of the $\ln{L}$ values is smaller between $\WHI$-based analysis and $R$-based ones, 
we can also identify positive residuals in the $\WHI$-based map at around
$(l, b) = (259\arcdeg, 24\arcdeg)$, ($256\arcdeg$, $33\fdg5$),
and ($236\arcdeg$, $37\fdg5$),
which corresponds to a coherent low-$T_\mathrm{d}$ and high-$\WHI$ (and $\Dem$) area.
These are the areas where the $\WHI$-based $\NH$ map predicts smaller gas column density.
We note that a region of apparently flat residuals 
at $(l, b) \sim (236\fdg5, 38\fdg5)$ (where a difference of predicted $\NH$ is the largest) is visible in all the three analyses,
even though the predicted $\NH$ is rather different there.
This is likely due to the interplay with a weak
({$\le$}8$\sigma$) $\gamma$-ray source FL8Y~J0946.2+0104
located at ($l$, $b$) $\sim$ ($235\fdg37$, $38\fdg56$).
The averages of
the normalizations (weighted inversely by the square of the error in each band)
for the neutral gas component, $c_{1}$ in Equation~(1), 
are $1.028 \pm 0.011$, $0.964 \pm0.012$, and $0.615 \pm 0.008$ 
for the $\WHI$-based, $R$-based, and $\tau_{353}$-based maps, respectively.

\begin{figure}[ht!]
\figurenum{6}
\gridline{
\fig{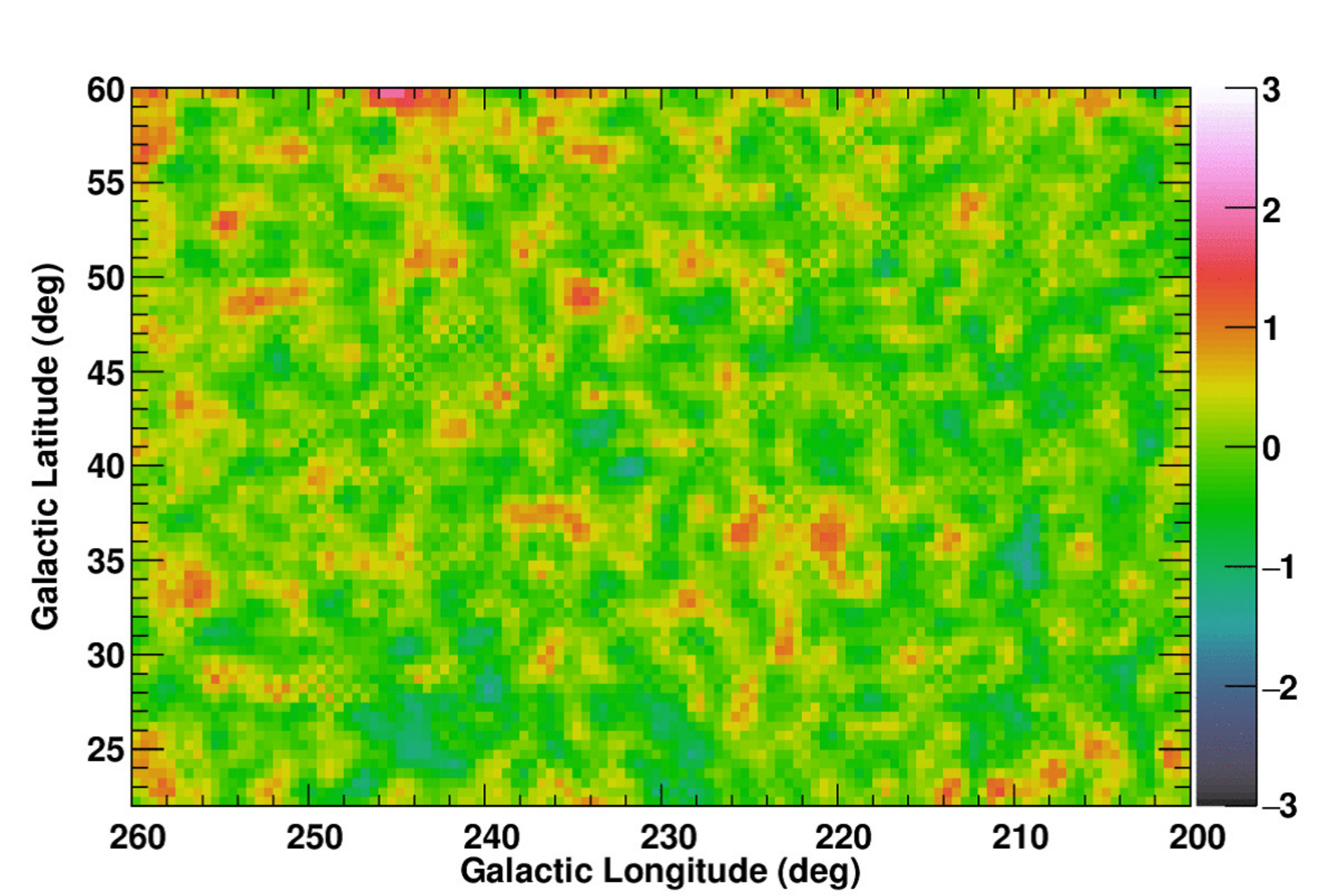}
{0.5\textwidth}{(a)}
\fig{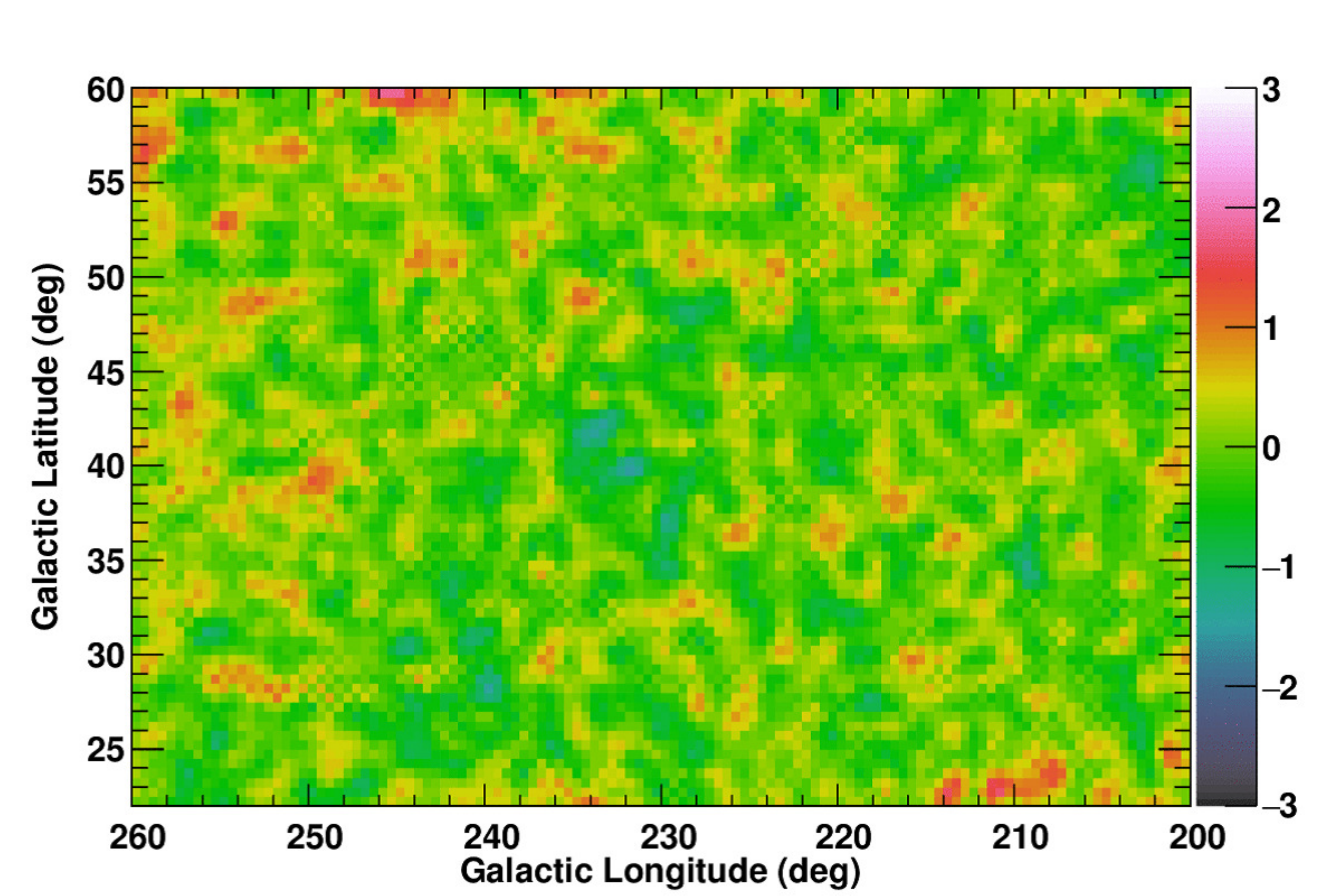}
{0.5\textwidth}{(b)}
}
\gridline{
\fig{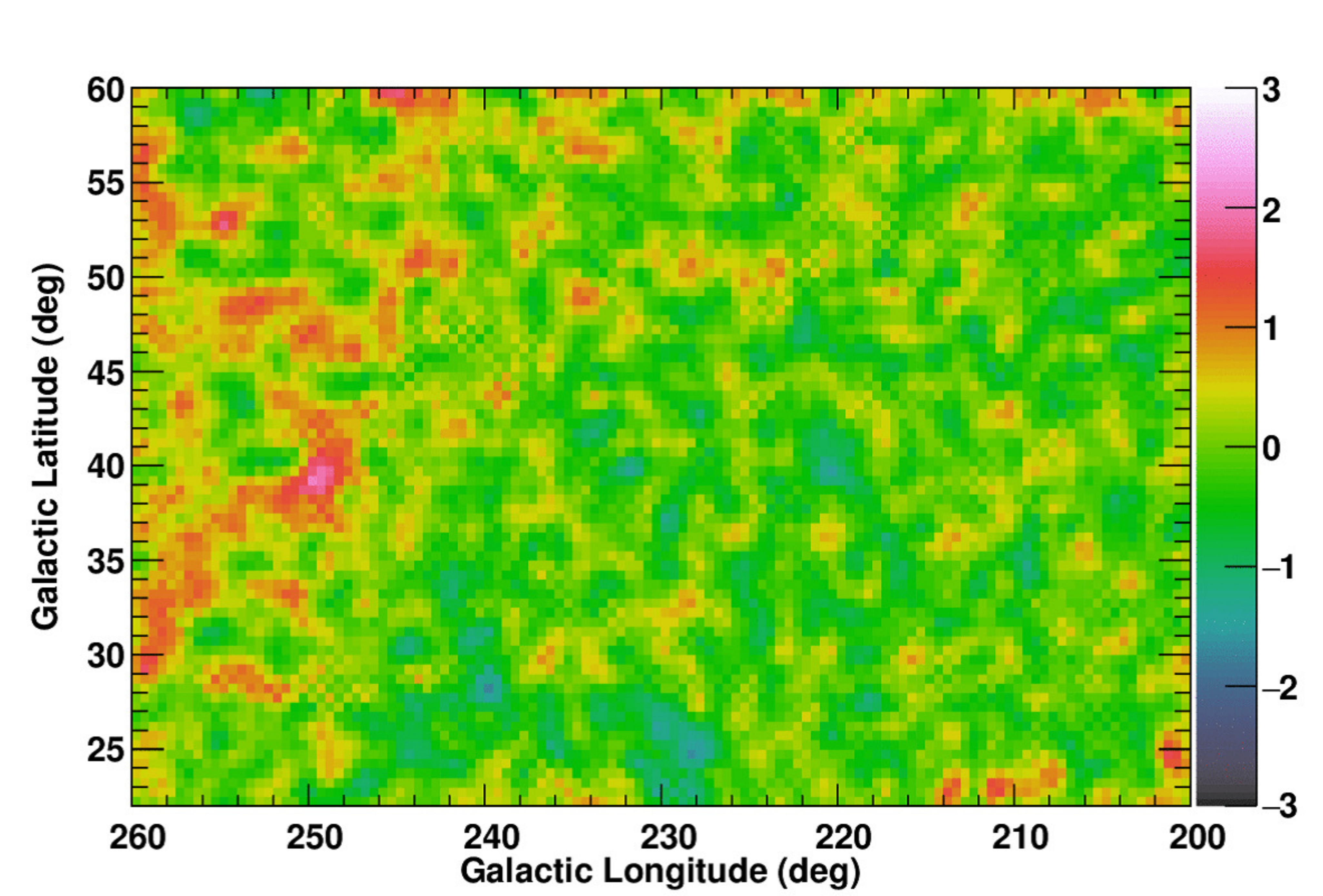}
{0.5\textwidth}{(c)}
}
\caption{
The residual maps (in units of sigma) obtained from the fit with the $\NH$ model template based on
(a) $\WHI$, (b) $R$, and (c) $\tau_{353}$. 
Although the fit has been performed
in $0\fdg25 \times 0\fdg25$ bins
above 400~MeV, all the data
have been re-binned in $0\fdg5 \times 0\fdg5$ pixels for display and smoothed with a k3a kernel
(1-2-1 two-dimensional boxcar smoothing) in the ROOT framework (\url{https://root.cern.ch}).
\label{fig:f6}
}
\end{figure}

The emission from the Sun and the Moon is subdominant in the region, 
but it could still have a small effect on the results. To test this effect, we redid the analysis 
with the model of the Sun and Moon emission used for the LAT 3FGL catalog \citep{Fermi3FGL}.
The change in the scale factors of the $\NH$ map is negligible 
($\le 1\%$) and the results are thus not affected by this component.

We also note that the normalization of the IC term
($c_{2}(E)$ in Equation~(1)) is nearly 0 for the $R$-based analysis,
likely (at least partially) because of the interplay with the isotropic term (the dominant component
\footnote{We note that fixing the IC normalization to 1 gives the IC flux to be
only 1/5 of the isotropic component.}
in our ROI).
Because the data prefer the fit with $c_{2}(E) \sim 0$, we maintain the results but will examine the systematic uncertainty by
employing alternative IC models and fixing $c_{2}(E)$ to 1.0 (Sections~4.1.2 and 5.2).

\clearpage

\subsubsection{Dust Temperature-Sorted Modeling}

As we saw in Section~2.1 (Figure~1), the correlation between $\WHI$ and $\Dem$ depends on 
$T_\mathrm{d}$, and the temperature dependence is significantly different in the cases of
$R$ or $\tau_{353}$. Although the model with $\NH$ proportional to $R$
is preferred to the one proportional to $\tau_\mathrm{353}$
(and the one proportional to $\WHI$) in terms of $\ln{L}$ by the $\gamma$-ray data analysis as described in Section~4.1,
the true $\NH$ distribution could be appreciably different from either of them;
see Section~2.1 for the prerequisites for $\Dem$ to be proportional to $\NH$ and also
\citet{Mizuno2016}, who reported the apparent $T_\mathrm{d}$ dependence of
$\NH/R$ and $\NH/\tau_\mathrm{353}$ in the MBM 53, 54, and 55 clouds and the Pegasus loop.

Therefore, we performed an analysis with the $T_\mathrm{d}$-sorted $\NH$ template maps.
We split the $\NH$ template map (constructed from $R$ or $\tau_{353}$) into four templates
based on $T_\mathrm{d}$,
for $T_\mathrm{d} \le 19~\mathrm{K}$, $T_\mathrm{d} = 19\mbox{--}20~\mathrm{K}$, 
$T_\mathrm{d} = 20\mbox{--}21~\mathrm{K}$, and $T_\mathrm{d} \ge 21~\mathrm{K}$
\footnote{For information, the relative values of the integral of $R \times$ solid angle are 7.1\%, 44.8\%, 41.4\%, and 7.0\%
for $T_\mathrm{d} \le 19~\mathrm{K}$, $T_\mathrm{d} = 19\mbox{--}20~\mathrm{K}$, 
$T_\mathrm{d} = 20\mbox{--}21~\mathrm{K}$, and $T_\mathrm{d} \ge 21~\mathrm{K}$, respectively,
and those of $\tau_{353} \times$ solid angle are 8.5\%, 47.4\%, 38.6\%, and 5.5\%
for $T_\mathrm{d} \le 19~\mathrm{K}$, $T_\mathrm{d} = 19\mbox{--}20~\mathrm{K}$, 
$T_\mathrm{d} = 20\mbox{--}21~\mathrm{K}$, and $T_\mathrm{d} \ge 21~\mathrm{K}$, respectively. 
They give the relative contribution to the $\gamma$-ray flux under the assumption of uniform CR intensity.
},
and fit the $\gamma$-ray data with Equation~(1) but using
$\sum_{i} c_{1,i}(E) \cdot \NHi(l, b)$ instead of
$c_{1}(E) \cdot \NH(l, b)$,
where $c_{1, i}(E)$ and $\NHi(l, b)$ represent
the scale factor and the template gas map for each of the four templates.
Under the assumption of a uniform CR intensity, $c_{1, i}(E)$ should not show $T_\mathrm{d}$ dependence
if the $\NH$/$\Dem$ ratio is constant.
To keep the fit stable while accommodating more free parameters, in this analysis (and that in Section~4.2.2)
we omitted the lowest-energy band
(where the angular resolution is poor) and the highest energy band (where the photon statistics 
is low)
so that the energy bands are restricted to the range 0.2--12.8~GeV.
The improvement in the fits, the likelihood test statistics $\mathrm{TS} \equiv 2 \Delta \ln{L}$, were
41.4 and 496.6 with 18 more degrees of freedom (giving statistical significances of 
3.2$\sigma$ and 20$\sigma$) for the $R$-based and $\tau_{353}$-based fit, respectively.
Therefore, the fit improvement was significant in both cases
\footnote{
TS with respect to the null hypothesis is asymptotically distributed as
a chi-square with the degrees of freedom equal to the difference
in the number of free parameters between two hypotheses
(\url{http://fermi.gsfc.nasa.gov/ssc/data/analysis/documentation/Cicerone/Cicerone_Likelihood/Likelihood_overview.html}).};
however the $R$-based analysis was still preferred.

Tables~1 and 2 show how the scaling factor
$c_{1, i}(E)$ depends on $T_\mathrm{d}$, and the averages over 0.2--12.8~GeV are
summarized in Figure~7. The $\tau_{353}$-based analysis reveals
a positive correlation (a lower scaling factor in $T_\mathrm{d} \le 20~\mathrm{K}$),
implying an overestimation of $\NH/\tau_{353}$ in the low $T_\mathrm{d}$ area.
Even though a negative correlation might be seen in the $R$-based analysis,
the dependence on $T_\mathrm{d}$ is less clear and the fit improvement over the analysis with the
single $\NH$ template map is moderate. Therefore careful examination of the systematic 
uncertainties is required.

\begin{figure}[ht!]
\figurenum{7}
\gridline{
\fig{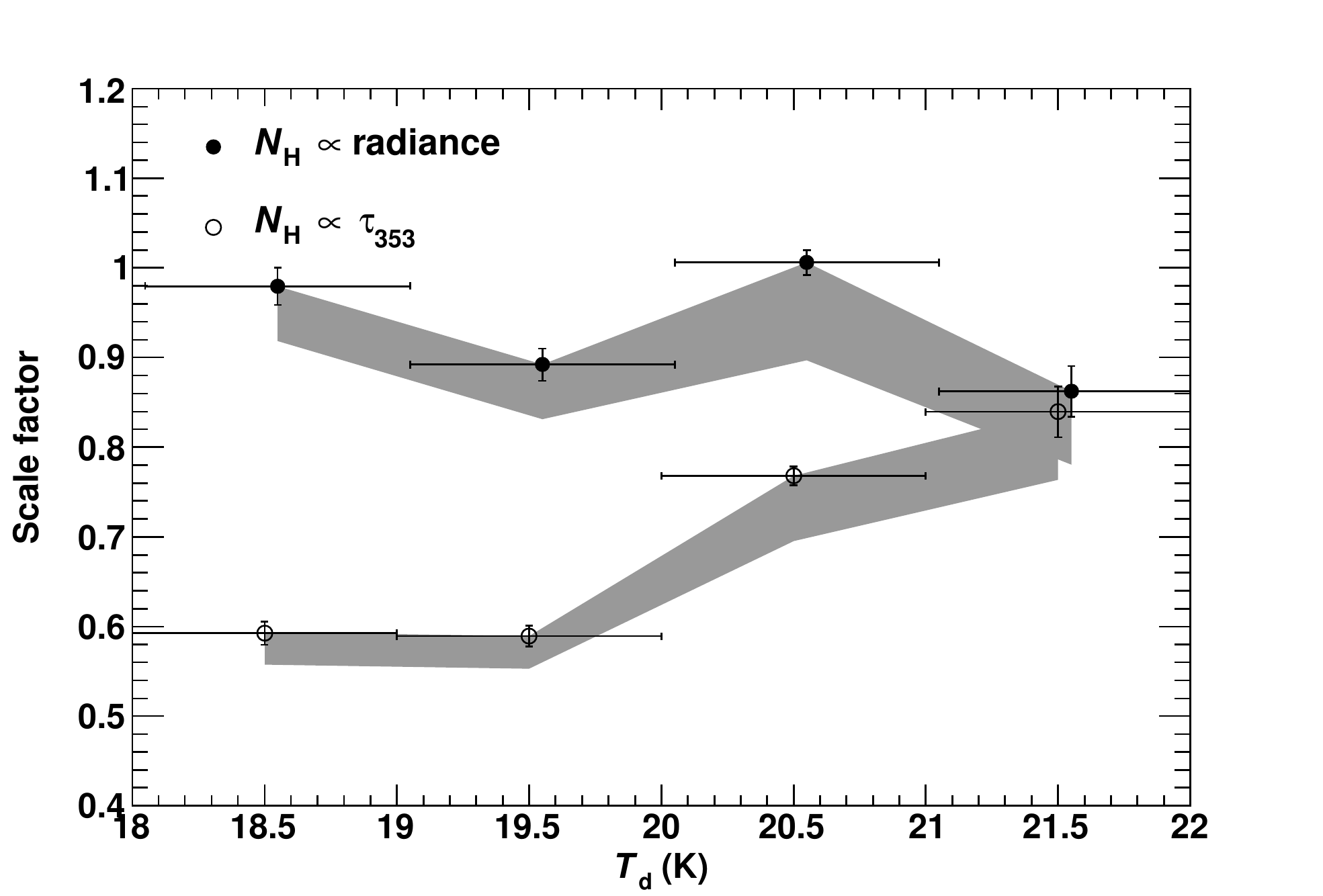}
{0.5\textwidth}{}
}
\caption{
Summary of the scale factors $c_{1,i}$ in Equation~(1) averaged over 0.2--12.8~GeV 
in each range of $T_\mathrm{d}$. The filled and open circles show the temperature dependence of the scale factors
for the $R$-based and $\tau_{353}$-based $\NH$ maps, respectively, and the gray bands show the
systematic uncertainty (see the text in Section~4.1.2 for details).
Points for the $R$-based analysis were shifted horizontally to the right by 0.05~K for display.
Although a small fraction of the pixels have $T_\mathrm{d}$ below 18~K or above 22~K, they are included
in the first and last data points, respectively.
\label{fig:f7}
}
\end{figure}

A possible systematic effect that might affect the $T_\mathrm{d}$ dependence is the uncertainty of the IC model.
Even if we adjust the IC spectrum by scaling it in each energy range, the spatial distribution
of our IC model might not be accurate.
This could affect the results shown in Figure~7 in two ways;
by changing the $T_\mathrm{d}$ dependence of the scaling factor
(i.e., the measured $T_\mathrm{d}$ dependence of $\NH/\Dem$) and by changing the 
values of the scaling factor
(i.e., the measured $\gamma$-ray emissivity or CR intensity).
If there is a small spatial variation of the (quasi-)isotropic component
that is dominant in our ROI, and this uncertainty is not absorbed by the IC model, the scale factors of the $\NH$ templates
will be affected as well.
In order to investigate such possibilities,
we tested several alternative IC models.
As described in Section~3.2, we used the IC model produced in the GALPROP run
54\_Yusifov\_z4kpc\_R30kpc\_Ts150K\_EBV2mag as our baseline model.
This configuration assumes a CR source distribution proportional to the pulsar distribution of
\citet{Yusifov2004}
and a CR halo size $z_{h}$ of 4~kpc.
As described in \citet{Fermi3rdQ}, \citet{Palma2012}, and \citet{FermiPaper2},
the CR source distribution and the halo height typically most strongly affect the
propagated CR spatial distribution (in the Galactocentric distance and the height from
the Galactic plane), and therefore the IC spatial distribution (in $l$ and $b$).
Therefore, we tested two additional CR source distributions,
the pulsar-based distribution by \citet{Lorimer2006}
and the supernova remnant (SNR)-based distribution by \citet{Case1998}, 
and one additional CR halo height,
10~kpc;
these are also available in \citet{FermiPaper2}, where the diffusion coefficient was adjusted when changing 
the CR source distribution and halo height to match the direct CR measurements at the Earth's surface.
The obtained IC maps show the smallest gradient in the Galactic longitude direction
with the SNR-based CR source distribution (the flattest distribution of our three choices)
and the smallest gradient in the Galactic latitude direction 
with $z_{h}=10~\mathrm{kpc}$ (the larger halo height of our two choices).
These tests result in a very small variation in the scaling factors (comparable to the statistical errors)
in the case of the $R$-based $\NH$ template maps.
This is due to the small normalization of the IC terms, which is nearly 0, as exemplified
by Table~1, and likely underestimates the systematic uncertainty.
Therefore, we also employed a fit using our baseline IC model
in which the normalization of the IC term was fixed to 1.0.
While the scale factors were found to be robust against fixing the IC normalization to 1.0 
for the case of the $\tau_{353}$-based analysis,
the scale factors for the $R$-based analysis decreased by $\sim$10\%. The systematic uncertainty due to the 
choice of IC model, evaluated as described above,
is shown by the shaded bands in Figure~7.
We found that the $T_\mathrm{d}$ dependence seen for the $\tau_{353}$-based analysis is robust against 
systematic uncertainties examined here,
while that for the $R$-based analysis is less significant 
(scale factors of each $T_\mathrm{d}$ range are roughly the same within errors)
if not zero.
We also 
tested the alternative Sun/Moon emission model (see Section~4.1.1) and
confirmed that the change in the $T_\mathrm{d}$ dependence of the scale factors
for the $R$-based analysis is $\le 1\%$ and negligible compared to the systematic uncertainty due to the IC modeling.

In Figure~7,
we observed a clear positive $T_\mathrm{d}$ dependence for $\NH$/$\tau_{353}$.
This trend implies an underestimation of the $\NH/\tau_{353}$ ratio in low-temperature areas
and cannot be interpreted as being due to the properties of CRs, because
the physical parameters that determine $\tau_{353}$
(e.g., dust--to--gas ratio or the dust cross section at 353~GHz)
do not affect the CR intensity.
The only possible explanation in terms of the CR properties is the exclusion of charged particles in dense
clouds with large magnetic fields expected in areas of low $T_\mathrm{d}$. 
However, CRs have been confirmed to
penetrate into dense cloud cores with $\WCO \ge 10~\mathrm{K~km~s^{-1}}$
\citep[e.g.,][]{Fermi2ndQ,Fermi3rdQ,FermiCham},
which corresponds to densities much larger than those of the clouds studied here with conventional 
values of $X_\mathrm{CO}$. 
With $X_\mathrm{CO} = \left( 1\mbox{--}2 \right) \times 10^{20}~\mathrm{cm^{-2}~K^{-1}~(km~s^{-1})^{-1}}$
inferred in nearby clouds \citep[e.g.,][]{Grenier2015},
we would have $\NH = \left( 2\mbox{--}4 \right) \times 10^{21}~\mathrm{cm^{2}}$ for 
$\WCO = 10~\mathrm{K~km~s^{-1}}$; this is larger than for the densest clouds in the region studied
(see Figures~1 and 2).
Indeed, $c_{1, i}(E)$ in Table~2 shows that the trend in Figure~7
is seen in both the low- and high-energy bands.
Therefore, the main cause of the $T_\mathrm{d}$ dependence found here is not the
properties of the CRs.
Instead, it suggests that the dust opacity 
increases as $T_\mathrm{d}$ decreases.

\floattable
\begin{deluxetable}{ccccccccc}[ht!]
\tablecaption{Results with the $R$-based $\NH$ maps sorted by $T_{d}$ 
in the northern region}
\tablecolumns{9}
\tablenum{1}
\label{tab:t1}
\tablewidth{0pt}
\tablehead{
\colhead{Energy} & \colhead{$c_{1,1}$} &\colhead{$c_{1,2}$} &\colhead{$c_{1,3}$} & \colhead{$c_{1,4}$} &
\colhead{$c_\mathrm{2}$} & \colhead{$c_\mathrm{3}$} & \colhead{$I_{\rm iso}$} & \colhead{$I_{\rm iso}$} \\
\colhead{(GeV)} & ($T_\mathrm{d} \le 19~\mathrm{K}$) & (19--20~K) & (20--21~K) & ($T_\mathrm{d} \ge 21~\mathrm{K}$) &
&& \colhead{(intensity\tablenotemark{a})} & \colhead{(index)}}
\startdata
0.2--0.4  & $0.89\pm0.04$ & $0.77\pm0.04$ & $0.96\pm0.03$ & $0.78\pm0.06$ & $\le0.20$     & $1.14\pm0.06$ &$40.7\pm0.4$  & $2.28\pm0.01$ \\
0.4--0.8  & $0.99\pm0.04$ & $0.89\pm0.03$ & $0.99\pm0.02$ & $0.87\pm0.05$ & $\le0.10$     & $1.31\pm0.10$ &$16.1\pm0.2$  & $2.27\pm0.02$ \\
0.8--1.6  & $1.00\pm0.05$ & $0.96\pm0.05$ & $1.05\pm0.04$ & $0.92\pm0.07$ & $\le0.31$     & $1.38\pm0.12$ &$6.02\pm0.17$ & $2.44\pm0.02$ \\
1.6--3.2  & $1.02\pm0.06$ & $0.97\pm0.05$ & $1.05\pm0.04$ & $0.91\pm0.07$ & $\le0.17$     & $1.20\pm0.17$ &$2.23\pm0.06$ & $2.31\pm0.04$ \\
3.2--6.4  & $0.99\pm0.10$ & $0.89\pm0.08$ & $1.03\pm0.07$ & $0.86\pm0.13$ & $\le0.38$     & $1.12\pm0.26$ &$0.99\pm0.04$ & $2.25\pm0.06$ \\
6.4--12.8 & $1.44\pm0.21$ & $1.17\pm0.18$ & $1.23\pm0.16$ & $1.04\pm0.27$ & $\le1.16$     & $1.73\pm0.43$ &$0.34\pm0.04$ & $2.44\pm0.10$ \\
\enddata
\tablenotetext{a}{The integrated intensity ($10^{-7}~{\rm ph~s^{-1}~cm^{-2}~sr^{-1}}$) in each band.}
\tablecomments{
The errors are the 1-sigma statistical uncertainties.
Each of the four scale factors ($c_{1,1}$, $c_{1,2}$, $c_{1,3}$, and $c_{1,4}$) gives the
normalization for a specified range of $T_\mathrm{d}$ of the neutral-gas template in each
energy bin.
$c_{2}$ and $c_{3}$ give the normalizations for the IC and the $\gamma$-ray emission from the Sun and Moon, respectively, in each energy bin.
$I_\mathrm{iso}$ is modeled using a power law with the integrated intensity and the photon index as free parameters.
}
\end{deluxetable}

\floattable
\begin{deluxetable}{ccccccccc}[ht!]
\tablecaption{Results with the $\tau_{353}$-based $\NH$ maps sorted by $T_{d}$ 
in the northern region}
\tablecolumns{9}
\tablenum{2}
\label{tab:t2}
\tablewidth{0pt}
\tablehead{
\colhead{Energy} & \colhead{$c_{1,1}$} &\colhead{$c_{1,2}$} &\colhead{$c_{1,3}$} & \colhead{$c_{1,4}$} &
\colhead{$c_\mathrm{2}$} & \colhead{$c_\mathrm{3}$} & \colhead{$I_{\rm iso}$} & \colhead{$I_{\rm iso}$} \\
\colhead{(GeV)} & ($T_\mathrm{d} \le 19~\mathrm{K}$) & (19--20~K) & (20--21~K) & ($T_\mathrm{d} \ge 21~\mathrm{K}$) &
&& \colhead{(intensity\tablenotemark{a})} & \colhead{(index)}}
\startdata
0.2--0.4  & $0.54\pm0.03$ & $0.49\pm0.03$ & $0.70\pm0.02$ & $0.76\pm0.06$ & $0.39\pm0.14$ & $0.88\pm0.07$ & $41.0\pm0.7$  & $2.25\pm0.01$ \\ 
0.4--0.8  & $0.60\pm0.02$ & $0.58\pm0.02$ & $0.77\pm0.02$ & $0.83\pm0.05$ & $\le0.24$     & $0.86\pm0.10$ & $17.4\pm0.3$  & $2.26\pm0.01$ \\
0.8--1.6  & $0.61\pm0.03$ & $0.65\pm0.02$ & $0.82\pm0.02$ & $0.91\pm0.06$ & $0.29\pm0.24$ & $1.00\pm0.13$ & $6.25\pm0.22$ & $2.43\pm0.02$ \\
1.6--3.2  & $0.60\pm0.04$ & $0.63\pm0.03$ & $0.80\pm0.03$ & $0.85\pm0.08$ & $\le0.34$     & $0.87\pm0.17$ & $2.46\pm0.07$ & $2.33\pm0.03$ \\
3.2--6.4  & $0.61\pm0.06$ & $0.62\pm0.06$ & $0.83\pm0.06$ & $0.85\pm0.14$ & $\le0.48$     & $0.89\pm0.25$ & $1.04\pm0.04$ & $2.27\pm0.05$ \\
6.4--12.8 & $0.91\pm0.14$ & $0.80\pm0.12$ & $0.96\pm0.13$ & $1.01\pm0.28$ & $\le1.34$     & $1.53\pm0.43$ & $0.35\pm0.04$ & $2.46\pm0.10$ \\
\enddata
\tablenotetext{a}{The integrated intensity ($10^{-7}~{\rm ph~s^{-1}~cm^{-2}~sr^{-1}}$) in each band.}
\tablecomments{
The errors are the 1-sigma statistical uncertainties.
Each of the four scale factors ($c_{1,1}$, $c_{1,2}$, $c_{1,3}$, and $c_{1,4}$) gives the
normalization for a specified range of $T_\mathrm{d}$ of the neutral-gas template in each
energy bin.
$c_{2}$ and $c_{3}$ give the normalizations for the IC and the $\gamma$-ray emission from the Sun and Moon, respectively, in each energy bin.
$I_\mathrm{iso}$ is modeled using a power law with the integrated intensity and the photon index as free parameters.
}
\end{deluxetable}

\clearpage

\subsubsection{Final Modeling}

As shown in Section~4.1.2, while the $\gamma$-ray data analysis reveals that the  $\NH/\tau_{353}$ ratio has a
positive $T_\mathrm{d}$ dependence, the $\NH/R$ ratio is rather constant. Because an
$R$-based analysis was still preferred in terms of $\ln{L}$ in $T_\mathrm{d}$-sorted modeling,
we used $R$ to construct our best estimate of the $\NH$ distribution.
Within the systematic uncertainties, no dependence on $T_\mathrm{d}$ is indicated (see Figure~7);
therefore, applying a simple correction such as a linear function of $T_\mathrm{d}$ 
to the conversion factor for $R$ ($38.4 \times 10^{26}~\mathrm{cm^{-2}~(W~m^{-2}~sr^{-1})^{-1}}$)
is not necessary.
The most noticeable feature in Figure~7 is an apparent
decrease in the scaling factor for $T_\mathrm{d} \ge 21~\mathrm {K}$ where
the conversion from $R$ to $\NH$ was calibrated (Section~4.1). To examine whether this marked decrease is robust,
we performed a further analysis using two $T_\mathrm {d}$-sorted $\NH$ maps, one with
$T_\mathrm {d} \le 21~\mathrm{K}$, and the other with $T_\mathrm {d} \ge 21~\mathrm{K}$.
The obtained scale factors for the neutral-gas template agree within 1\% for the
low and high $T_\mathrm{d}$ maps
when averaged over the entire energy band.
We also tested other thresholds of $T_\mathrm{d}$ (19.0~K and 20.0~K) and confirmed that
the scaling factors agree within 1\% for the low and high $T_\mathrm{d}$ maps.
These results indicate that there is no strong statistical evidence for a
$T_\mathrm{d}$ dependence and therefore the single $R$-based $\NH$ map is preferred.
Therefore we adopt it as
our best estimate of the $\NH$ distribution.
Finally, we fit the $\gamma$-ray data using this map (in the same way as in Section~4.1.1)
but with finer energy bins, in order to
study the spectral shape in more detail; each band was divided into two sub-bands.
We summarize
the best-fit model parameters and the obtained spectral components in Table~3 and Figure~8, respectively.

\begin{figure}[ht!]
\figurenum{8}
\gridline{
\fig{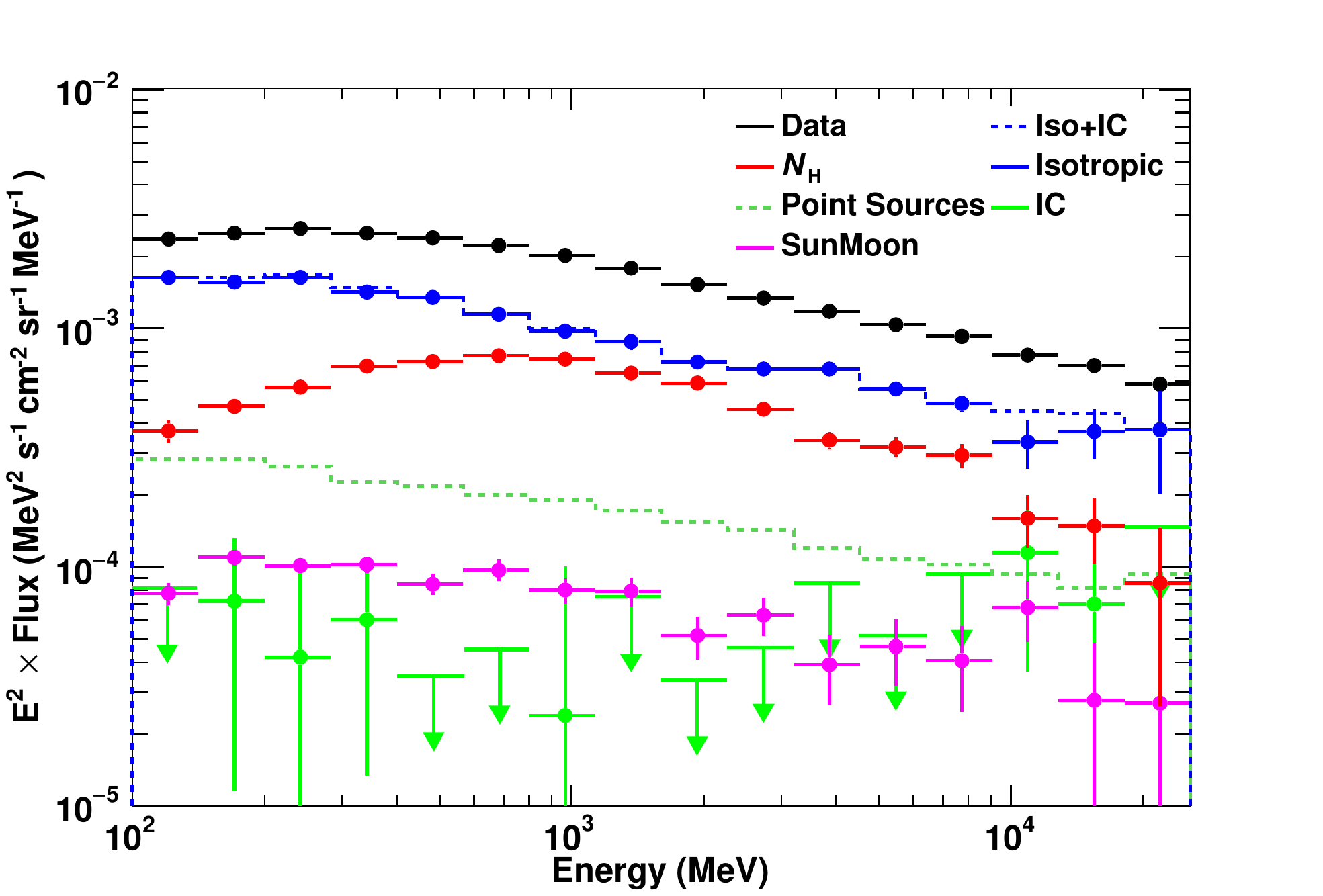}
{0.5\textwidth}{}
}
\caption{
The spectrum of each component obtained from the fit with the single, $R$-based $\NH$ map in the northern region.
The sum of the isotropic and IC components is also shown for reference.
The expected flux of the $\NHII$ model is 
${\sim}100$ times smaller than the flux of $\NH$ model and is below the vertical axis range.
\label{fig:f8}
}
\end{figure}

\floattable
\begin{deluxetable}{ccccccc}[ht!]
\tablecaption{Results of the fit with the $R$-based single $\NH$ map
in the northern region
}
\tablecolumns{6}
\tablewidth{0pt}
\tablenum{3}
\label{tab:t3}
\tablehead{
\colhead{Energy} & \colhead{$c_{1}$} & \colhead{($E^{2} \cdot c_{1} \cdot q_{\gamma}$)\tablenotemark{a}} & \colhead{$c_{2}$} & 
\colhead{$c_{3}$} & \colhead{$I_\mathrm{iso}$} & \colhead{$I_\mathrm{iso}$} \\
\colhead{(GeV)} & & & & 
& \colhead{(norm\tablenotemark{b})} & \colhead{(index)} 
}
\startdata
0.10--0.14 & $0.93\pm0.10$ & 1.11 & $\le 0.29$    & $0.88\pm0.09$ & $48.1\pm1.4$  & $1.93\pm0.05$ \\
0.14--0.20 & $0.87\pm0.06$ & 1.41 & $0.27\pm0.23$ & $1.17\pm0.09$ & $32.7\pm1.0$  & $2.06\pm0.05$ \\
0.20--0.28 & $0.84\pm0.04$ & 1.71 & $0.18\pm0.24$ & $1.15\pm0.08$ & $24.1\pm0.5$  & $2.26\pm0.03$ \\
0.28--0.40 & $0.91\pm0.03$ & 2.12 & $0.28\pm0.22$ & $1.26\pm0.10$ & $14.8\pm0.4$  & $2.30\pm0.04$ \\
0.40--0.57 & $0.93\pm0.03$ & 2.22 & $\le 0.18$    & $1.15\pm0.12$ & $9.93\pm0.17$ & $2.25\pm0.04$ \\
0.57--0.80 & $1.00\pm0.03$ & 2.37 & $\le 0.26$    & $1.47\pm0.15$ & $5.97\pm0.14$ & $2.28\pm0.05$ \\
0.80--1.13 & $1.04\pm0.05$ & 2.30 & $0.18\pm0.49$ & $1.34\pm0.18$ & $3.56\pm0.21$ & $2.28\pm0.06$ \\
1.13--1.60 & $0.99\pm0.05$ & 2.00 & $\le 0.52$    & $1.45\pm0.20$ & $2.29\pm0.17$ & $2.52\pm0.08$ \\
1.60--2.26 & $1.05\pm0.04$ & 1.84 & $\le 0.26$    & $1.06\pm0.22$ & $1.33\pm0.05$ & $2.30\pm0.10$ \\
2.26--3.20 & $0.98\pm0.05$ & 1.42 & $\le 0.39$    & $1.43\pm0.26$ & $0.87\pm0.04$ & $2.23\pm0.11$ \\
3.20--4.53 & $0.91\pm0.08$ & 1.05 & $\le 0.80$    & $0.99\pm0.32$ & $0.62\pm0.04$ & $2.48\pm0.13$ \\
4.53--6.40 & $1.12\pm0.11$ & 0.99 & $\le 0.54$    & $1.34\pm0.42$ & $0.36\pm0.02$ & $2.20\pm0.18$ \\
6.40--9.05 & $1.36\pm0.16$ & 0.92 & $\le 1.10$    & $1.34\pm0.53$ & $0.22\pm0.02$ & $2.52\pm0.23$ \\
9.05--12.8 & $0.97\pm0.24$ & 0.50 & $1.54\pm1.05$ & $2.57\pm0.73$ & $0.11\pm0.03$ & $1.87\pm0.38$ \\
12.8--18.1 & $1.17\pm0.36$ & 0.46 & $1.08\pm1.38$ & $1.22\pm0.91$ & $0.08\pm0.02$ & $2.01\pm0.39$ \\
18.1--25.6 & $0.87\pm0.60$ & 0.27 & $\le 2.64$    & $1.40\pm1.24$ & $0.06\pm0.03$ & $2.36\pm0.39$ \\
\enddata
\tablenotetext{a}{The emissivity ($c_{1} \times q_{\gamma}$) multiplied by $E^{2}$ where $E=\sqrt{E_\mathrm{min}E_\mathrm{max}}$ in each energy bin 
in units of $10^{-24}~\mathrm{MeV^{2}~s^{-1}~cm^{-2}~sr^{-1}~MeV^{-1}}$.}
\tablenotetext{b}{The integrated intensity ($10^{-7}~\mathrm{ph~s^{-1}~cm^{-2}~sr^{-1}}$) in each band.}
\tablecomments{
The errors are the 1-sigma statistical uncertainties.
$c_{1}$, $c_{2}$, and $c_{3}$ give the normalizations for the neutral-gas template,
IC, and the $\gamma$-ray emission from the Sun and Moon, respectively, in each energy bin.
$I_\mathrm{iso}$ is modeled using a power law with the integrated intensity and the photon index as free parameters.
For convenience, the best-fit value of the emissivity multiplied by $E^{2}$ is also tabulated.
}
\end{deluxetable}

\clearpage

\subsection{Southern Region}

\subsubsection{Initial Modeling with a Single Gas Map}

We then proceeded to model the southern region in which the same analysis procedures 
were used as for the northern region.
Since the ecliptic plane does not run through the ROI, we fixed the scale factors
for the Sun and Moon emission template to 1.0.
The $\NH$ model maps were prepared as described in Section~2.2:
We used
$\NH(\mathrm{cm^{-2}}) = 1.82 \times 10^{18} \cdot \WHI(\mathrm{K~km~s^{-1}})$, or
$\NH(\mathrm{cm^{-2}}) = 32.0 \times 10^{26} \cdot R(\mathrm{W~m^{-2}~sr^{-1}})$, or
$\NH(\mathrm{cm^{-2}}) = 122 \times 10^{24} \cdot \tau_{353}$.
As a side effect of masking the Orion-Eridanus superbubble,
the $\NH$ template map has a similar spatial distribution
(larger intensity toward the Galactic center) to that of the IC model and they are degenerate with each other.
As we progressed in the iterative method,
we observed that the $\NH$ template is given progressively lower
scale factor ($c_{1}$) while the IC component is given higher one ($c_{2}$) in low-energy bands, although the contribution
of point sources to the total $\gamma$-ray flux is almost unchanged. To mitigate this, we employed a
(semi-)global fitting as a preparatory stage. We first adopted wider energy bins allowing overlaps
(70.7--282.8~MeV, 141.4--565.7~MeV, etc.), and we included point sources iteratively until the fit improvement is saturated
as we did for the analysis of the northern region. Allowing overlapping the energy bins makes the fit
results more stable by encouraging spectral inter-bin consistency.
We then fixed the IC normalization to the best-fit value and repeated the analysis with the original energy bins
(100--200~MeV, 200--400~MeV, etc.). This ``two-stage" analysis is employed hereafter for the southern region.

The obtained log-likelihoods, 
summed over individual energy ranges in 0.1--25.6~GeV
with the $R$-based and $\tau_{353}$-based $\NH$ maps are 19.4 and $-$74.6, respectively,
when compared to that of the $\WHI$-based $\NH$ map.
Therefore, the $R$-based $\NH$ map is preferred compared 
to the $\tau_{353}$-based map.
Like the northern region, this conclusion is unchanged against the significance threshold of point source model
or the lowest energy threshold.
The averages of
the normalization for the neutral gas component, $c_{1}$ in Equation~(1), 
are $0.946 \pm 0.008$, $0.946 \pm 0.007$, and $ 0.690 \pm 0.005$
for the $\WHI$-based, $R$-based, and $\tau_{353}$-based maps, respectively.
The residual maps are summarized in Figure~9.

\begin{figure}[ht!]
\figurenum{9}
\gridline{
\fig{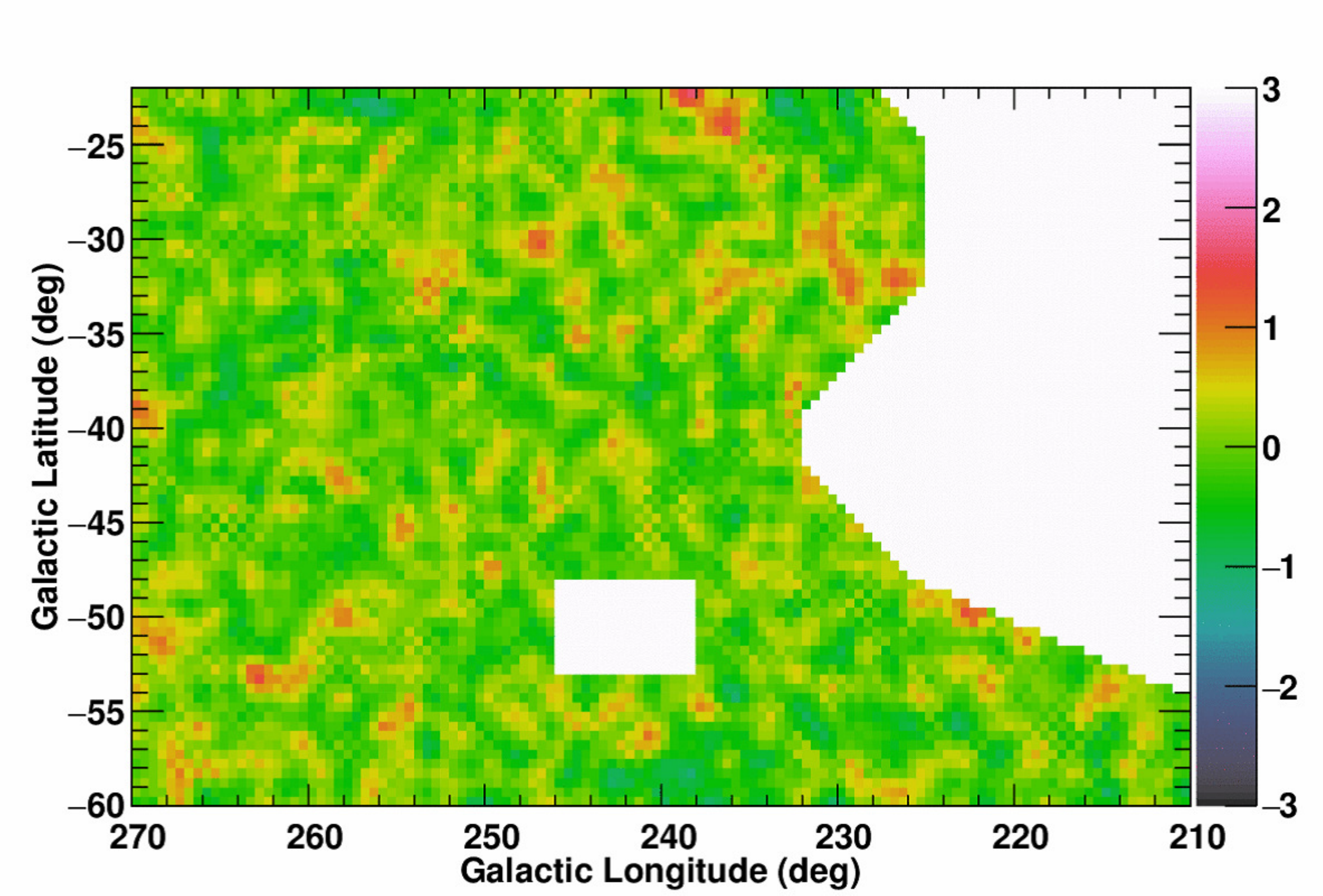}
{0.5\textwidth}{(a)}
\fig{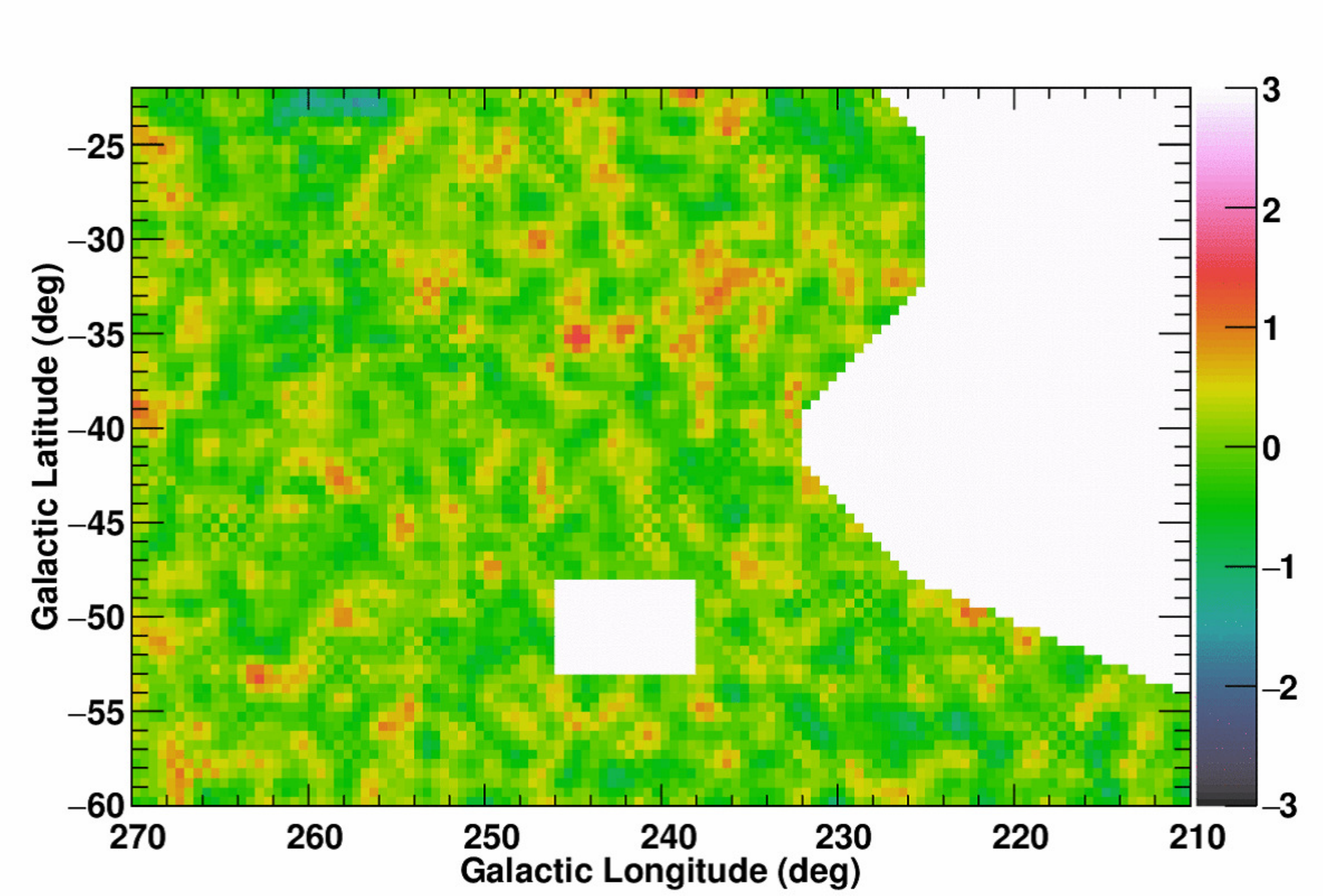}
{0.5\textwidth}{(b)}
}
\gridline{
\fig{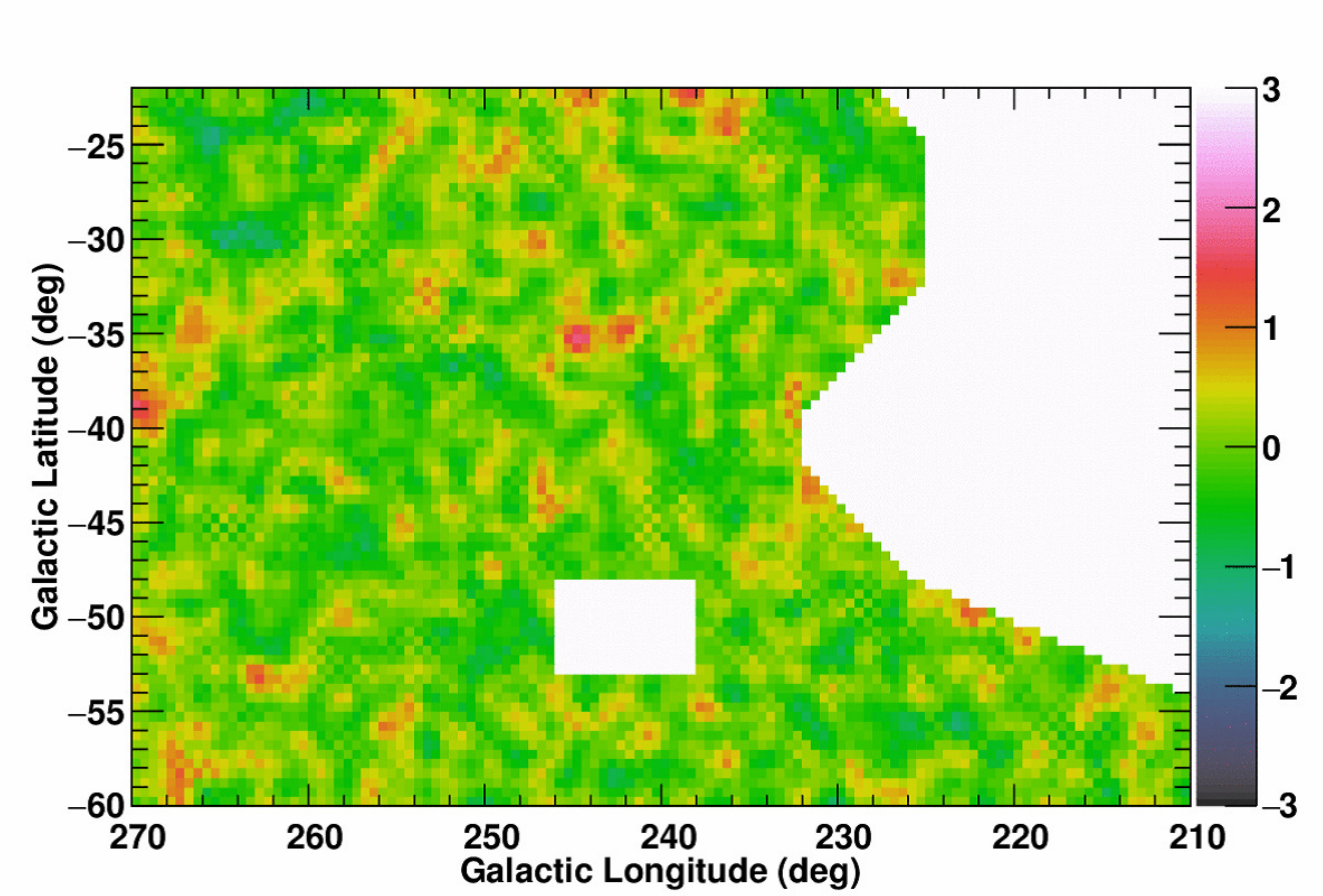}
{0.5\textwidth}{(c)}
}
\caption{
The same as Figure~6 but for the southern region instead of the northern region.
\label{fig:f9}
}
\end{figure}

\clearpage

\subsubsection{Dust Emission-Sorted Modeling}

As we saw in Section~2.1 (Figure~2), the correlation between $\WHI$ and $\Dem$ shows 
a change of the slope in particular for $\tau_{353}$
while the $T_\mathrm{d}$ dependence is small;
unlike the case of the northern region, the $\WHI$--$\Dem$ relationship depends
on $\Dem$ rather than $T_\mathrm{d}$.
Even though the $R$-based $\NH$ map is preferred for the three models in terms of $\ln{L}$, 
the true $\NH$ distribution could be appreciably
different and a possible nonlinear relationship between $\NH$ and $\Dem$ should be examined.
Therefore, we performed an analysis with $R$-sorted and $\tau_{353}$-sorted $\NH$ maps
(instead of $T_\mathrm{d}$-sorted $\NH$ maps applied for the northern region). 
We split the $\NH$ template map (constructed from $R$) into three templates based on $R$,
for $R \le 12$, $R = 12\mbox{--}20$, 
and $R \ge 20$ in units of $10^{-8}~\mathrm{W~m^{-2}~sr^{-1}}$
\footnote{For information, the relative values of the integral of $R \times$ solid angle 
(proportional to the relative flux with uniform CR intensity)
are 65.8\%, 25.5\%, and 8.7\%
for $R \le 12$, $R = 12\mbox{--}20$, 
and $R \ge 20$ in units of $10^{-8}~\mathrm{W~m^{-2}~sr^{-1}}$, respectively.
},
and fit the $\gamma$-ray data with Equation~(1), using
$\sum_{i} c_{1,i}(E) \cdot \NHi(l, b)$ instead of
$c_{1}(E) \cdot \NH(l, b)$, where $c_{1, i}(E)$ and $\NHi(l, b)$ represent
the scale factor and template gas map for each of the three templates.
For the $\tau_{353}$-based $\NH$ template, we split it into three as
$\tau_{353} \le 4$, $\tau_{353} = 4\mbox{--}6$, 
and $\tau_{353} \ge 6$ in units of $10^{-6}$
\footnote{For information, the relative values of the integral of $\tau_{353} \times$ solid angle 
(proportional to the relative flux with uniform CR intensity)
are 67.1\%, 17.4\%, and 15.5\%
for $\tau_{353} \le 4$, $\tau_{353} = 4\mbox{--}6$, 
and $\tau_{353} \ge 6$ in units of $10^{-6}$, respectively.
}.
With a plausible assumption of a uniform CR intensity,
$c_{1, i}(E)$ is expected to trace the $\NH$/$\Dem$ ratio.
We used data in the range of 0.2--12.8~GeV to avoid a possible unstable fit in the lowest (0.1--0.2~GeV) and highest 
(12.8--25.6~GeV) energy bands;
then, we obtained a value of TS of
22.0 and 101.8 for $R$ and $\tau_{353}$, respectively,
with 18 more degrees of freedom.
This indicates a 7.4$\sigma$ improvement when the $\tau_{353}$ model for
$\NH$ is used,
while the improvement of the fit is not significant for $R$.
We also note that $\tau_{353}$-sorted modeling is still not favored compared to the
analysis using the single $\NH$ map based on $R$ in terms of $\ln{L}$.
The $R$- and $\tau_{353}$-dependence of the scaling factors is summarized in 
Tables~4 and 5, and the averages over 0.2--12.8~GeV are
summarized in Figure~10.

As for the northern region, we examined the systematic uncertainty due to the choice of the IC model
and plotted the results in Figure~10; the outer polygonal area shows the full uncertainty evaluated by
trying all six IC models, and the inner, shaded area shows the variation where the model with the worst $\ln{L}$ was 
excluded. 
One may also argue that the obtained scale factors of the IC term and the isotropic emission
intensity are greately different between the northern and southern ROIs (Tables~1, 2, 4, and 5),
and that the IC normalization is nearly 0 (which is not physical)
with the $R$-based $\NH$ template fit for the northern region. 
This indicates
that our model does not completely describe the $\gamma$-ray data.
For example, the IC spatial template may not agree with the true distribution,
or there may be
a small variation of the (quasi-) isotropic component. 
If the IC spatial template is not representing the underlying distribution of the 
IC emission observed by the LAT, it will most likely be ingested in the isotropic component contributions. Therefore
it is the total IC and isotropic that matters for these particular ROIs, and
the sum of the IC term and isotropic emission is similar between two ROIs (Figures~8 and 11).
Therefore, most of uncertainties of the IC term and the isotropic emission are mutually
absorbed, and we believe that the effect on the neutral gas component is properly examined
by employing several IC models as described above and in Section~4.1.2.
We also note that while the differences of log-likelihoods are very small (19.4)
between the $R$-based model and the $\WHI$-based one, 
the specific choice of the template does not affect the gas (and CR) properties very much;
the normalizations of the neutral gas component are almost identical between the two gas 
models as described in Section~4.2.1.

\begin{figure}[ht!]
\figurenum{10}
\gridline{
\fig{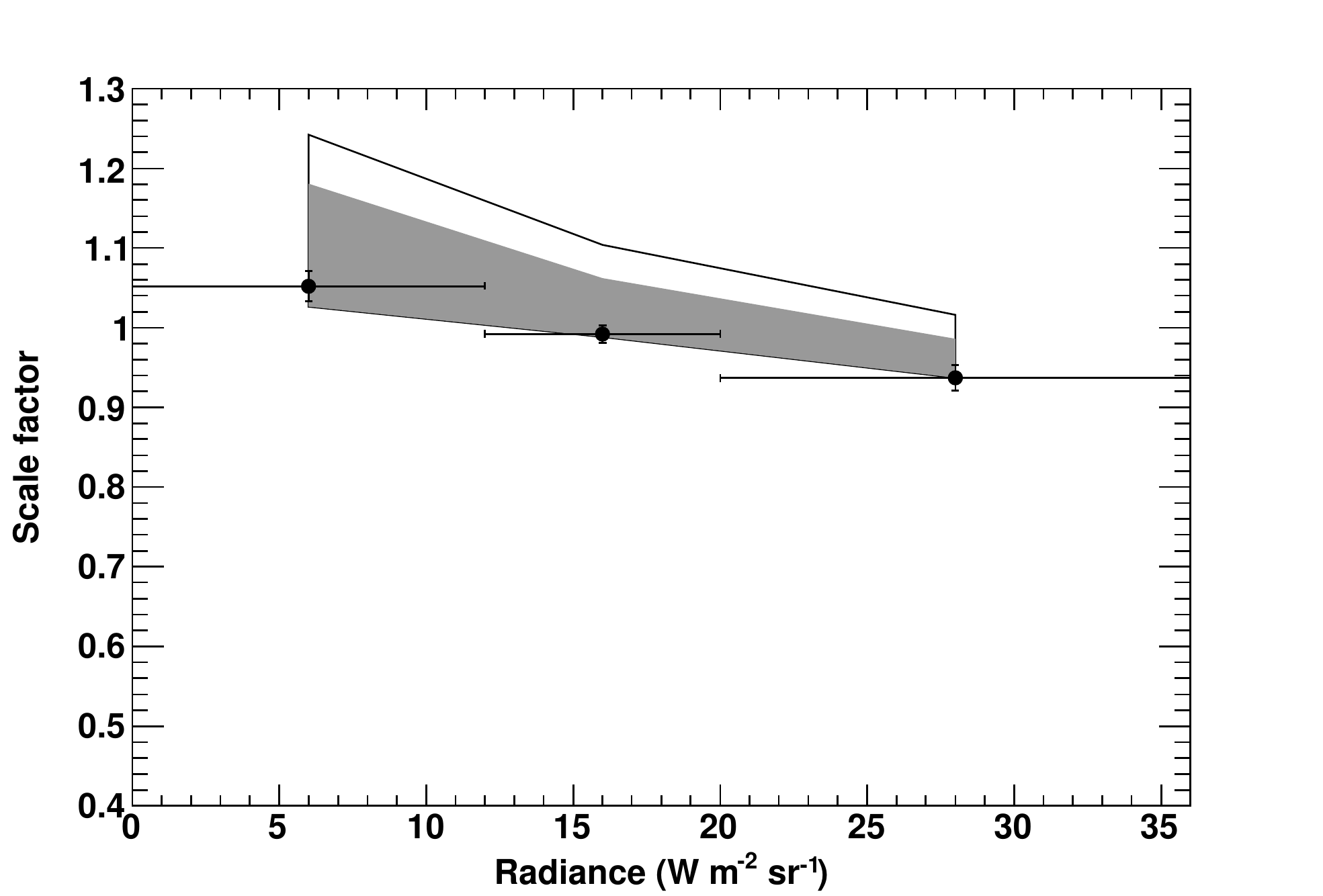}
{0.5\textwidth}{(a)}
\fig{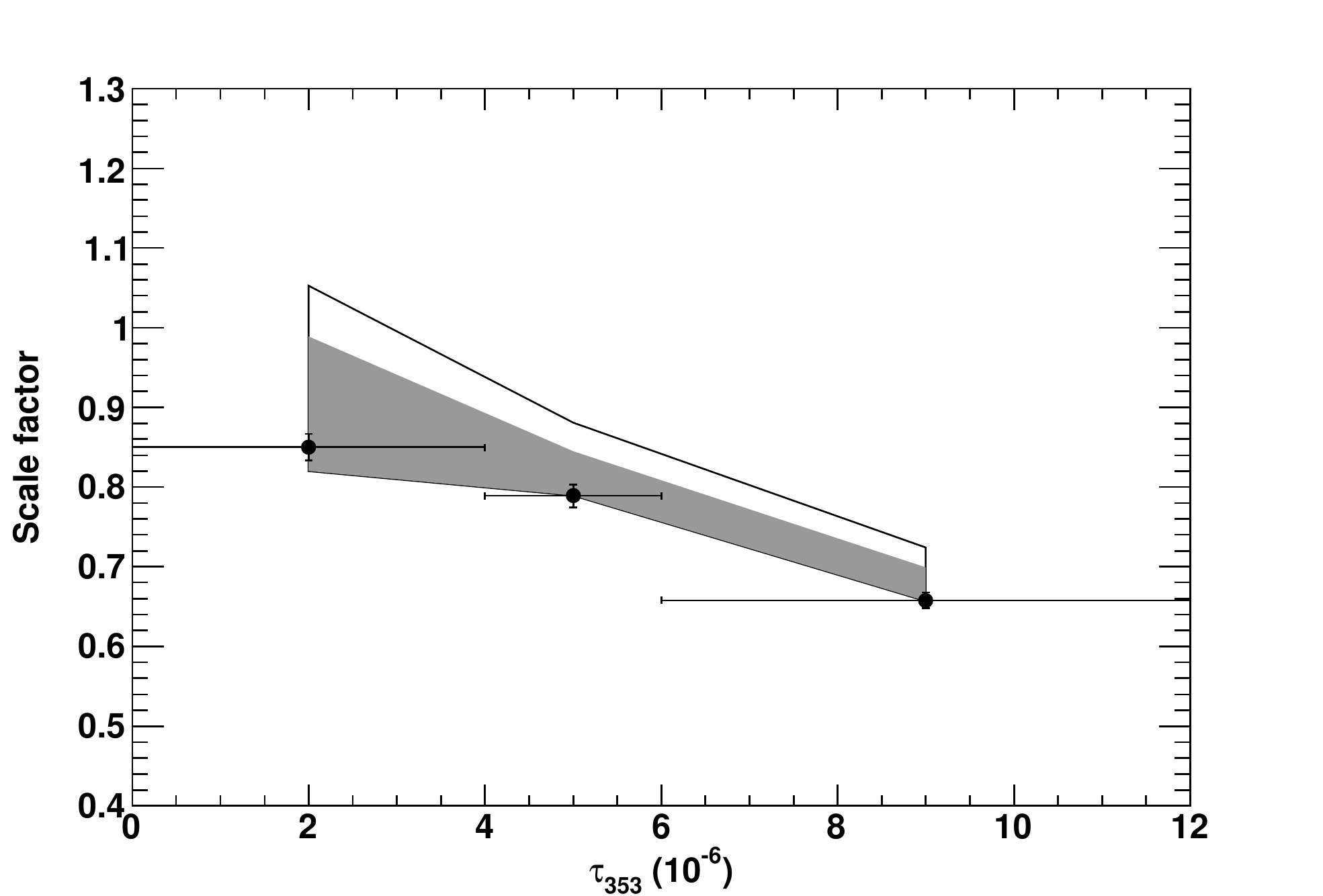}
{0.5\textwidth}{(b)}
}
\caption{
Summary of the scale factor $c_{1,i}$ in Equation~(1) averaged over 0.2--12.8~GeV in each range of 
$R$ (a) and $\tau_{353}$ (b).
The outer polygonal area shows the systematic uncertainty evaluated using all six IC models, and 
the inner, shaded area shows the
uncertainty where the model with the worst $\ln{L}$ was excluded.
Although a small fraction of the pixels show $\Dem$ above the horizontal axis ranges,
they are included in the last data points.
\label{fig:f10}
}
\end{figure}

In Figure~10,
we observe a clear negative $\tau_{353}$ dependence of the scale factor that is robust against the systematics
due to the choice of IC model. This trend
implies an overestimation of the $\NH/\tau_{353}$ ratio in the high-density area.
As discussed in Section~4.1.2, this cannot be interpreted as being due to the properties of CRs.
Instead, it is likely due to the dust properties, 
such as the change of the dust cross section 
by dust grain evolution
\citep[e.g.,][]{Roy2013}.
For the case of $R$-sorted analysis, there seems to be a negative dependence. However,
the change of $\NH$/$R$ ratio (inversely proportional to the scale factor) is much smaller 
than that for $\tau_{353}$
whatever IC model is employed, and the fit improvement is not significant. 
Therefore an
$R$-based single $\NH$ map is suggested to reproduce the $\NH$ distribution inferred by $\gamma$-ray data.

\floattable
\begin{deluxetable}{ccccccc}[ht!]
\tablecaption{Results with the $R$-based $\NH$ maps sorted by $R$ 
in the southern region \label{tab:rad_R}}
\tablecolumns{7}
\tablenum{4}
\label{tab:t4}
\tablewidth{0pt}
\tablehead{
\colhead{Energy} & \colhead{$c_{1,1}$} &\colhead{$c_{1,2}$} &\colhead{$c_{1,3}$} &
\colhead{$c_\mathrm{2}$} & \colhead{$I_{\rm iso}$} & \colhead{$I_{\rm iso}$} \\
\colhead{(GeV)} & ($R\tablenotemark{a} \le 12$) & (12--20) & ($R \ge 20$) & 
& \colhead{(norm\tablenotemark{b})} & \colhead{(index)}}
\startdata
0.2--0.4  & $1.00\pm0.04$ & $0.88\pm0.02$ & $0.88\pm0.03$ & $2.72\pm0.10$ & $23.6\pm0.3$  & $2.35\pm0.02$ \\
0.4--0.8  & $1.04\pm0.03$ & $1.01\pm0.02$ & $0.91\pm0.03$ & $2.32\pm0.14$ & $9.97\pm0.16$ & $2.33\pm0.03$ \\
0.8--1.6  & $1.11\pm0.04$ & $1.09\pm0.02$ & $1.02\pm0.04$ & $2.53\pm0.19$ & $3.41\pm0.08$ & $2.46\pm0.05$ \\
1.6--3.2  & $1.07\pm0.06$ & $1.01\pm0.04$ & $1.02\pm0.05$ & $2.56\pm0.26$ & $1.18\pm0.05$ & $2.25\pm0.08$ \\
3.2--6.4  & $1.01\pm0.11$ & $1.09\pm0.07$ & $0.87\pm0.09$ & $2.37\pm0.38$ & $0.57\pm0.03$ & $2.14\pm0.11$ \\
6.4--12.8 & $1.36\pm0.22$ & $1.15\pm0.13$ & $0.96\pm0.17$ & $3.07\pm0.58$ & $0.19\pm0.02$ & $2.83\pm0.20$ \\
\enddata
\tablenotetext{a}{$R$ is given in units of $10^{-8}~\mathrm{W~m^{-1}~sr^{-1}}$}
\tablenotetext{b}{The integrated intensity ($10^{-7}~{\rm ph~s^{-1}~cm^{-2}~sr^{-1}}$) in each band.}
\tablecomments{
The errors are the 1-sigma statistical uncertainties.
Each of the three scale factors ($c_{1,1}$, $c_{1,2}$, and $c_{1,3}$) gives the
normalization for a specified range of $R$ of the neutral-gas template in each
energy bin.
$c_{2}$ is the IC template normalization for each energy bin obtained at the first stage of the fitting
(see text).
$I_\mathrm{iso}$ is modeled using a power law with the integrated intensity and the photon index as free parameters.
}
\end{deluxetable}

\floattable
\begin{deluxetable}{ccccccc}[ht!]
\tablecaption{Results with the $\tau_{353}$-based $\NH$ maps sorted by $\tau_{353}$ 
in the southern region \label{tab:tau_Tau}}
\tablecolumns{7}
\tablenum{5}
\label{tab:t5}
\tablewidth{0pt}
\tablehead{
\colhead{Energy} & \colhead{$c_{1,1}$} &\colhead{$c_{1,2}$} &\colhead{$c_{1,3}$} &
\colhead{$c_\mathrm{2}$} & \colhead{$I_{\rm iso}$} & \colhead{$I_{\rm iso}$} \\
\colhead{(GeV)} & ($\tau_{353}\tablenotemark{a} \le 4$) & (4--6) & ($\tau_{353} \ge 6$) & 
& \colhead{(norm\tablenotemark{b})} & \colhead{(index)}}
\startdata
0.2--0.4  & $0.84\pm0.04$ & $0.67\pm0.03$ & $0.64\pm0.02$ & $3.01\pm0.10$ & $22.7\pm0.3$  & $2.33\pm0.02$ \\
0.4--0.8  & $0.83\pm0.03$ & $0.82\pm0.03$ & $0.63\pm0.02$ & $2.85\pm0.13$ & $9.35\pm0.15$ & $2.31\pm0.03$ \\
0.8--1.6  & $0.87\pm0.03$ & $0.85\pm0.03$ & $0.71\pm0.02$ & $3.15\pm0.18$ & $3.15\pm0.08$ & $2.47\pm0.05$ \\
1.6--3.2  & $0.89\pm0.05$ & $0.78\pm0.04$ & $0.70\pm0.03$ & $3.06\pm0.25$ & $1.07\pm0.04$ & $2.27\pm0.09$ \\
3.2--6.4  & $0.80\pm0.09$ & $0.91\pm0.07$ & $0.60\pm0.05$ & $2.83\pm0.37$ & $0.54\pm0.03$ & $2.16\pm0.12$ \\
6.4--12.8 & $1.00\pm0.18$ & $0.87\pm0.15$ & $0.63\pm0.10$ & $3.54\pm0.54$ & $0.18\pm0.02$ & $2.33\pm0.21$ \\
\enddata
\tablenotetext{a}{$\tau_{353}$ is given in units of $10^{-6}$}
\tablenotetext{b}{The integrated intensity ($10^{-7}~{\rm ph~s^{-1}~cm^{-2}~sr^{-1}}$) in each band.}
\tablecomments{
The errors are the 1-sigma statistical uncertainties.
Each of the three scale factors ($c_{1,1}$, $c_{1,2}$, and $c_{1,3}$) gives the
normalization for a specified range of $\tau_{353}$ of the neutral-gas template in each
energy bin.
$c_{2}$ is the IC template normalization for each energy bin obtained at the first stage of the fitting
(see text).
$I_\mathrm{iso}$ is modeled using a power law with the integrated intensity and the photon index as free parameters.
}
\end{deluxetable}

\clearpage

\subsubsection{Final Modeling}

We employed the $\gamma$-ray data as a robust tracer of the total neutral gas distribution.
We therefore can apply a correction to the $\NH$ model template based on $\gamma$-ray data in principle.
To prove this concept, and to see if this affects the choice of dust tracer ($R$ or $\tau_{353}$),
we started with
the uncorrected $\NH$ map that is proportional to $\tau_{353}$ (denoted as $\NHtau$)
and modified the gas column density to take into account the observed $\tau_{353}$-dependence (Figure~10).
We assumed that the $\NH$ is proportional to $\tau_{353}$ up to a particular value ($\tau_\mathrm{bk}$) and
deviates from the proportionality linearly above that. Then, we can apply the correction to the $\NH$ model
using the empirical function below:
\begin{eqnarray}
\NHmod =
\left\{
\begin{array}{l}
\NHtau~(\tau_{353} < \tau_\mathrm{bk})~~, \\
\NHbk + (1-0.1 \cdot C) \cdot (\NHtau-\NHbk)
~(\tau_{353} \ge \tau_\mathrm{bk})~~,
\end{array}
\right.
\end{eqnarray}
where $\NHbk$ is the (uncorrected) gas column density (proportional to $\tau_{353}$) at $\tau_\mathrm{bk}$.
$C=1$ corresponds to a 10\% decrease in $\NH$ above $\NHbk$.
We carried out a grid scan ($\tau_\mathrm{bk}$=2, 3, 4, 5, 6 in unit of $10^{-6}$ and $C$=2, 3, 4, 5, 6)
and found that $\tau_\mathrm{bk}=4$ and $C=4$ gives the best representation of the \textit{Fermi}-LAT data.
This configuration increases the scale factor of the neutral gas component by 20\% and makes it agree with that from the
$R$-based one within 15\%.

The corrected $\NH$ model based on $\tau_{353}$, however, still gives smaller $\ln{L}$ compared to the single $\NH$ model based on $R$.
Because we found that the $R$-dependence of the scaling factors of the neutral gas component is small if not zero,
and the fit improvement is not significant over the single, $R$-based $\NH$ template,
we adopted this $R$-based $\NH$ model as our best estimate of the $\NH$ distribution.
We, therefore, fit the $\gamma$-ray data using this map with finer energy bands 
to study the spectral shape in more detail as we did for the northern region.
The best-fit model parameters and the obtained spectral components are summarized in Table~6 and Figure~11, respectively.

\begin{figure}[ht!]
\figurenum{11}
\gridline{
\fig{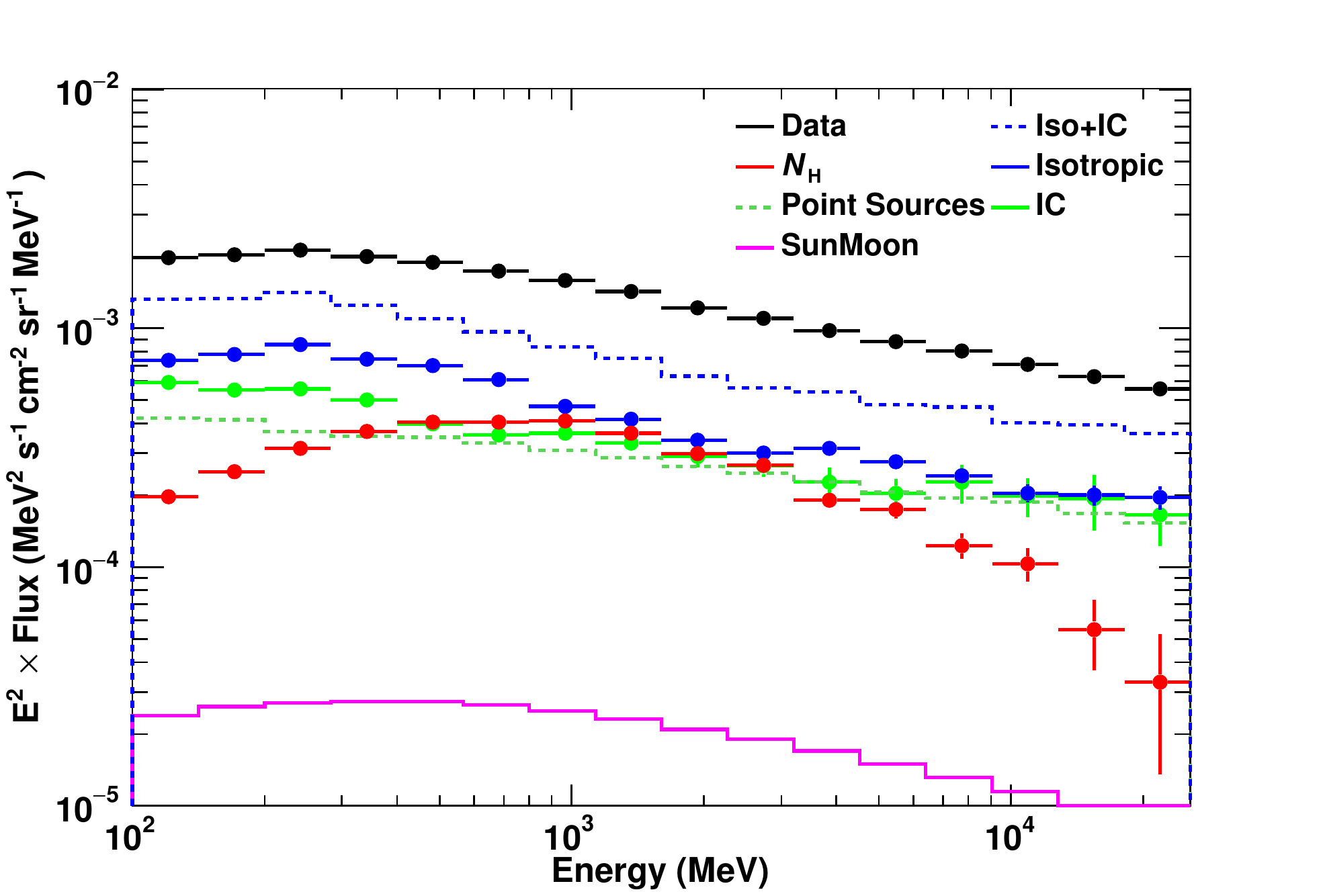}
{0.5\textwidth}{}
}
\caption{
The spectrum of each component obtained from the fit based on the single, $R$-based $\NH$ map in the southern region.
The sum of the isotropic and IC components is also shown for reference.
The expected flux of the $\NHII$ model is 
${\sim}50$ times smaller than the flux of $\NH$ model and is below the vertical axis range.
\label{fig:f11}
}
\end{figure}

\floattable
\begin{deluxetable}{cccccc}[ht!]
\tablecaption{Results of the fit with the $R$-based single $\NH$ map
in the southern region
}
\tablecolumns{5}
\tablenum{6}
\label{tab:t6}
\tablewidth{0pt}
\tablehead{
\colhead{Energy} & \colhead{$c_{1}$} & \colhead{($E^{2} \cdot c_{1} \cdot q_{\gamma}$)\tablenotemark{a}} & \colhead{$c_{2}$} & 
\colhead{$I_\mathrm{iso}$} & \colhead{$I_\mathrm{iso}$} \\
\colhead{(GeV)} & & & 
& \colhead{(norm\tablenotemark{b})} & \colhead{(index)} 
}
\startdata
0.10--0.14 & $0.91\pm0.04$ & 1.08 & $2.38\pm0.10$ & $26.7\pm0.4$  & $1.81\pm0.08$ \\
0.14--0.20 & $0.86\pm0.03$ & 1.40 & $2.38\pm0.10$ & $20.1\pm0.3$  & $2.00\pm0.07$ \\
0.20--0.28 & $0.86\pm0.02$ & 1.76 & $2.65\pm0.10$ & $15.6\pm0.1$  & $2.38\pm0.05$ \\
0.28--0.40 & $0.91\pm0.02$ & 2.11 & $2.65\pm0.10$ & $9.63\pm0.10$ & $2.28\pm0.06$ \\
0.40--0.57 & $0.97\pm0.02$ & 2.34 & $2.34\pm0.14$ & $6.39\pm0.08$ & $2.28\pm0.07$ \\
0.57--0.80 & $1.00\pm0.02$ & 2.36 & $2.34\pm0.14$ & $3.96\pm0.06$ & $2.33\pm0.09$ \\
0.80--1.13 & $1.08\pm0.02$ & 2.39 & $2.63\pm0.19$ & $2.16\pm0.04$ & $2.39\pm0.12$ \\
0.13--1.60 & $1.05\pm0.03$ & 2.13 & $2.63\pm0.19$ & $1.35\pm0.03$ & $2.64\pm0.15$ \\
1.60--2.26 & $1.00\pm0.03$ & 1.76 & $2.56\pm0.26$ & $0.78\pm0.02$ & $2.60\pm0.20$ \\
2.26--3.20 & $1.08\pm0.04$ & 1.57 & $2.56\pm0.26$ & $0.48\pm0.02$ & $2.11\pm0.25$ \\
3.20--4.53 & $0.97\pm0.06$ & 1.12 & $2.44\pm0.37$ & $0.36\pm0.01$ & $2.33\pm0.27$ \\
4.53--6.40 & $1.16\pm0.09$ & 1.03 & $2.44\pm0.37$ & $0.22\pm0.01$ & $2.52\pm0.34$ \\
6.40--9.05 & $1.08\pm0.13$ & 0.72 & $3.06\pm0.57$ & $0.14\pm0.01$ & $2.05\pm0.44$ \\
9.05--12.8 & $1.18\pm0.19$ & 0.61 & $3.06\pm0.57$ & $0.08\pm0.01$ & $1.82\pm0.57$ \\
12.8--18.1 & $0.81\pm0.27$ & 0.32 & $3.44\pm0.90$ & $0.06\pm0.01$ & $1.72\pm0.65$ \\
18.1--25.6 & $0.63\pm0.37$ & 0.19 & $3.44\pm0.90$ & $0.04\pm0.01$ & $3.92\pm0.78$ \\
\enddata
\tablenotetext{a}{The emissivity ($c_{1} \times q_{\gamma}$) multiplied by $E^{2}$ where $E=\sqrt{E_\mathrm{min}E_\mathrm{max}}$ in each energy bin 
in units of $10^{-24}~\mathrm{MeV^{2}~s^{-1}~cm^{-2}~sr^{-1}~MeV^{-1}}$.}
\tablenotetext{b}{The integrated intensity ($10^{-7}~\mathrm{ph~s^{-1}~cm^{-2}~sr^{-1}}$) in each band.}
\tablecomments{
The errors are the 1-sigma statistical uncertainties.
$c_{1}$ and $c_{2}$ give the normalization for the neutral-gas template and
IC, respectively, in each energy bin (the best-fit values obtained at the first stage of the fitting are given for the latter).
$I_\mathrm{iso}$ is modeled using a power law with the integrated intensity and the photon index as free parameters.
For convenience, the best-fit value of the emissivity ($c_{1} \times q_{\gamma}$) multiplied by $E^{2}$ is also tabulated.
}
\end{deluxetable}

\clearpage

\section{Discussion}

\subsection{ISM}

In Section~4, we used the GeV $\gamma$-rays observed by \textit{Fermi}-LAT as robust tracers of the ISM gas
under the assumption of a uniform CR intensity
and obtained the $\NH$ distributions inferred by the $\gamma$-ray data.
Trends of the scale factor for $\NH$ templates
($T_\mathrm{d}$ dependence and $\tau_{353}$ dependence in the northern and southern regions, respectively)
are commonly seen between low- and high-energy bands (see Tables 1, 2, 4, and 5); this
supports the uniformity of the CR intensity in each ROI.
We found that the $\NH$ template based on the $R$ data best matches the $\gamma$-ray observations
in both northern and southern regions and in the following discussion, we assume
$R$ is a good tracer of $\NH$.
The obtained relationships between 
$\WHI$ and $\NH$ are shown in Figure~12
together with maps of the excess gas column density above $\NHIthin$.
We point out that the $\WHI$/$\tau_{353}$ ratio (and $\NH$/$\tau_{353}$ ratio)
strongly depends on $T_\mathrm{d}$ in the northern region, while in the southern 
region this dependence is weaker.
The dust optical depth $\tau_{353}$ depends on $T_\mathrm{d}$ and the dust spectral index,
$\beta$; the two properties are tightly connected \citep{Planck2014a}.
The anti-correlation between $T_\mathrm{d}$ and $\beta$ is apparent in the northern region
while the southern region presents more dispersion.
Differences in these dust properties suggest different grain evolution \citep[e.g.,][]{Jones2013,Kohler2015}.
How this grain evolution is related to the observed $\tau_{353}$ dependence of the 
$\NH$/$\tau_{353}$ ratio in the southern region
is not clear.
In the following, we will focus on discussing implications of Figure~12.

\begin{figure}[ht!]
\figurenum{12}
\gridline{
\fig{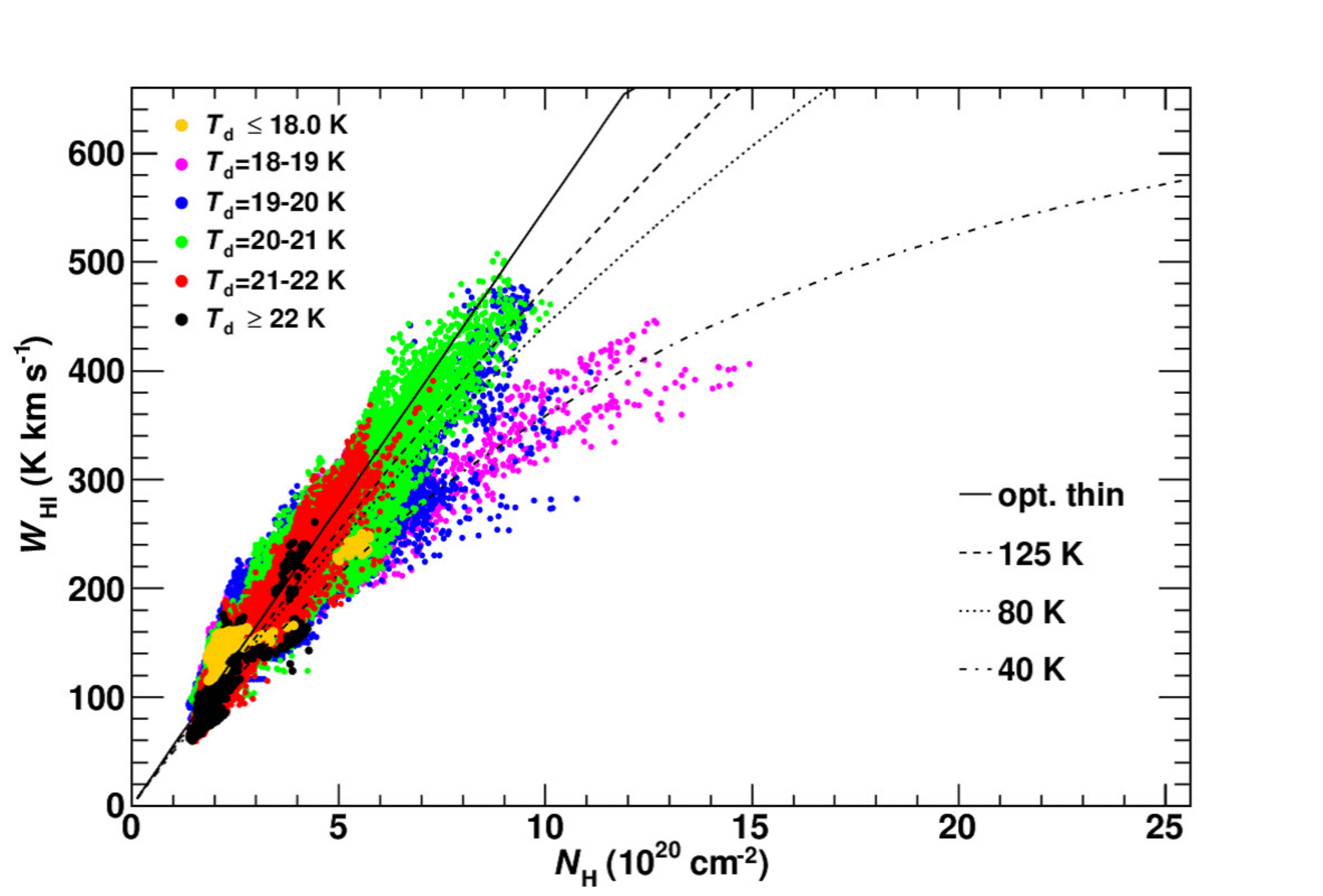}
{0.5\textwidth}{(a)}
\fig{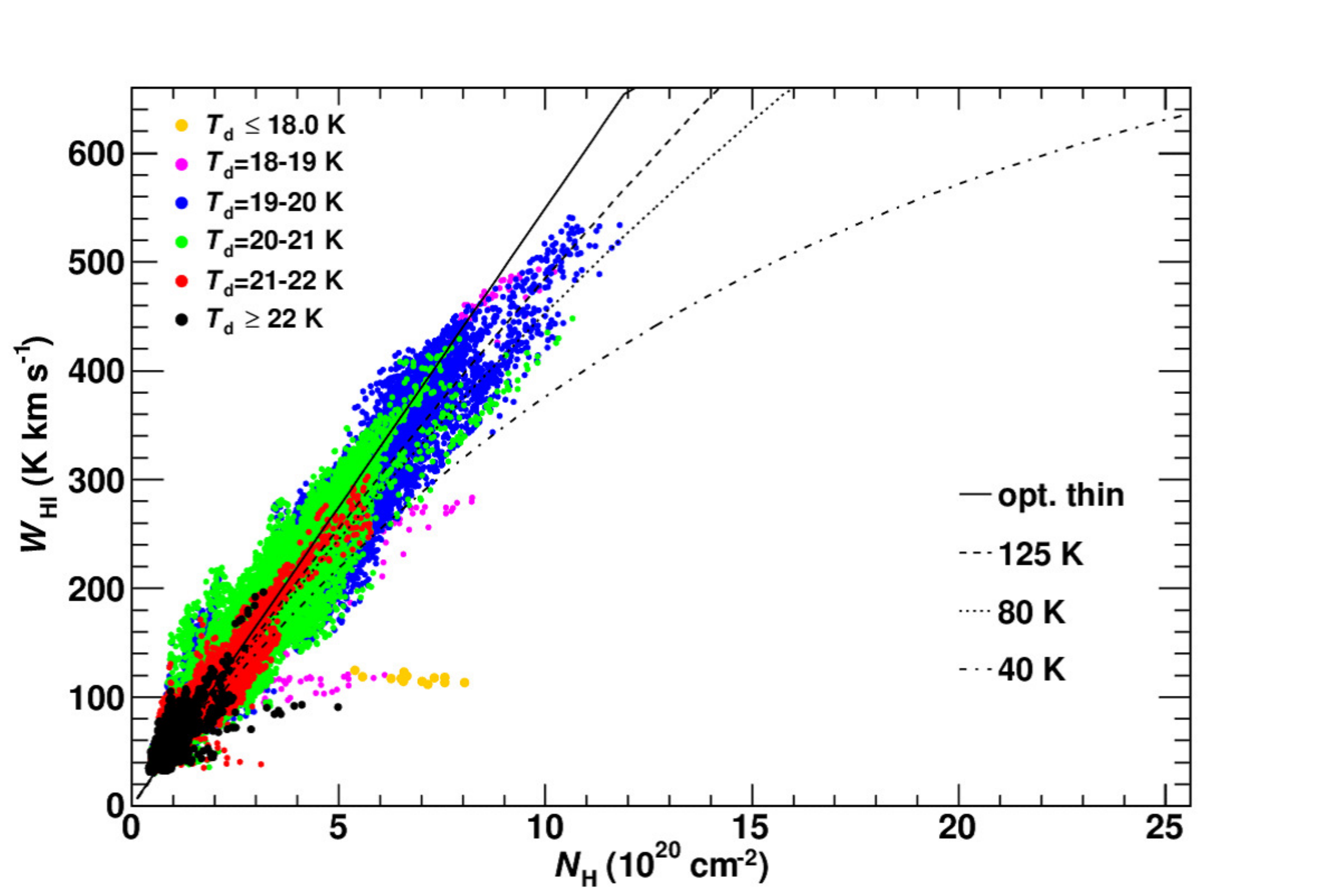}
{0.5\textwidth}{(b)}
}
\gridline{
\fig{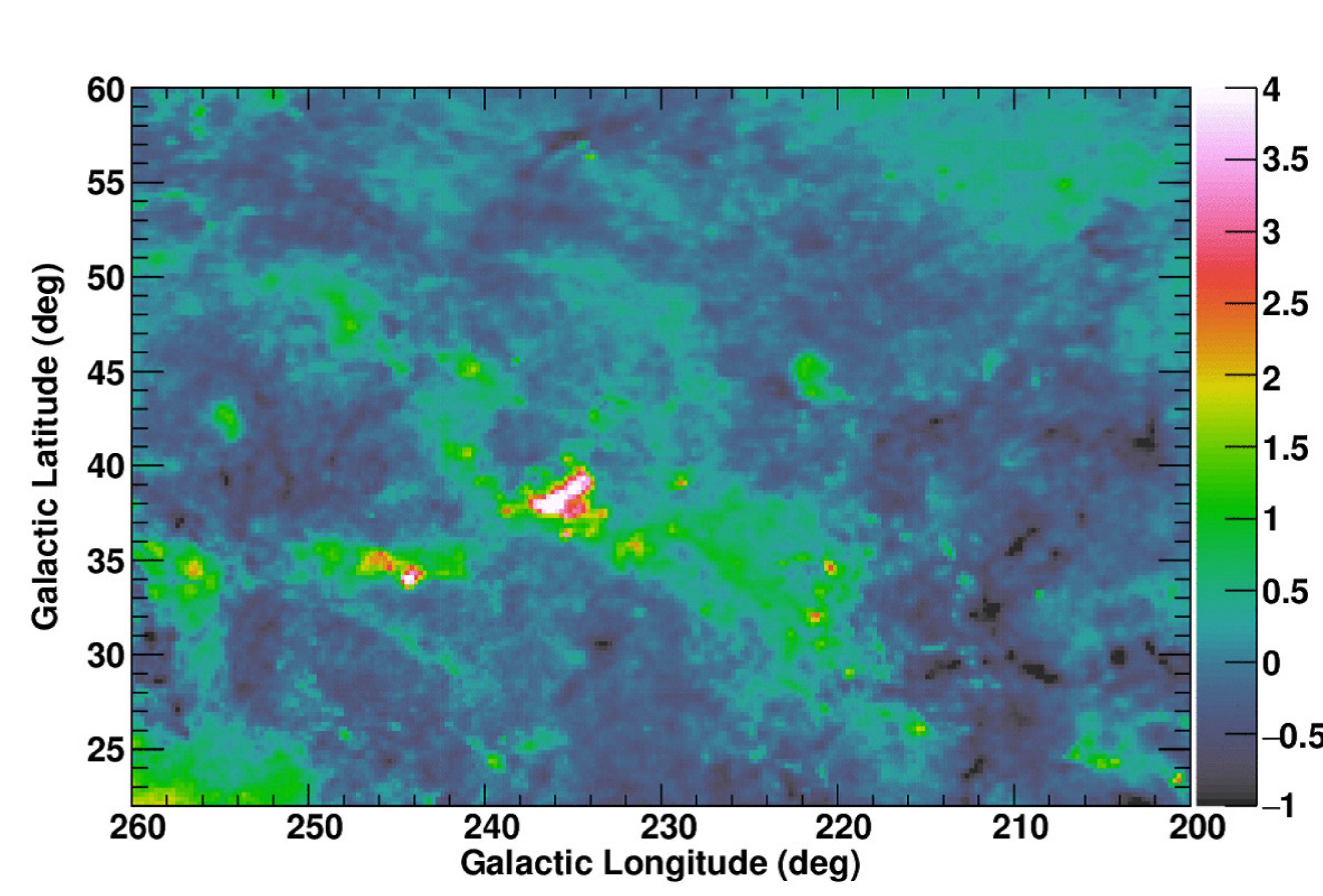}
{0.5\textwidth}{(c)}
\fig{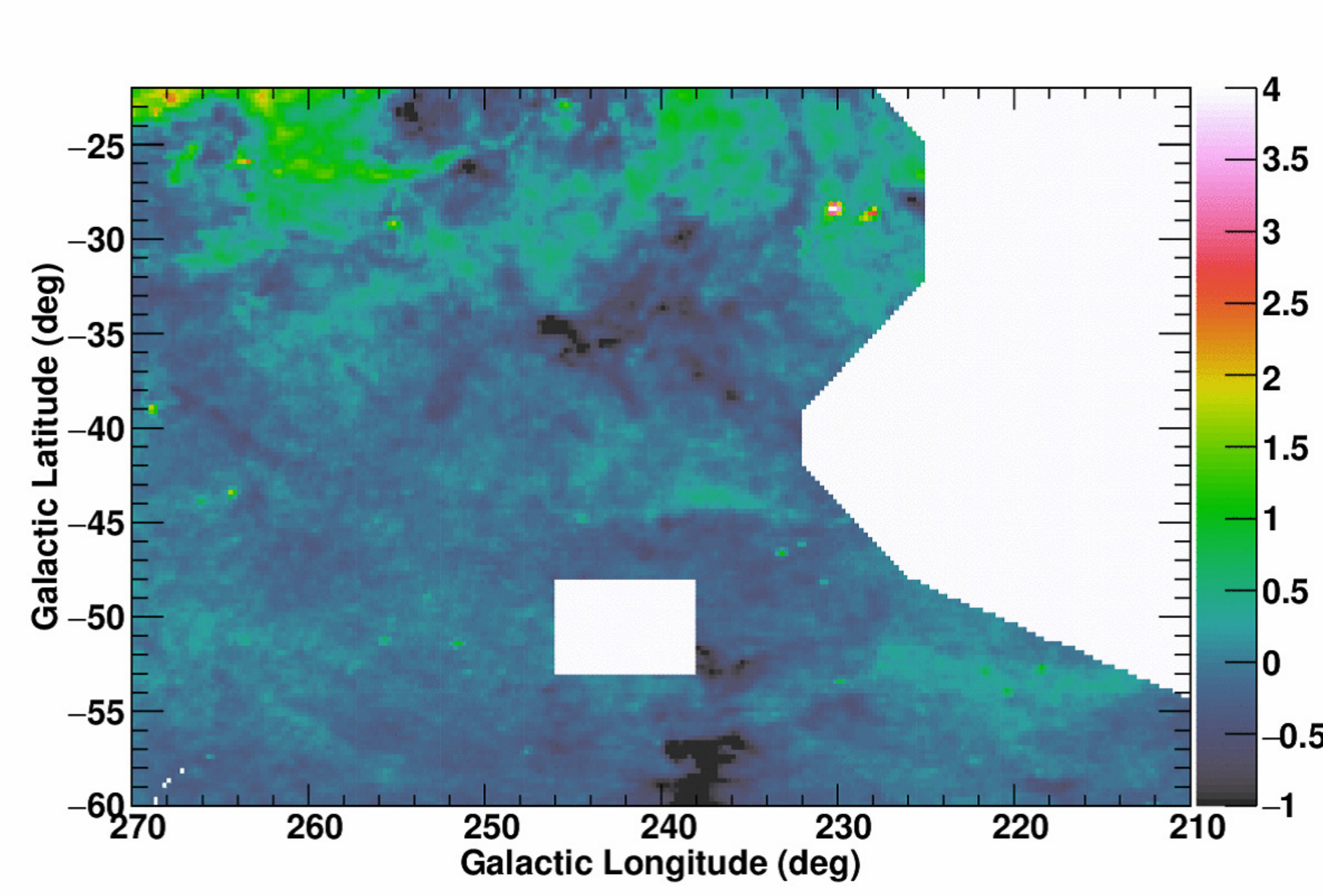}
{0.5\textwidth}{(d)}
}
\caption{
The correlation between $\WHI$ and $\NH$ inferred from the $\gamma$-ray data analysis
in the (a) northern and (b) southern regions, 
and the excess gas column density map (defined as $\NH-\NHIthin$) 
in units of $10^{20}~\mathrm{cm^{-2}}$ in the (c) northern and (d) southern regions.
The model curves for several choices of $T_\mathrm{S}$ are overlaid on panels (a) and (b).
\label{fig:f12}
}
\end{figure}

In the northern region (panel~(a)), the $\NH/\WHI$ ratio in the area of low $T_\mathrm{d}$ (18--19~K)
and high $\WHI$ ($\ge 300~\mathrm{K~km~s^{-1}}$)
is greater than those in other areas. This area corresponds to 
the excess gas column density at around
$(l, b) = (236\arcdeg, 38\fdg5)$ seen in panel~(c). It also corresponds to
the positive residual at
around $(l, b) = (236\arcdeg, 37\fdg5)$ seen in the $\WHI$-based analysis 
(see also Figure~6(a) and Section~4.1.1).
Because our ROI does not include strong CO emission (see Appendix~C) and 
the data for the area of interest spans a wide range of $\WHI$,
these gas-related emissions are likely due to
optically thick $\HI$ and we will consider this scenario hereafter. 
The brightness temperature for the $\HI$ emission at the velocity $v$ is given by
\begin{equation}
T_\mathrm{b}(v) = [T_\mathrm{S}(v)-T_\mathrm{bg}] \cdot [1-\exp(-\THI(v))]~~,
\end{equation}
where $T_{\rm bg}$ is the background continuum radiation temperature, and 
$T_\mathrm{S}(v)$ and $\THI(v)$ are, respectively, a harmonic mean of the spin temperature at velocity $v$ on the line of sight
and an integration of the optical depth at this velocity. Then, if we approximate the $\HI$ emission spectrum
by a single boxcar spectrum on the line of sight
with a spectral width of $\Delta V_{\HIs}$, $T_\mathrm{S}(v)$ and $\THI(v)$ can be expressed by single values
independent of $v$ and thus the 
$\WHI$ can be correlated to $\NH$ as a function of $T_\mathrm{S}$ such that (e.g., \citet{Fukui2015})
\begin{equation}
\WHI (\mathrm{K~km~s^{-1}}) = [T_\mathrm{S}({\rm K})-T_\mathrm{bg}({\rm K})]
\cdot \Delta V_{\HIs}({\rm km~s^{-1}}) \cdot [1-\exp(-\THI)]~~,
\end{equation}
and 
\begin{equation}
\THI = \frac{N_{\rm H_{tot}}({\rm cm^{-2}})}{1.82 \times 10^{18}}
\cdot \frac{1}{T_\mathrm{S}({\rm K})}
\cdot \frac{1}{\Delta V_{\HIs} ({\rm km~s^{-1}})}~~,
\end{equation}
where $\Delta V_{\HIs}$ is 
defined as $\WHI$/(peak $\HI$ brightness temperature).
In Figure~12(a),
assuming that all of the gas is atomic,
we overlaid the model curves for several choices of $T_\mathrm{S}$ with
$\Delta V_{\HIs}={\rm 18~km~s^{-1}}$ (the median linewidth in the northern region).
As inferred from Figure~12(a), while most of the region is compatible with being optically thin,
the area with $T_\mathrm{d} = 18\mbox{--}19~\mathrm{K}$ gives, on average, $T_\mathrm{S} \sim 40~\mathrm{K}$.

In the southern region (Figure~12(b)), while most of the data lie along a mildly curved line, 
those in regions with $T_\mathrm{d} \le 18~\mathrm{K}$
show a flat profile with $\WHI \sim 100~\mathrm{K~km~s^{-1}}$ in the plot. 
This corresponds to a spot seen in the dust data
at ($l$, $b$) $\sim$ 
($230\arcdeg$, $-28\fdg5$) (see also Figure~12(d));
it is also seen in the residual map for the $\WHI$-based analysis (Figure~9(a)).
A plausible interpretation is that the spot consists of
CO-dark $\Htwo$ \citep[e.g.,][]{Smith2014}
because the flat profile means that the column density of $\HI$ is nearly constant.
Finally, we overlaid the model curves for several choices of $T_\mathrm{S}$ with
$\Delta V_{\HIs}={\rm 21~km~s^{-1}}$ (the median velocity of the southern region with several areas masked);
the majority of data agree, on average, with the model curve for $T_\mathrm{S} = 125~\mathrm{K}$,
supporting the choice of $T_\mathrm{S}$ by \citet{FermiHI}. This value also agrees well
with the average value found in the local ISM, $T_\mathrm{S} = 140~\mathrm{K}$ \citep{FermiHI2}. 
We also note that there is a scatter around the model curve for $T_\mathrm{S} = 125~\mathrm{K}$.
Because $\WHI$ and the optically-thin approximation gives the lower limit of $\NH$, the spread
is likely because of the uncertainty of the $\Dem/\NH$ ratio, and thus could introduce over- and under-estimation
of the $\NH$. For example, negative values of the excess gas column density are seen around
$(l, b) = (240\arcdeg, -35\arcdeg)$ in Figure~12(d), and they correspond to the residual
(i.e., underestimate of $\NH$) in Figure~9(b) and (c). This is a drawback of employing $\Dem$ as a tracer
of the total gas column density, and the small scatter around model curves in Figure~12 should not be
taken at face value.

We point out that the $\NH$ in Figure~12 is proportional to $R$. Although this is
our best estimate of $\NH$ based on the correlation between the $\gamma$-ray data and gas templates,
we did not measure the $\NH$ distribution on pixel scales. Accordingly, overinterpretation
of Figure~12 (e.g., estimating the $T_\mathrm{S}$ and the excess gas column density on very small scales)
should be avoided.

The average column density of the neutral gas ($\overline{\NH}$) 
in the northern region is obtained as ${\sim}3.4 \times 10^{20}~\mathrm{cm^{-2}}$
based on either $\WHI$ (with optically thin approximation) or
$R$ (with the conversion factor determined in Section~2.2).
On the other hand, that based on $\tau_{353}$ is
${\sim}4.3 \times 10^{20}~\mathrm{cm^{-2}}$, indicating that an $\NH \propto \tau_{353}$ model
(not favored by $\gamma$-ray analysis) overestimates the gas column density by ${\sim}30\%$.
In the southern region, again, while the values of $\overline{\NH}$ inferred by $\WHI$ and $R$
are similar (${\sim}2.2 \times 10^{20}~\mathrm{cm^{-2}}$), that based on $\tau_{353}$ is ${\sim}15\%$ higher.
This supports our earlier statement that the use of
the $\gamma$-ray data is crucial to accurately determine the $\NH$ distribution.


\clearpage

\subsection{CRs in the Local Environment}

Next, we discuss the $\HI$ emissivity spectra obtained in this study, which are summarized 
in Figure~13(a) and (b).
To investigate possible systematic uncertainties due to the choice of the $\NH$ template and the IC model,
in the northern region (where the normalization of the IC model is ${\sim}0$ for all six models) 
we bracketed the spectrum with that
obtained using the $\NH$ model based on $\WHI$ (i.e., the pure optically thin $\HI$ scenario) 
and with that
obtained using the $\NH$ model based on $R$ but with the normalization of the IC model fixed to 1.0.
For the southern region, we bracketed the spectrum those obtained using all six IC models.
We also took into account the LAT effective area uncertainty
\footnote{\url{http://fermi.gsfc.nasa.gov/ssc/data/analysis/LAT_caveats.html}};
we assumed the uncertainty to be 10\% below 200~MeV (where we used only events of PSF classes 2 and 3)
and 5\% above 200~MeV.
The fractional uncertainties of the spectrum due to the modeling ($\NH$ and IC
for the northern and southern regions, respectively) and that due to the effective area uncertainty
are summed in quadrature and shown as shaded bands;
they are 11\%--13\% below 200~MeV and $\la 10\%$ above 200~MeV.
For comparison,
we plotted the model curve for
the LIS that we adopted with $\epsilon_\mathrm{M}$ of 1.84 in the same figure.
In order to approximately indicate the uncertainty in the emissivity model
(primarily due to the uncertainty in the elemental composition of the CRs and the cross sections
other than for proton--proton (p--p) collisions), 
we also plotted the model curve for
$\epsilon_{\rm M}=1.45$
\citep[the lowest value referred to in][]{Mori2009},
which gives 15\%-20\% lower emissivity.
We also plotted the emissivity spectrum of the same ROI measured by
\citet{FermiHI} using six months of \textit{Fermi}-LAT data.
Our results favor the model curve with $\epsilon_\mathrm{M}=1.84$ and agree well with those by \citet{FermiHI},
but here we cover a wider energy range and investigate northern/southern regions separately.
In other words, the analysis presented here shows for the first time that 
the $\HI$ emissivities are consistent between the northern and southern regions at the 10\% level,
supporting the hypothesis that the CR intensity is uniform in the vicinity of the solar system.
The integral emissivities above 100~MeV are
$(1.58\pm0.04) \times10^{-26}~\mathrm{photons~s^{-1}~sr^{-1}~H\mbox{-}atom^{-1}}$
and 
$(1.59\pm0.02) \times10^{-26}~\mathrm{photons~s^{-1}~sr^{-1}~H\mbox{-}atom^{-1}}$
in the northern and southern regions, respectively,
and those above 300~MeV are
$(0.68\pm0.01) \times10^{-26}~\mathrm{photons~s^{-1}~sr^{-1}~H\mbox{-}atom^{-1}}$
and 
$(0.69\pm0.01) \times10^{-26}~\mathrm{photons~s^{-1}~sr^{-1}~H\mbox{-}atom^{-1}}$
in the northern and southern regions, respectively, with an additional systematic error of $\sim10\%$
due to the modeling and the effective area uncertainties (see above).
We also note that our emissivities agree (within ${\le}10\%$) with the results of by \citet{Shen2019}, where
the authors analyzed the same northern region employing 
a template-fitting method 
with the assumption of a uniform $T_\mathrm{S}$ for the atomic gas phase
and discussed the local CR spectrum
based on recent p--p interaction models and AMS--02 data. In other words, our analysis supports their findings
by showing that $T_\mathrm{S}=125~\mathrm{K}$ is compatible with most of the $\NH$ distribution 
and that the CR spectrum is uniform.

As in the case of discussing the ISM gas densities (Section~5.1), evaluating the gas model using the $\gamma$-ray data
is crucial to accurately constrain the $\HI$ emissivity and CR intensity. If we use the $\NH \propto \tau_{353}$ models,
the scale factors of the $\HI$ emissivity ($\propto$ CR intensity) are 30\%--40\% lower 
(see Sections~4.1.1 and 4.2.1).
Other source of uncertainty on the CR intensity are the
hadronic interaction cross section and the elemental
composition of CRs as indicated by two curves ($\epsilon_{\rm M} = 1.84$ and 1.45) in Figure~13(a).
If we adopt $\epsilon_{\rm M} = 1.45$, we would need ${\sim}25$\% higher proton LIS flux, which might be incompatible
with the proton flux directly measured at the Earth. Given the uncertainty and the fact that the directly measured
CRs do not necessarily represent the LIS, we do not deny such a possibility. See discussions in,
e.g., \citet{LIS1}, \citet{LIS2}, and \citet{Shen2019}.
In Figure~13(a), one may also
recognize that the model overestimates the data below a few 100 MeV
while it predicts lower flux above 1~GeV. This might indicate a possible spectral break of the proton LIS.
For example, \citet{LIS1} reported a break at a few GeV
based on \citet{FermiHI2} that gives a similar spectrum to ours
(see also Figure~13(b)). To reach a robust conclusion, constraining the electron LIS using radio synchrotron emission
\citep[e.g.,][]{LIS2} and 
an accurate determination of the emissivity spectrum below a few 100 MeV is crucial. 
Because the analysis suffers from coupling with
the point sources through IC model
in low-energy bands (see Section~4.1.1), we defer such a study to future projects using gas-rich areas.

\begin{figure}[ht!]
\figurenum{13}
\gridline{
\fig{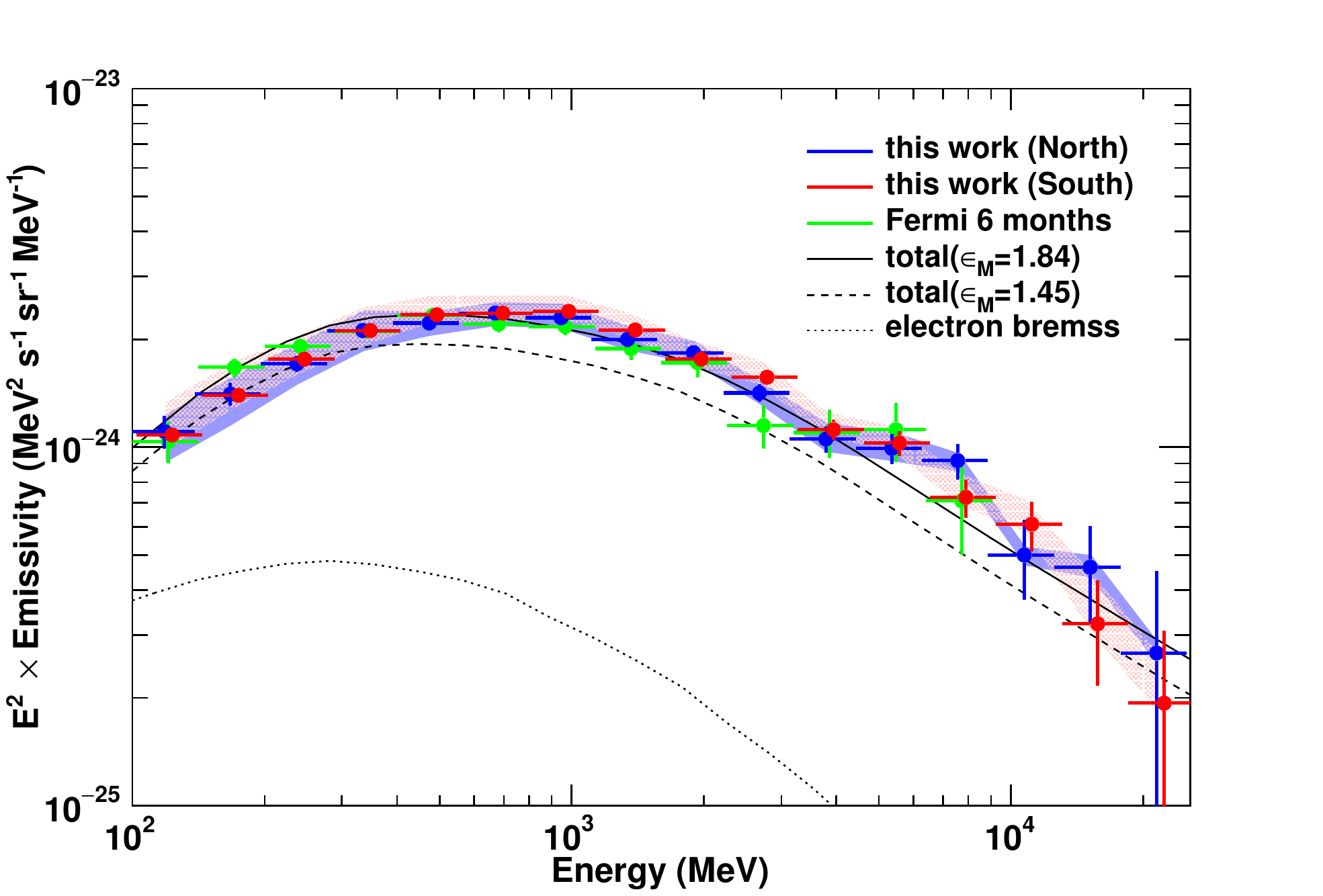}
{0.5\textwidth}{(a)}
\fig{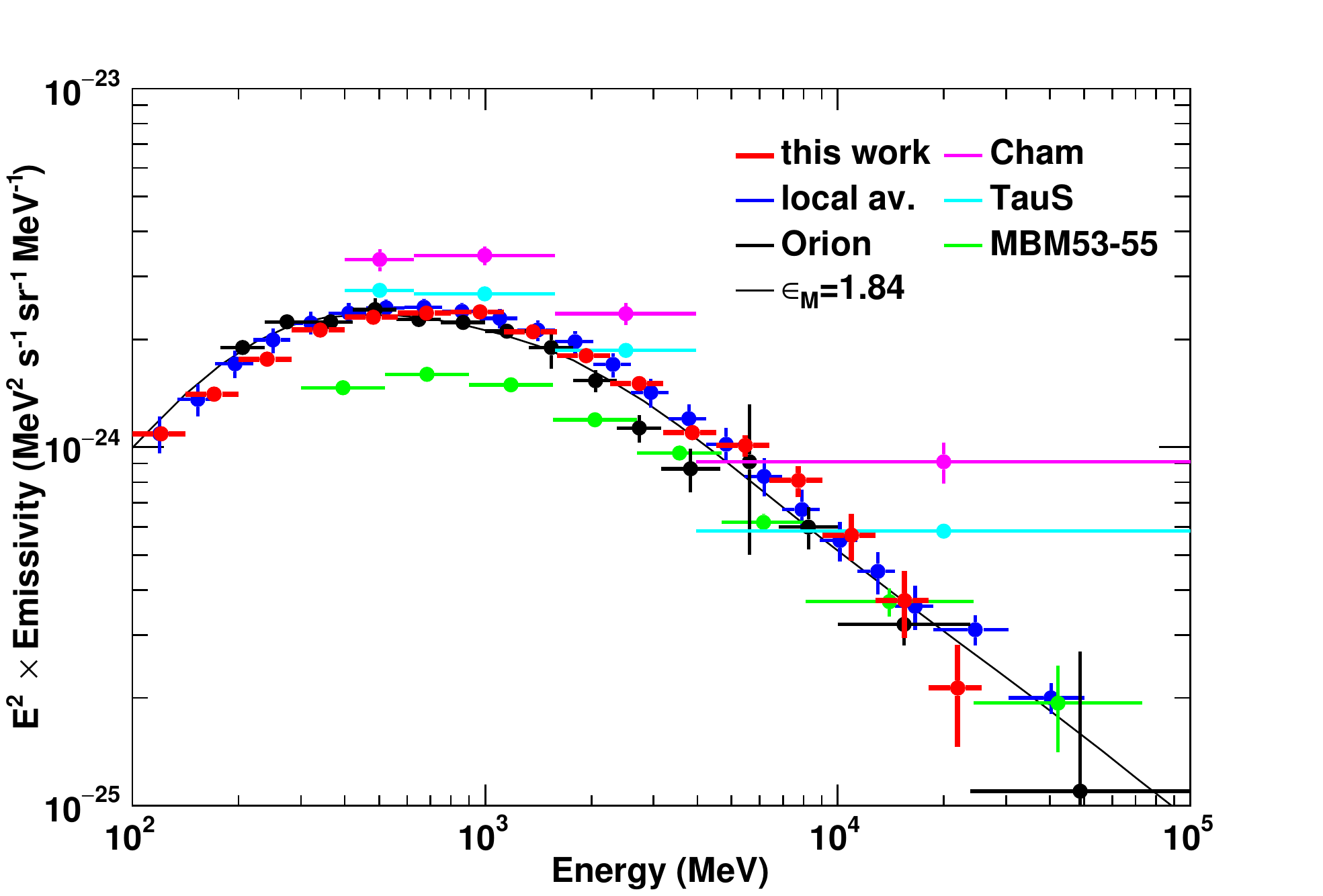}
{0.5\textwidth}{(b)}
}
\caption{
(a) Summary of the $\HI$ emissivity spectra of the northern and southern regions.
They are compared with the model curves based on the LIS for
$\epsilon_{\rm M}=$ 1.45 and 1.84, and the result of the relevant study by \textit{Fermi}-LAT
based on six months of observation \citep{FermiHI}.
The contribution of the electron bremsstrahlung is also shown.
The shaded bands show the systematic uncertainties of the spectrum
(see the text in Section~5.2 for details).
(b) The average of the $\HI$ emissivity spectra obtained in this study compared with previous
\textit{Fermi}-LAT results for high latitude areas. Errors are statistical only.
\label{fig:f13}
}
\end{figure}

Finally, we compare our $\HI$ emissivity spectrum (the average of the northern and southern regions) with several other \textit{Fermi}-LAT studies
of nearby clouds: the average spectrum found in the local ISM in $10\arcdeg \le |b| \le 70\arcdeg$ by \citet{FermiHI2},
that toward the Orion molecular clouds by \citet{FermiOrion}, that toward the Chamaeleon molecular clouds by \citet{Planck2015},
that toward the South Taurus cloud by \citet{Remy2017}, and that toward the MBM 53, 54, 55 clouds and the Pegasus loop by \citet{Mizuno2016},
as summarized in Figure~13(b).
Although the spectral shape does not change significantly over the samples examined here,
the peak-to-peak variation of the normalization is by
a factor of $\sim$2 even in nearby clouds. Given the diffusive nature of the Galactic CRs,
and the lack of significant change of the spectral shape, the variation is mostly attributable to
uncertainties of the gas column density,
particularly due to assumptions of the value of $T_\mathrm{S}$.
For example, 
as discussed by \citet{Planck2015}, the emissivity toward the Chamaeleon clouds
agrees with that found by \citet{FermiHI2} within $\sim20\%$ 
if we assume $T_\mathrm{S}=140~\mathrm{K}$ for $\HI$ clouds,
although the $\gamma$-ray fit favors higher $T_\mathrm{S}$.
A small gradient at the 20\% level could be possible and of interest to understand
the CR generation and propagation in the vicinity of the solar system, 
and a systematic study of nearby clouds is therefore important.
In such future studies, one should overcome the uncertainty on the $\NH$ distribution
by using the $\HI$, CO, and dust data together with GeV $\gamma$-rays as a robust tracer of the ISM gas.

\clearpage

\section{Summary and Future Prospects}

We performed a detailed study of the ISM and CRs 
in the mid-latitude region of the third quadrant
using the \textit{Fermi}-LAT data in the 0.1--25.6~GeV range and other interstellar gas tracers such as
the HI4PI survey and the \textit{Planck} dust model.
Even though this region was analyzed in an early publication 
of the \textit{Fermi}-LAT collaboration using six months of data,
the analysis was significantly improved
using eight years of \textit{Fermi}-LAT data with the aid of newly available gas tracers
and with the northern and southern regions treated separately.
We used $\gamma$-rays as as a robust tracer of the ISM gas and examined possible variations
of the $\NH$/$\Dem$ ratio. We tested several IC models and confirmed that the effect on the
$\NH$/$\Dem$ ratio is at the 10\% level, and also confirmed that 
the uncertainty of the Sun/Moon emission model does not affect
the gas component.
We found that dust opacity at 353~GHz increases in low $T_\mathrm{d}$ or high density areas 
for the northern and southern regions, respectively.
On the other hand, the $\gamma$-ray analysis preferred the $R$-based $\NH$ template
in both northern and southern regions, 
and we adopted 
$R$-based $\NH$ models as our best estimate of the $\NH$ distribution.
While most of the gas can be interpreted as being $\HI$ of
$T_\mathrm{S} = 125~\mathrm{K}$ or higher, an area of optically thick $\HI$ of
$T_\mathrm{S} \sim 40~\mathrm{K}$ was revealed
and possible CO-dark $\Htwo$ was identified.
The measured integrated $\gamma$-ray emissivities above 100~MeV were found to be
$(1.58\pm0.04)\times10^{-26}~\mathrm{photons~s^{-1}~sr^{-1}~H\mbox{-}atom^{-1}}$
and 
$(1.59\pm0.02)\times10^{-26}~\mathrm{photons~s^{-1}~sr^{-1}~H\mbox{-}atom^{-1}}$
in the northern and southern regions, respectively, 
supporting the existence of uniform CR intensity in the vicinity of the solar system.
Although our emissivity agrees with the calculation
using the LIS model based on the directly measured CR proton spectrum
with $\epsilon_{\rm M}=1.84$, we caution that the uncertainty of the $\gamma$-ray emissivity model
is still at the 20\% level.
The choice of the ISM gas tracer and the correction of the $\NH$ model using $\gamma$-ray data are crucial
to accurately measure the ISM gas distribution and investigate the CR intensity.
As discussed by \citet{Mizuno2016}, the $\NH/\Dem$ ratio was found to depend on $T_\mathrm{d}$ 
in the MBM~53, 54, and 55 clouds and the Pegasus loop through $\gamma$-ray data analysis.
Now we find, through this study, that the $\NH/\tau_{353}$ ratio depends also on $\tau_{353}$
as predicted by several dust evolution models.
In the present study we demonstrated that, in order to accurately measure the ISM gas distribution and study the
CR intensity and spectrum, the dependence on both $T_\mathrm{d}$ and $\Dem$ needs to be taken into account and a
detailed examination of the $\WHI$--$\Dem$ relationship is required. This work may serve as a reference for
future studies of nearby $\HI$/CO clouds.
\\

The \textit{Fermi} LAT Collaboration acknowledges generous ongoing support
from a number of agencies and institutes that have supported both the
development and the operation of the LAT as well as scientific data analysis.
These include the National Aeronautics and Space Administration and the
Department of Energy in the United States, the Commissariat \`a l'Energie Atomique
and the Centre National de la Recherche Scientifique / Institut National de Physique
Nucl\'eaire et de Physique des Particules in France, the Agenzia Spaziale Italiana
and the Istituto Nazionale di Fisica Nucleare in Italy, the Ministry of Education,
Culture, Sports, Science and Technology (MEXT), High Energy Accelerator Research
Organization (KEK) and Japan Aerospace Exploration Agency (JAXA) in Japan, and
the K.~A.~Wallenberg Foundation, the Swedish Research Council and the
Swedish National Space Board in Sweden.

Additional support for science analysis during the operations phase is gratefully
acknowledged from the Istituto Nazionale di Astrofisica in Italy and the Centre
National d'\'Etudes Spatiales in France. This work performed in part under DOE
Contract DE-AC02-76SF00515.

We would like to thank the referee for his/her valuable comments. This work was partially supported by JSPS Grants-in-Aid for Scientific Research (KAKENHI) Grant Numbers JP17H02866 (T.~M.) and JP26800160 (K.~H.), and Core Research Energetic Universe in Hiroshima University.

Some of the results in this paper have been derived using the HEALPix
\citep{Gorski2005} package.

\vspace{5mm}
\facilities{Fermi,Planck,WMAP,Parkes,Effelsberg}

\software{\textit{Fermi} Science Tools (\url{http://fermi.gsfc.nasa.gov/ssc/data/analysis/software/}), 
GALPROP \citep{Galprop1,Galprop2}, HEALPix \citep{Gorski2005},  ROOT (\url{https://root.cern.ch})}

\clearpage

\appendix

\section{Velocity-Sorted $\WHI$ Maps}

The velocity-sorted $\WHI$ maps are summarized in Figures~14 and 15 for the northern and southern regions, respectively.
As shown in the panels (a) and (d) of Figure~14, most of the gas is local (the velocity $|v| \le 35~\mathrm{km~s^{-1}}$) 
in the northern region.
Three bright radio continuum sources ($\WHI \ge 50~\mathrm{K~km~s^{-1}}$)
are seen in $|v| \ge 70~\mathrm{km~s^{-1}}$ at
($l$, $b$) around ($246\fdg1$, $39\fdg9$), ($233\fdg2$, $43\fdg8$), and ($208\fdg6$, $44\fdg5$).
They were removed by filling the source areas with the average of the peripheral pixels: 
values in a circular region
with radius $0\fdg4$ are filled with the average of pixels in a ring with
inner radius of $0\fdg4$ and outer radius of $0\fdg5$ in the $\WHI$ map.
The parameters (position and radius) are summarized in Table~7.

\begin{figure}[ht!]
\figurenum{14}
\gridline{
\fig{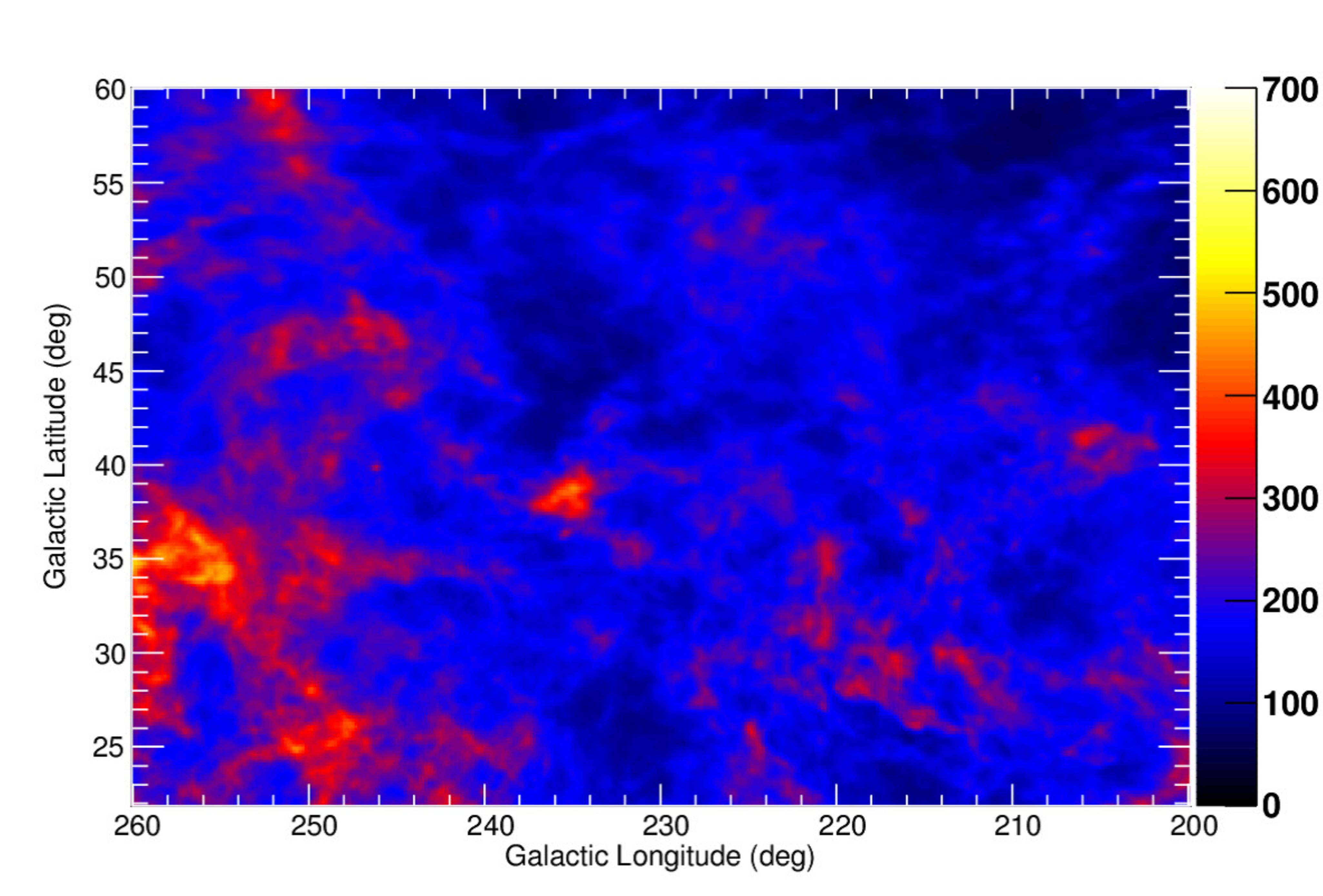}
{0.33\textwidth}{(a)}
\fig{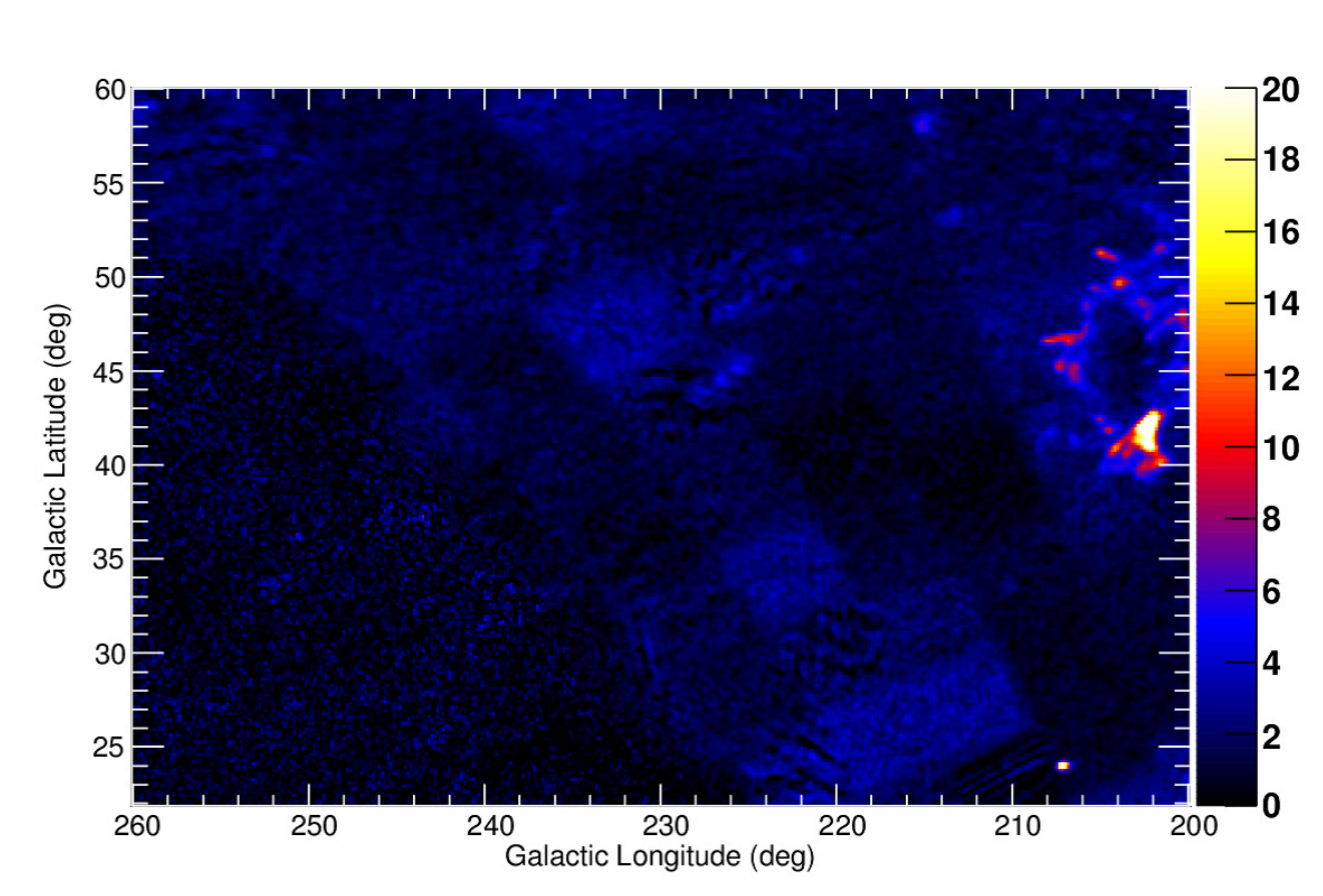}
{0.33\textwidth}{(b)}
\fig{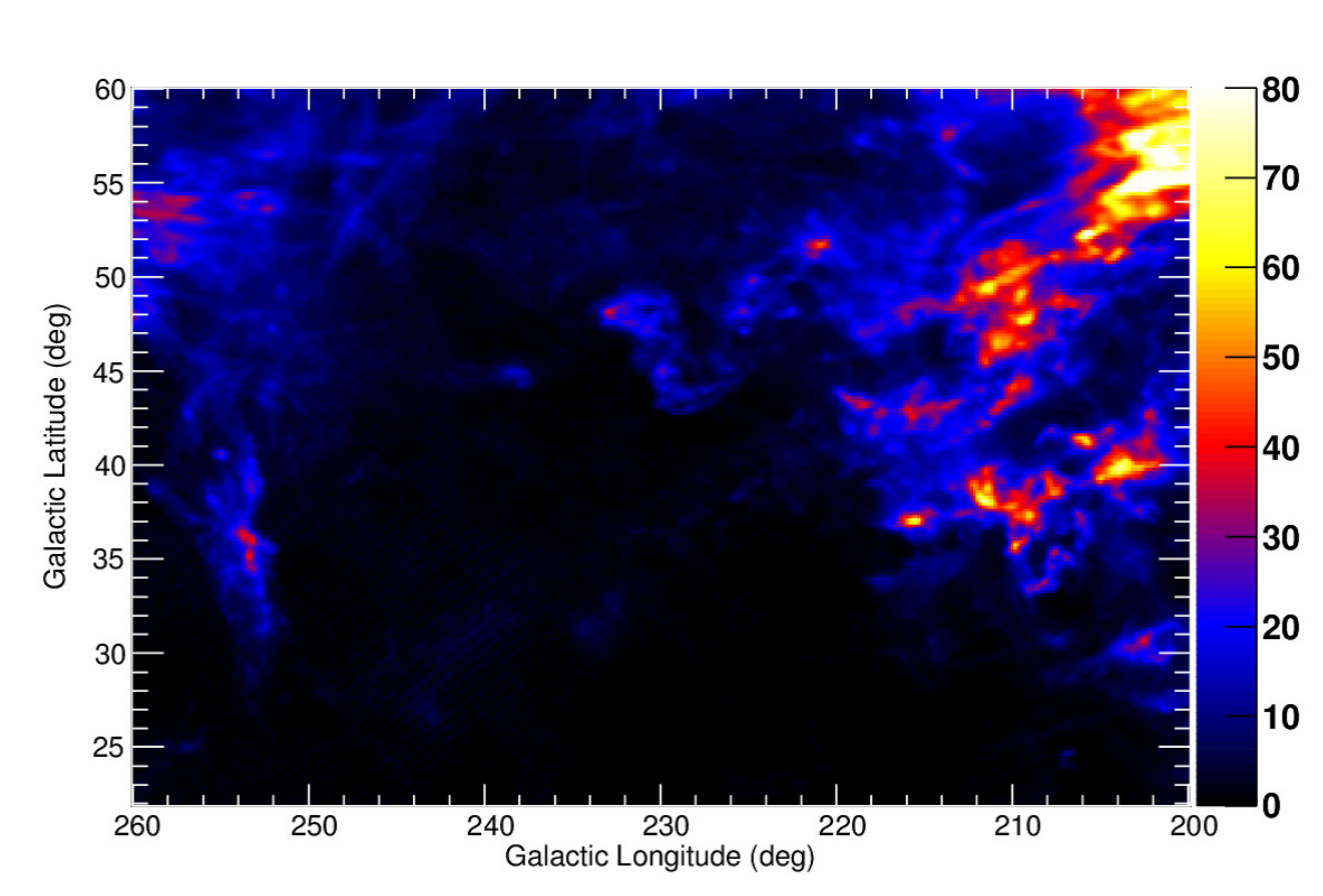}
{0.33\textwidth}{(c)}
}
\gridline{
\fig{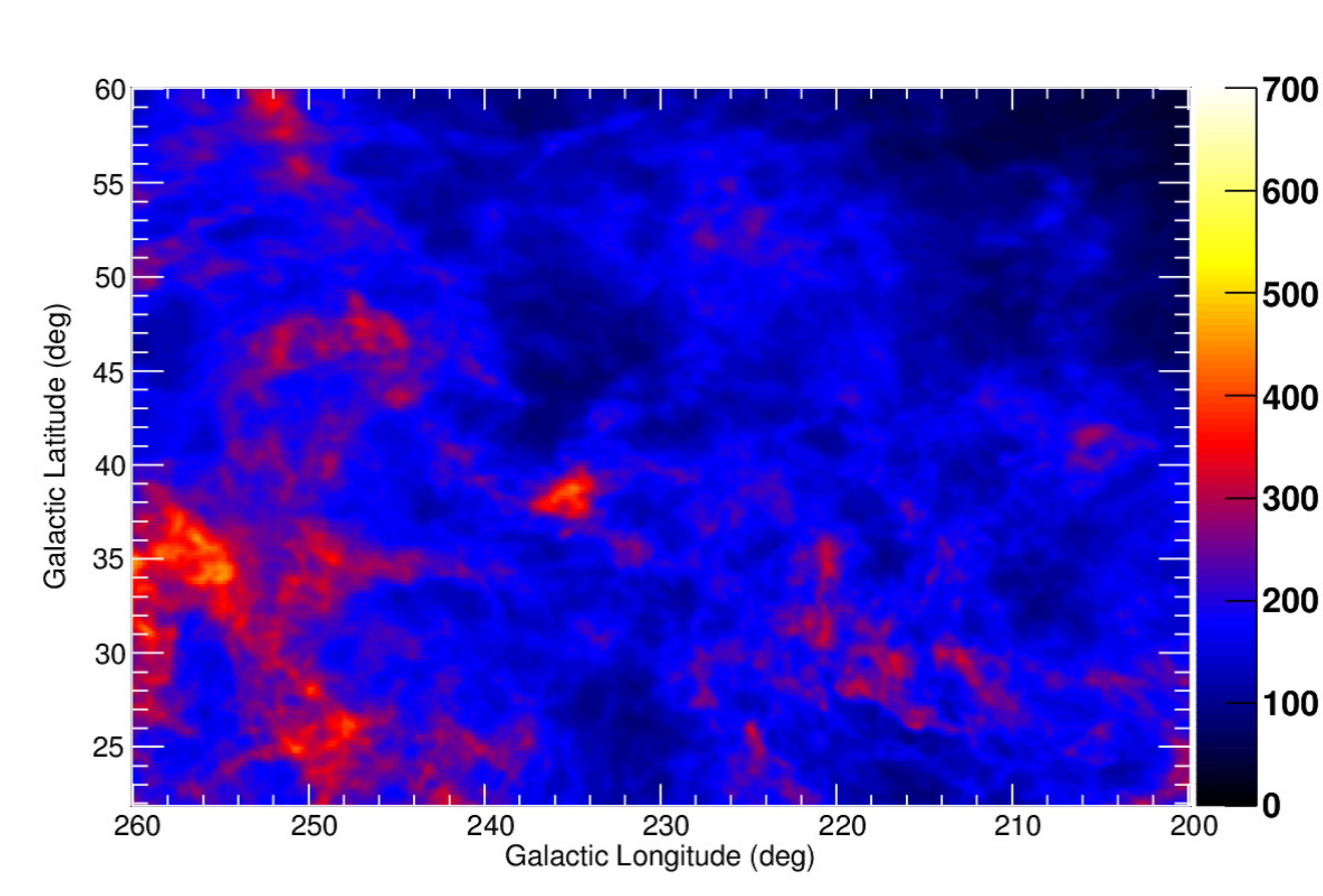}
{0.33\textwidth}{(d)}
\fig{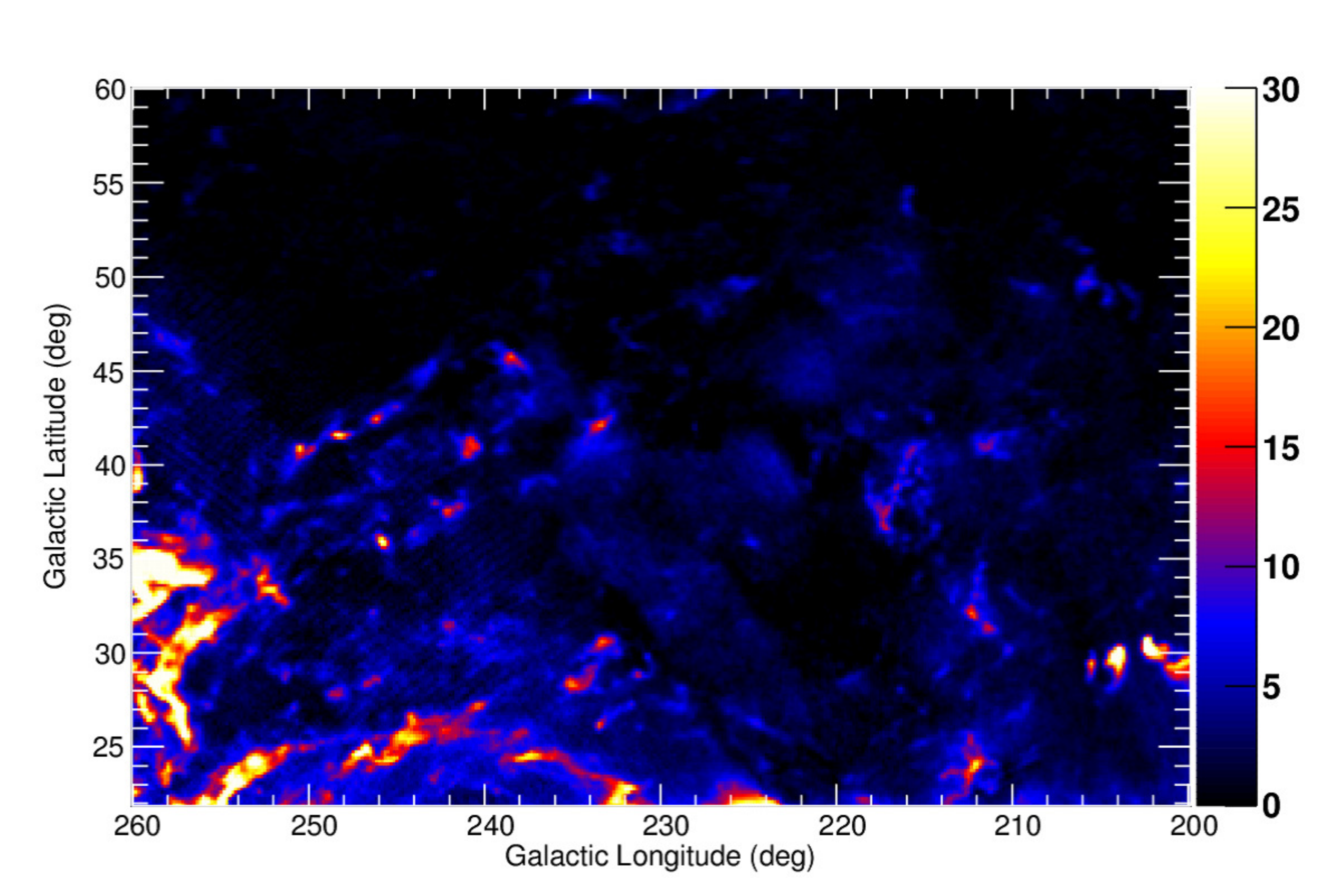}
{0.33\textwidth}{(e)}
\fig{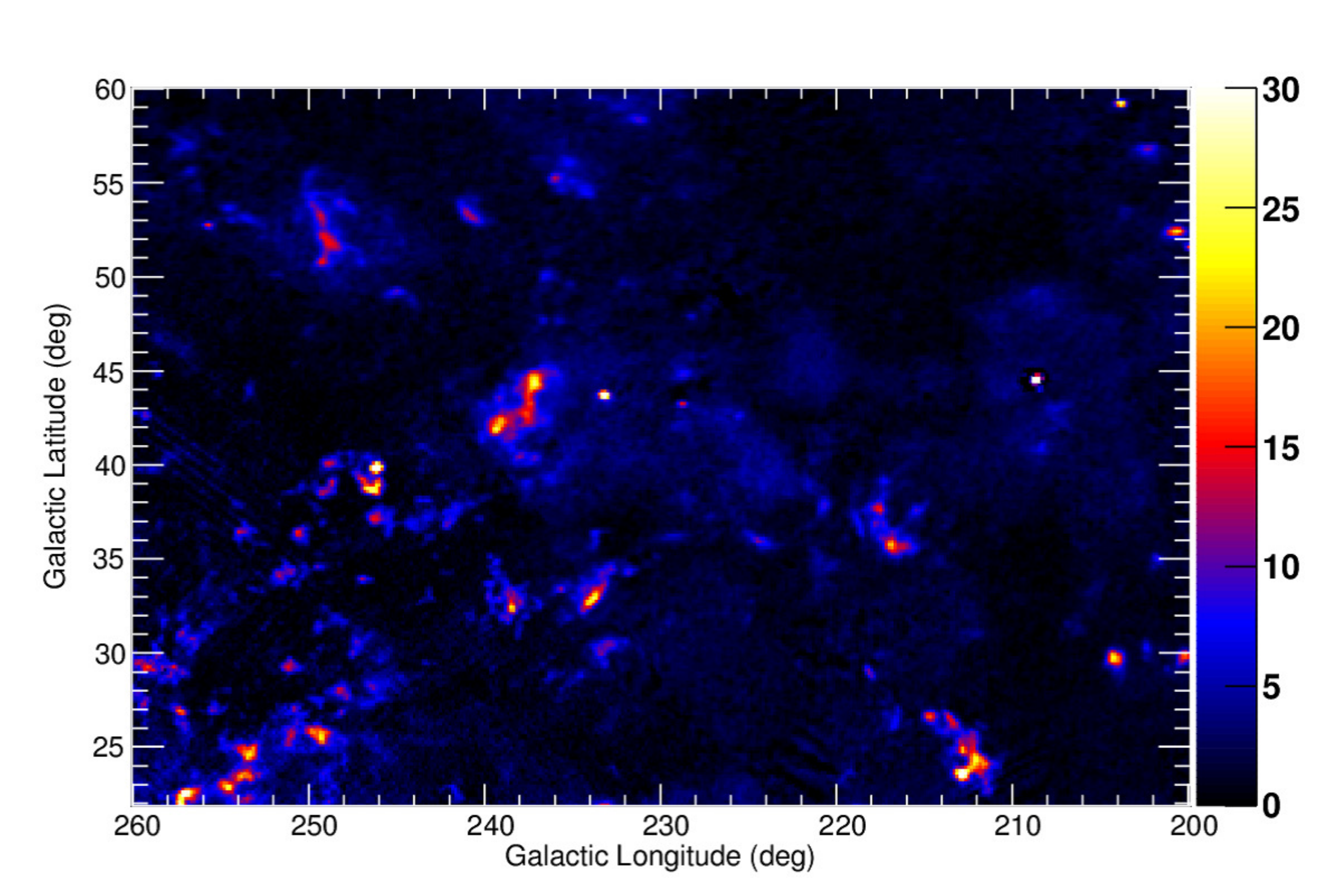}
{0.33\textwidth}{(f)}
}
\caption{
$\WHI$ maps ($\mathrm{K~km~s^{-1}}$)
in the northern region (a) integrated over the whole velocity range (from $-600$ to $+600~\mathrm{km~s^{-1}}$)
and in the velocities 
(b) from $-600$ to $-70~\mathrm{km~s^{-1}}$, 
(c) from $-70$ to $-35~\mathrm{km~s^{-1}}$, 
(d) from $-35$ to $35~\mathrm{km~s^{-1}}$, 
(e) from $+35$ to $+70~\mathrm{km~s^{-1}}$, and
(f) from $+70$ to $+600~\mathrm{km~s^{-1}}$.
\label{fig:f101}
}
\end{figure}

\begin{figure}[ht!]
\figurenum{15}
\gridline{
\fig{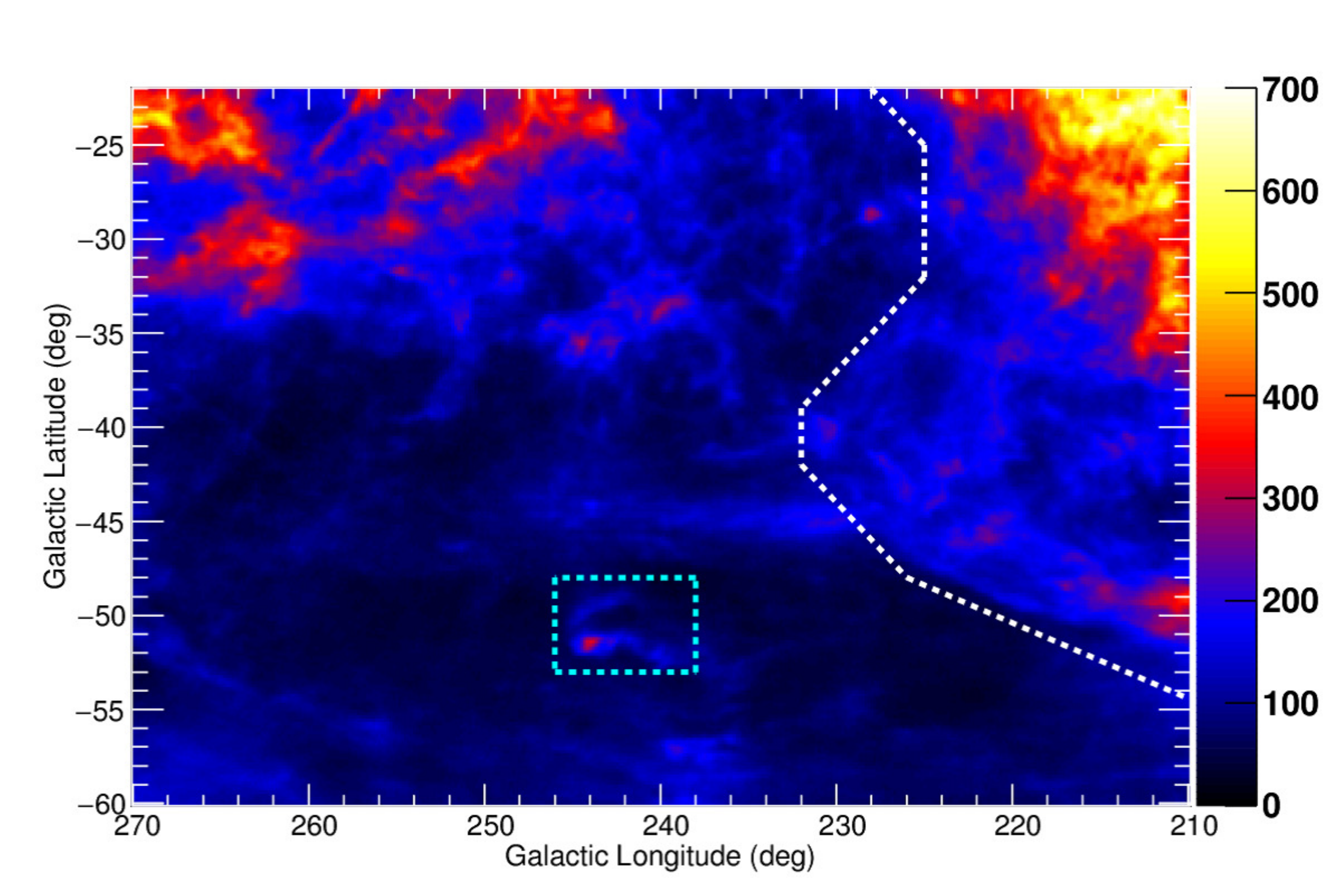}
{0.33\textwidth}{(a)}
\fig{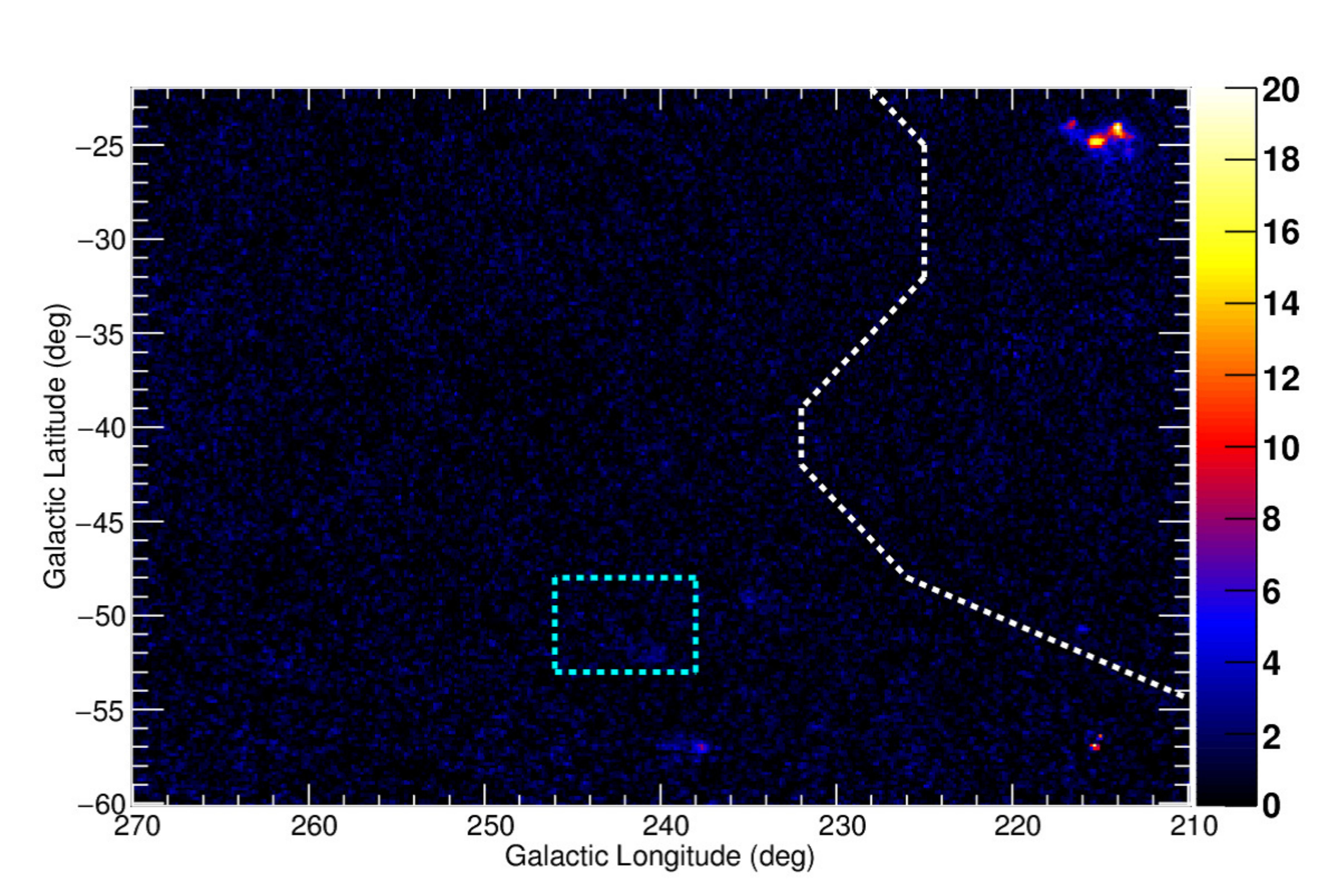}
{0.33\textwidth}{(b)}
\fig{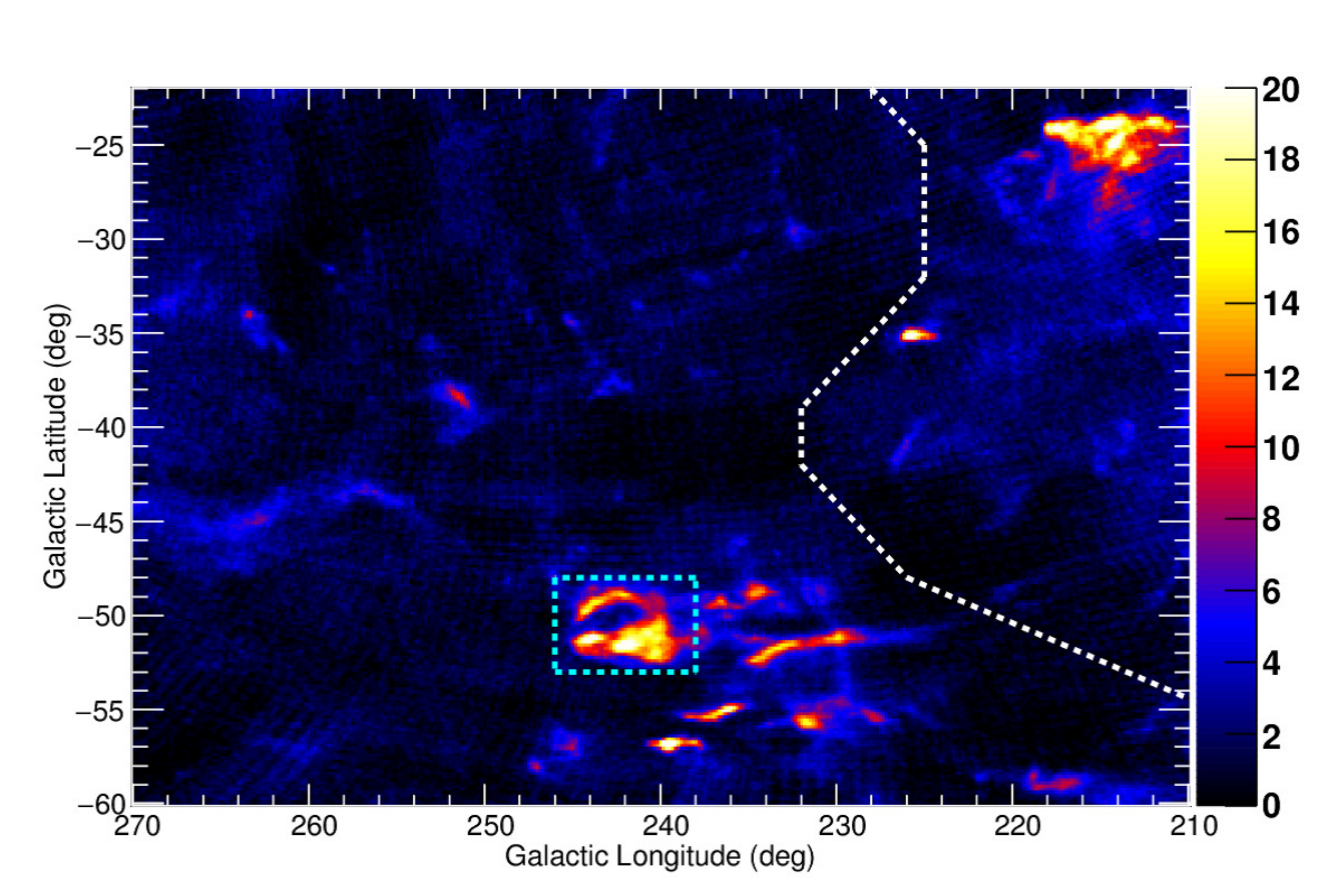}
{0.33\textwidth}{(c)}
}
\gridline{
\fig{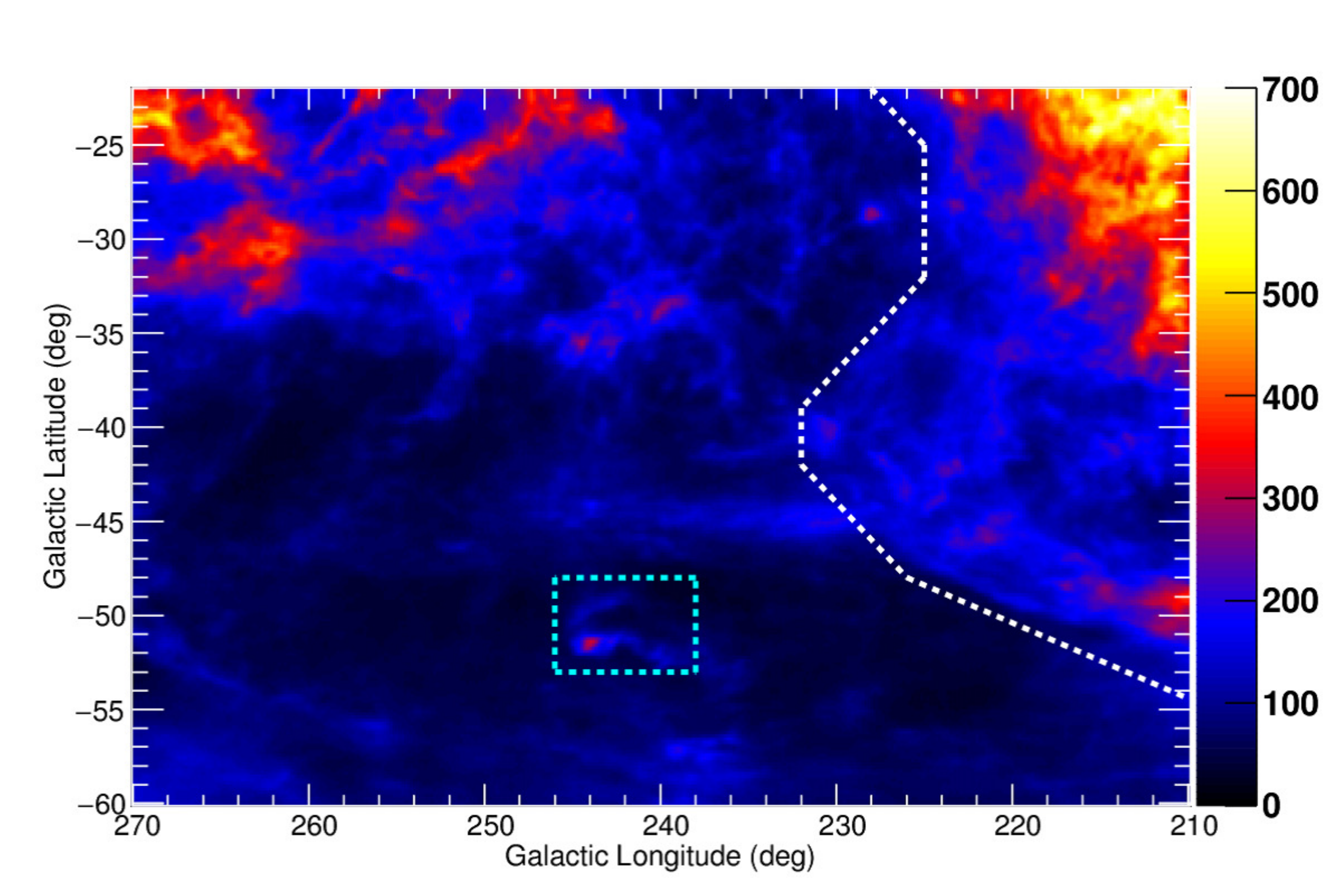}
{0.33\textwidth}{(d)}
\fig{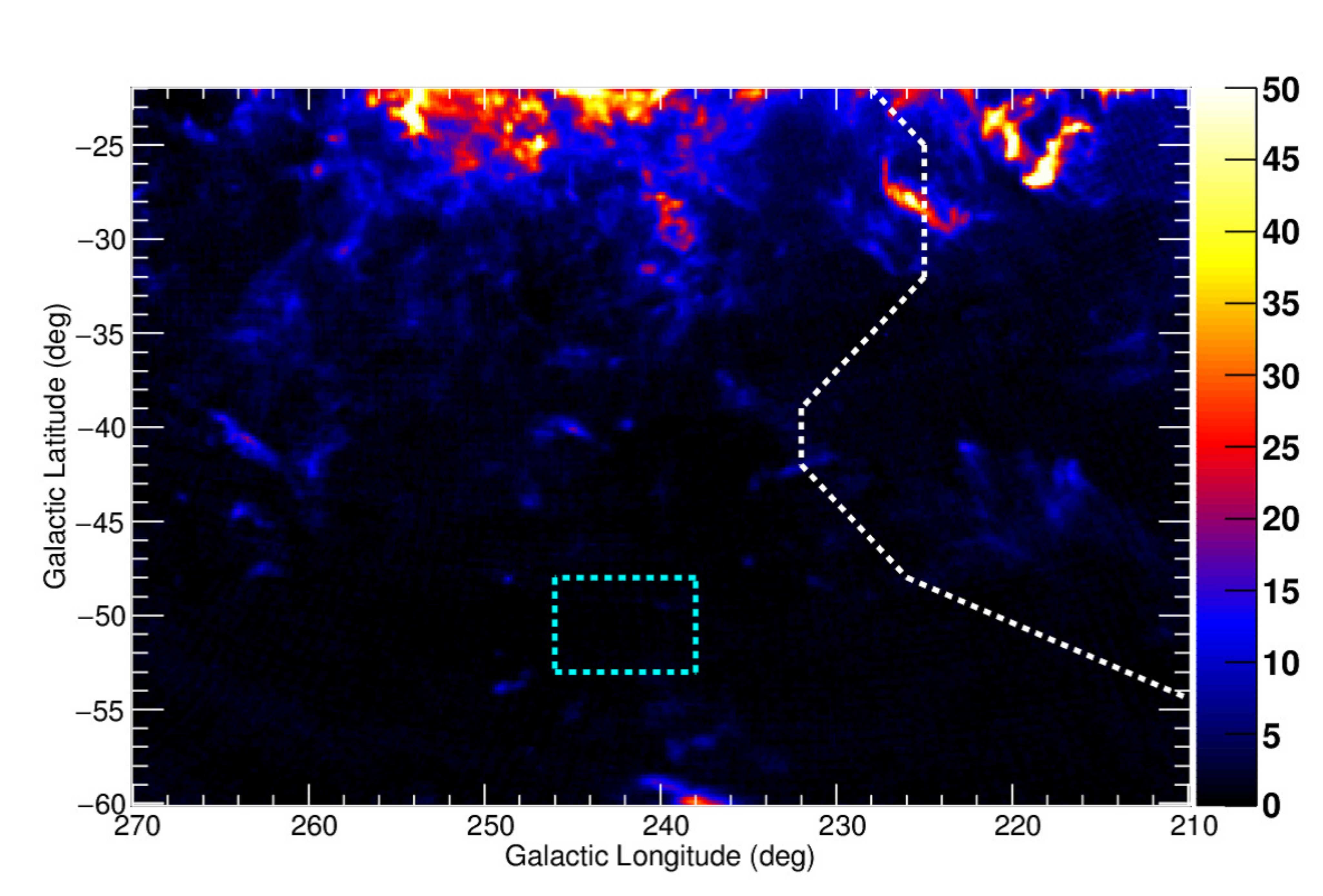}
{0.33\textwidth}{(e)}
\fig{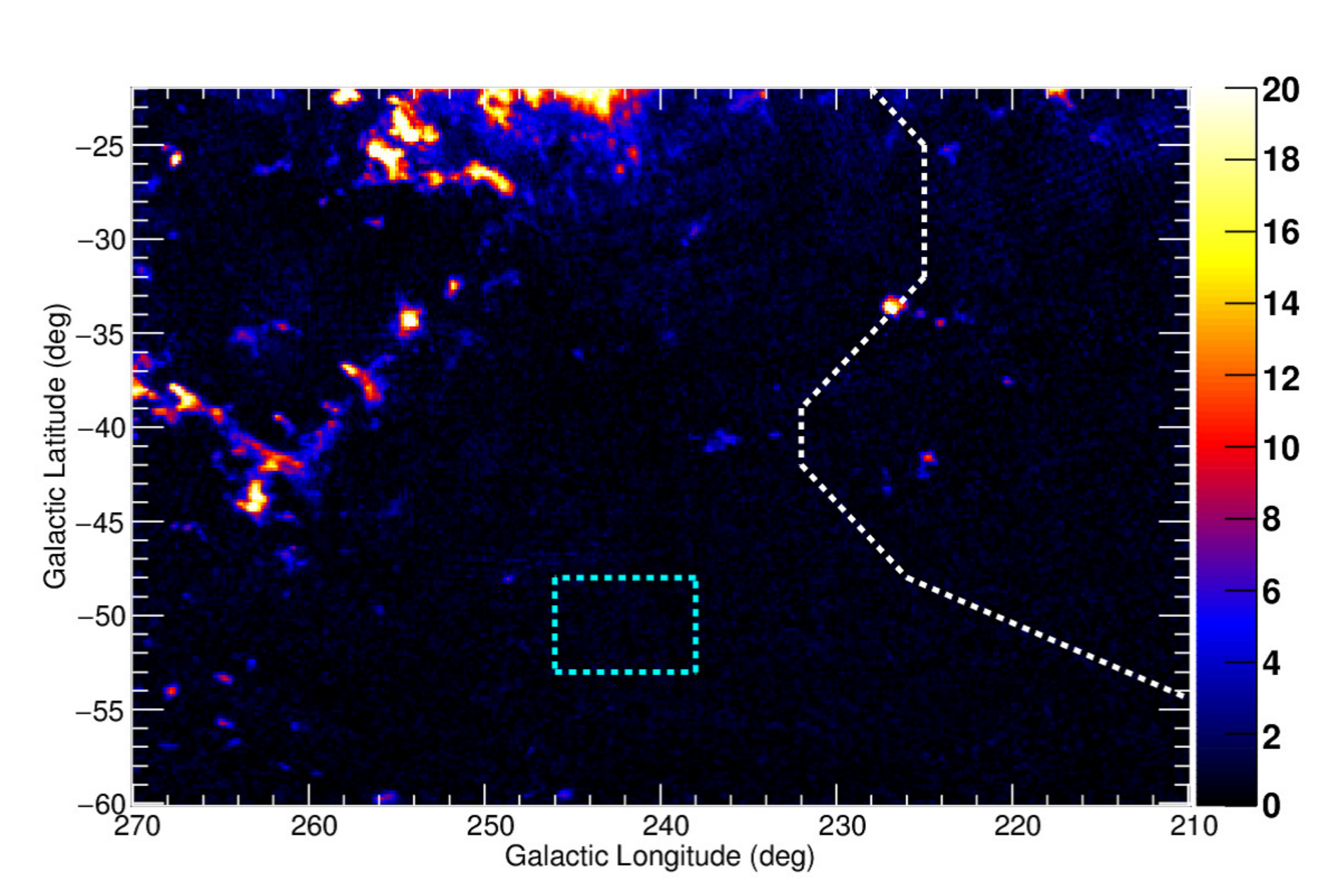}
{0.33\textwidth}{(f)}
}
\caption{
The same as Figure~15 but for the southern region.
The intermediate velocity cloud
and the Orion-Eridanus superbubble are masked by areas shown by dotted lines (see text for details).
\label{fig:f102}
}
\end{figure}

In the southern region, while most of the gas is in local ($|v| \le 35~\mathrm{km~s^{-1}}$),
we can identify a coherent structure in $238\arcdeg \le l \le246\arcdeg$ and 
$-53\arcdeg \le b \le -48\arcdeg$ in $v$ from -70 to -35~$\mathrm{km~s^{-1}}$;
the structure is likely to be an
intermediate velocity cloud \citep[e.g.,][]{Wakker2001}.
This area has a large scatter in $\WHI$ while $\Dem$ is rather constant
likely because of the contamination of the clouds. 
Because the scatter in $\WHI$ would affect
the $\WHI$--$\Dem$ correlation (Section~2.2),
we masked the area in 
the study of the $\WHI$--$\Dem$ relationship and $\gamma$-ray data analysis.
We can also identify another coherent structure ($\HI$ cloud) in $v$ from $-$70 to $-$35~$\mathrm{km~s^{-1}}$
in $200\arcdeg \le l \le 215\arcdeg$ and
$35\arcdeg \le b \le 60\arcdeg$ in the northern region. Because the structure shows
the $\WHI$--$\Dem$ relationship similar to those in other regions, 
we did not mask the area to maximize the photon statistics in $\gamma$-ray data analysis.
The relative contribution of the $\HI$ cloud to the $\gamma$-ray flux
(assuming uniform CR intensity) and the mass of the ISM gas (assuming the same distance)
can be evaluated by integrating $\WHI$ in the ROI. The relative contribution of this structure
(integrated from $-$70 to $-$35~$\mathrm{km~s^{-1}}$ in 
$200\arcdeg \le l \le 215\arcdeg$ and $35\arcdeg \le b \le 60\arcdeg$) to the whole emission of
$\WHI$ in the northern region was found to be only $\le 2~\%$;
therefore the effects on the evaluated CR emissivity and $\NH$ are, if any, negligible. 
The contributions of local clouds
(integrated from $-$35 to 35~$\mathrm{km~s^{-1}}$ in each ROI) to the whole emission are more than 93\% and 94\%
for the northern and southern regions, respectively.

The Orion-Eridanus superbubble \citep[e.g.,][]{Ochsendorf2015} can be identified as filamentary structures
in $\HI$ 21-cm and $\mathrm{H}_{\alpha}$ lines. To visualize the superbubble, we made a $\WHI$ map
in $v=-1$ to $8~\mathrm{km~s^{-1}}$ \citep{Brown1995}
and an $\mathrm{H}_{\alpha}$ map \citep{Finkbeiner2003} in Figure~16.
Since the CR and ISM properties of the structure could be appreciably different from those of other regions,
the area of the superbubble was masked in the study of the $\WHI$--$\Dem$ relationship and $\gamma$-ray data analysis
with a polygon defined by
($l$, $b$) = ($228\arcdeg$, $-22\arcdeg$), ($225\arcdeg$, $-25\arcdeg$),
($225\arcdeg$, $-32\arcdeg$), ($232\arcdeg$, $-39\arcdeg$), ($232\arcdeg$, $-42\arcdeg$), 
($226\arcdeg$, $-48\arcdeg$), and ($210\arcdeg$, $-54.4\arcdeg$).
Indeed, the masked area shows a different $\WHI$--$\Dem$ relation from other areas in the ROI.

\floattable
\begin{deluxetable}{ccccc}[ht!]
\tablecaption{
Radio and infrared sources excised and interpolated 
in the $\WHI$ map and \textit{Planck} dust maps.}
\tablecolumns{5}
\tablenum{7}
\label{tab:t7}
\tablewidth{0pt}
\tablehead{
\multicolumn{2}{c}{Position} & \colhead{$r_{1}$} & \colhead{$r_{2}$} &\colhead{Object type} \\
\cline{1-2}
\colhead{$l$ (deg)} & \colhead{$b$ (deg)} & \colhead{(deg)} & \colhead{(deg)} &\colhead{}
}
\startdata
255.5 & $52.8$ & 0.12 & 0.15 & infrared source \\
246.1 & $39.9$ & 0.4 & 0.5 & radio source \\
241.7 & $-36.5$ & 0.12 & 0.15 & infrared source \\
241.2 & $-35.9$ & 0.12 & 0.15 & infrared source \\
238.0 & $-54.6$ & 0.12 & 0.15 & infrared source \\
233.2 & $43.8$ & 0.4 & 0.5 & radio source \\
221.4 & $45.1$ & 0.12 & 0.15 & infrared source \\
214.1 & $47.8$ & 0.12 & 0.15 & infrared source \\
208.7 & $44.5$ & 0.12 & 0.15 & infrared source \\
208.6 & $44.5$ & 0.4 & 0.5 & radio source \\
\enddata
\tablecomments{
Values in a circular region with a radius of $r_{1}$ are filled with the average
of pixels in a ring with inner radius of $r_{1}$ and outer radius of $r_{2}$
for each position.
}
\end{deluxetable}

\begin{figure}[ht!]
\figurenum{16}
\gridline{
\fig{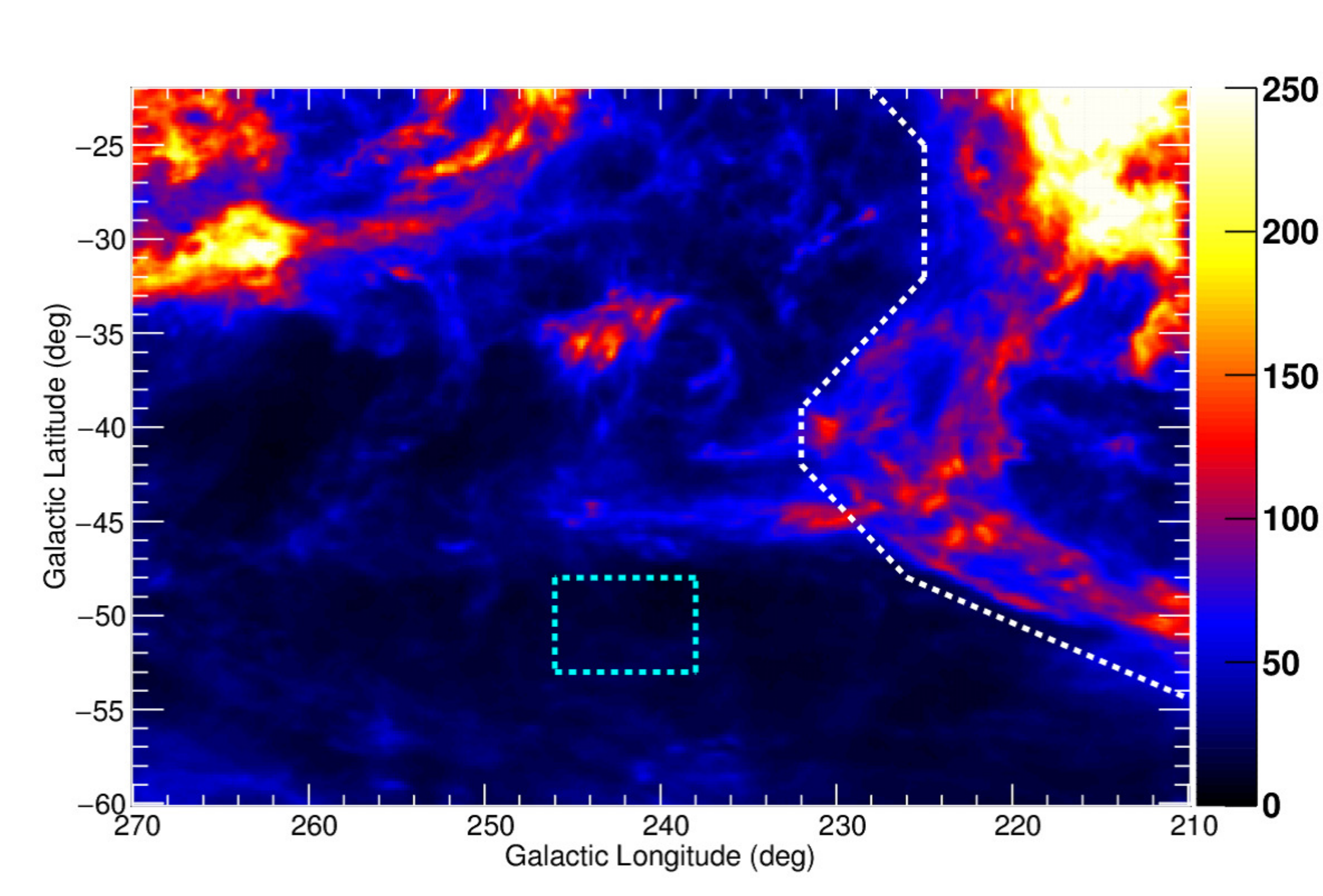}
{0.5\textwidth}{(a)}
\fig{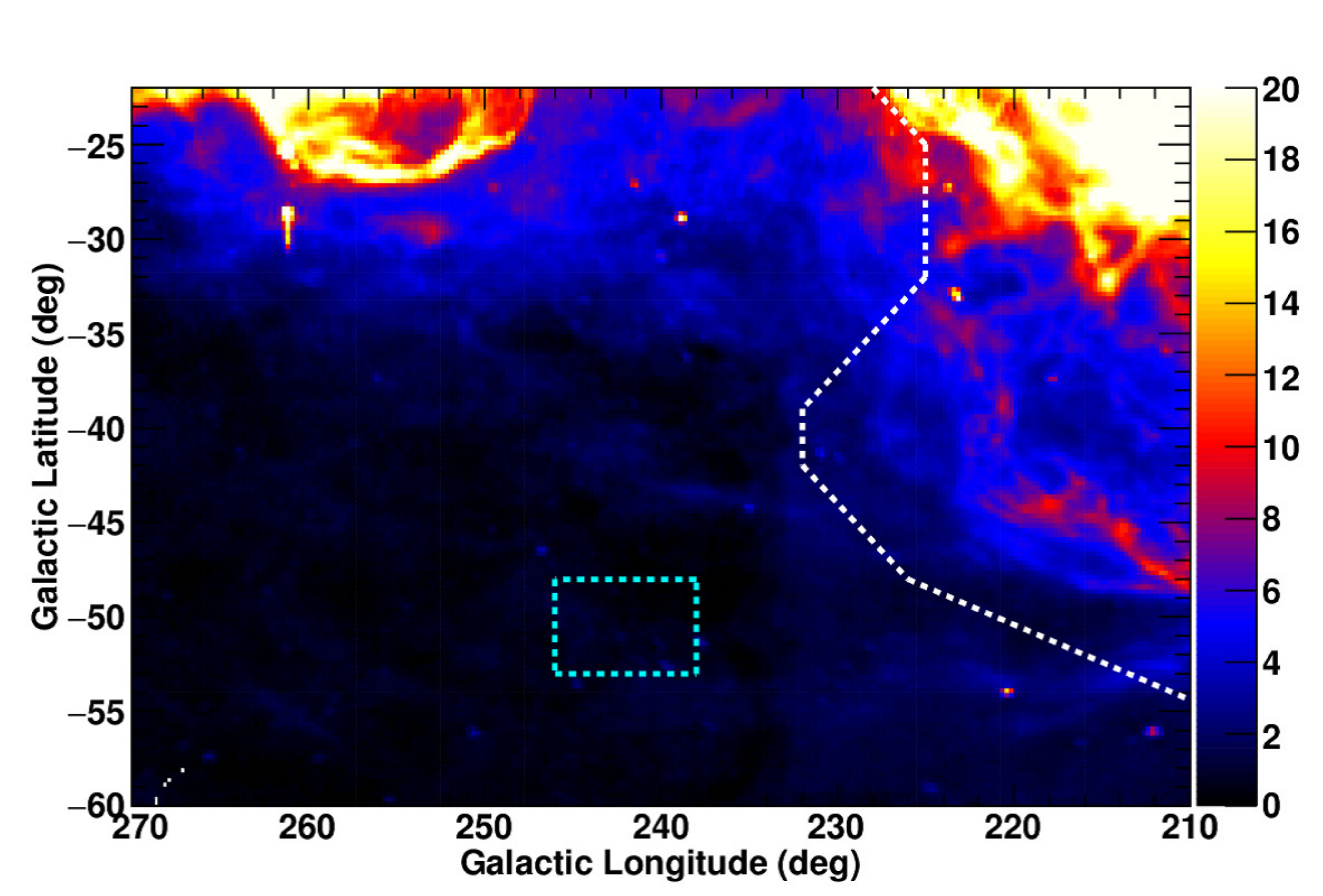}
{0.5\textwidth}{(b)}
}
\caption{
$\WHI$ map in the southern region in the velocities from $-1$ to $8~\mathrm{km~s^{-1}}$
\citep{Brown1995} (a) and $\mathrm{H}_{\alpha}$ map \citep{Finkbeiner2003} (b).
The Orion-Eridanus superbubble can be identified as filamentary structures in those maps
and is masked by the polygon shown as a dotted white line.
\label{fig:103}
}
\end{figure}

\clearpage

\section{Infrared Sources}

We compared the $R$ and $\tau_{353}$ maps and identified several spots of high $R/\tau_{353}$ ratio
($\ge 0.05~\mathrm{W~m^{-2}~sr^{-1}}$) and high $R$ 
($\ge 16 \times 10^{-8}$ and $10 \times 10^{-8}$ in units of $\mathrm{W~m^{-2}~sr^{-1}}$ in the northern and southern regions, respectively).
Their positions are ($l$, $b$) = ($241\fdg7$, $-36\fdg5$), ($241\fdg2$, $-35\fdg9$),
($238\fdg0$, $-54\fdg6$), ($214\fdg1$, $47\fdg8$), ($255\fdg5$, $52\fdg8$), 
and ($208\fdg7$, $44\fdg5$).
We identified them as infrared sources and removed them
by filling the source areas in the $R$, $\tau_{353}$, and $T_\mathrm{d}$ maps
with the average of the peripheral pixels: 
values in a circular region
with radius $0\fdg12$ are filled with the average of the pixels in a ring with
inner radius of $0\fdg12$ and outer radius of $0\fdg15$.
The parameters (position and radius) are also summarized in Table~7.

\clearpage

\section{Planck CO Map}

We also examined the Planck type~3 CO map \citep{Planck2014b} and confirmed that 
there is no strong CO 2.6~mm emission in our ROI. 
In Figure~17(a), we identified emission of moderate intensity (peak intensity $\sim 4~\mathrm{K~km~s^{-1}}$)
at ($l$, $b$) $\sim$ ($221\fdg4$, $45\fdg1$).
It is also seen in the
$R$ and $\tau_{353}$ maps and likely to be an infrared source,
and was removed from the dust maps by filling in with the average value of peripheral pixels.
The source is also listed in Table~7. Other bright CO 2.6~mm emission at around
($l$, $b$) = ($211\arcdeg$, $-36\fdg5$) is inside the mask of the Orion-Eridanus superbubble,
as shown in panel~(b).

\begin{figure}[ht!]
\figurenum{17}
\gridline{
\fig{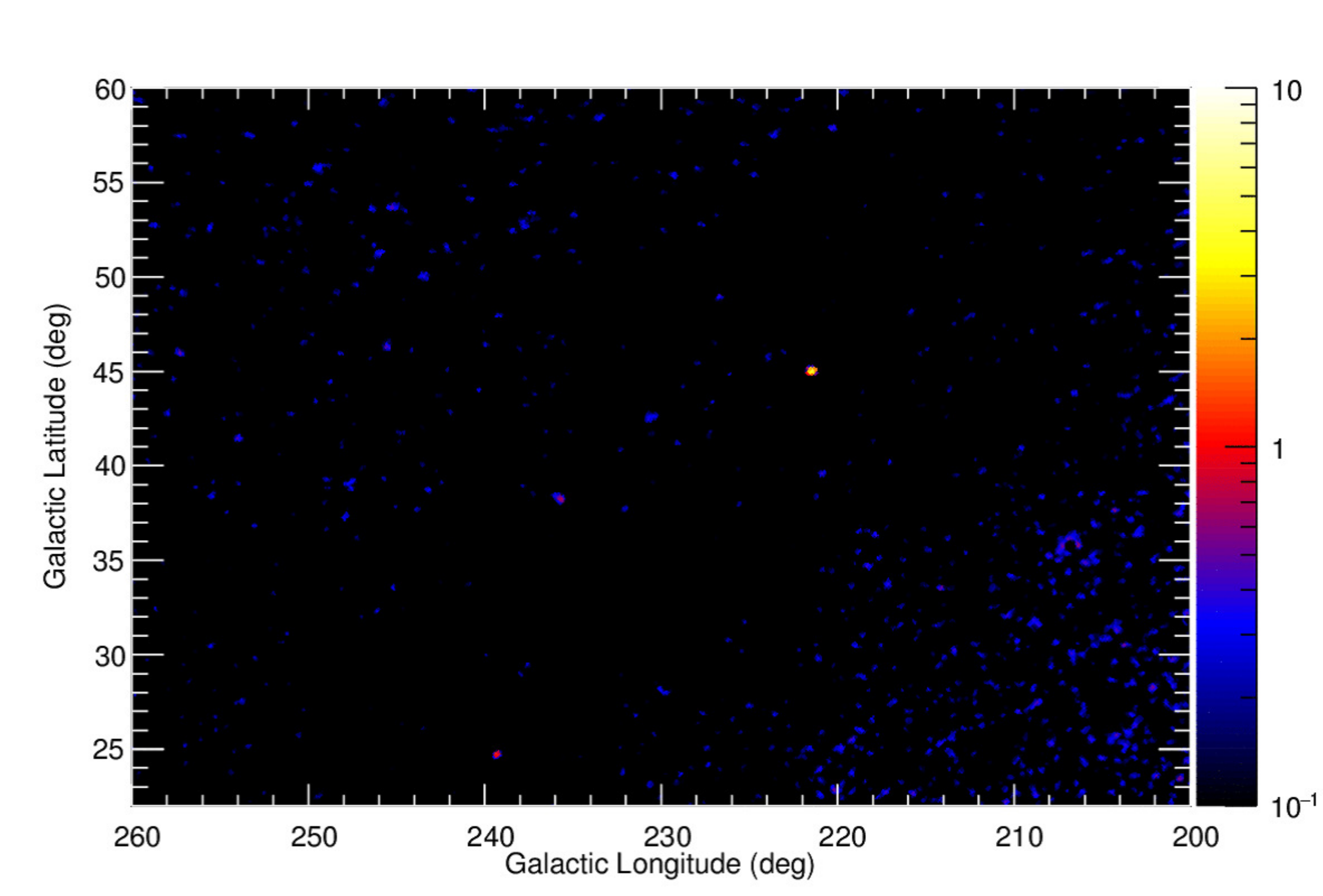}
{0.5\textwidth}{(a)}
\fig{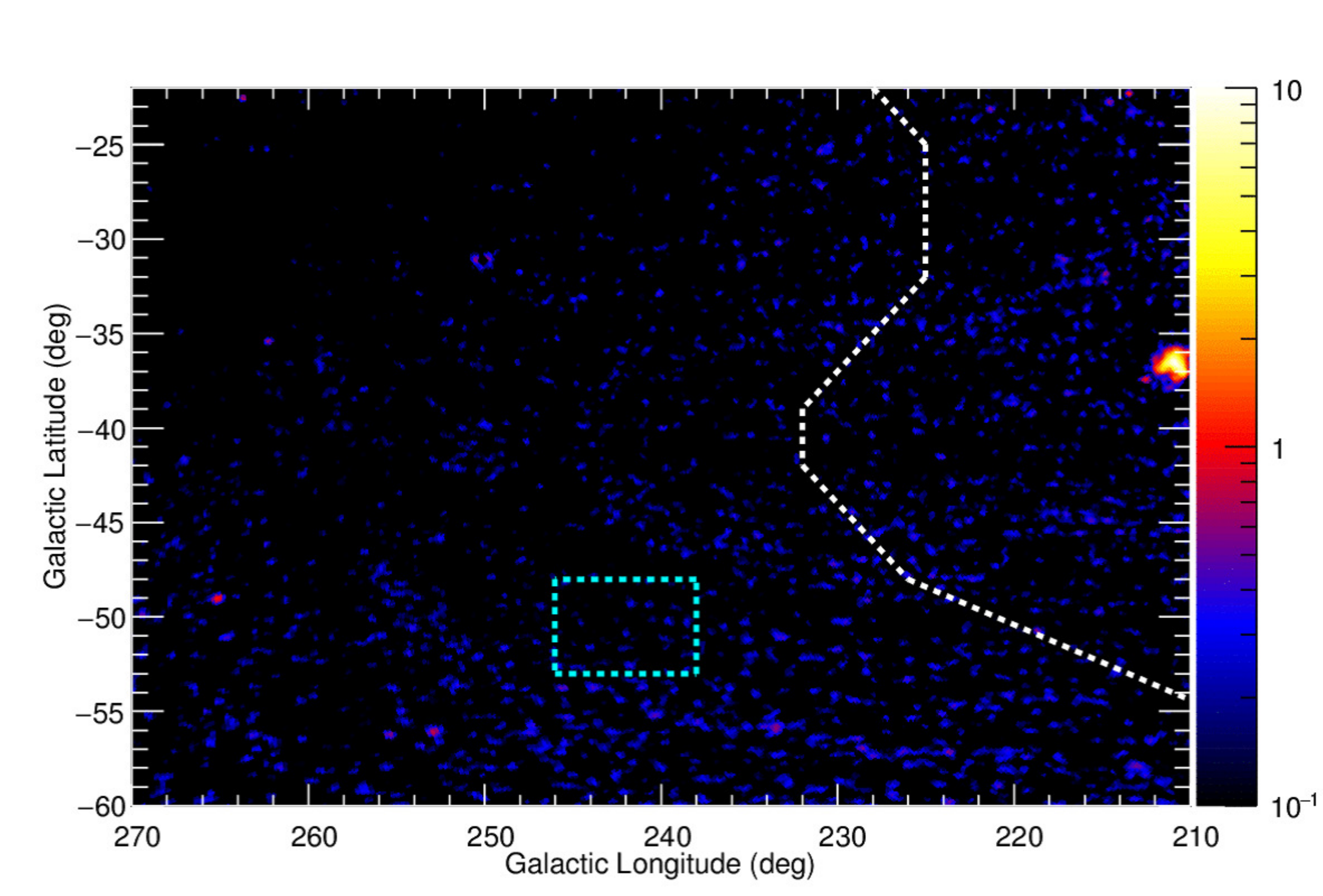}
{0.5\textwidth}{(b)}
}
\caption{
Planck type~3 $\WCO$ maps ($\mathrm{K~km~s^{-1}}$) in the (a) northern and (b) southern regions.
The spot of moderate intensity at ($l$, $b$) $\sim$ ($221\fdg4$, $45\fdg1$) was removed from 
the dust maps used in the study. Other bright CO 2.6~mm emission seen in the southern region
is inside the area that is masked (see Appendix~A).
\label{fig:104}
}
\end{figure}

\clearpage

\section{$\WHI$--$\Dem$ Correlation With Finer $T_\mathrm{d}$ Bins}

In Section~2.1, we examined the $T_\mathrm{d}$ dependence of the
$\WHI$--$\Dem$($R$ or $\tau_{353}$) relationship in the six $T_\mathrm{d}$ bins,
where the data are grouped in 1~K ranges of $T_\mathrm{d}$ 
(see Figures~1 and 2).
In Figures~18 and 19 we show the same plots, but in which the data are grouped in 0.5~K ranges of 
$T_\mathrm{d}$ to reduce the overlapping of points in the plot at the expense of separating
data into two plots to cover the whole $T_\mathrm{d}$ range.

\begin{figure}[ht!]
\figurenum{18}
\gridline{
\fig{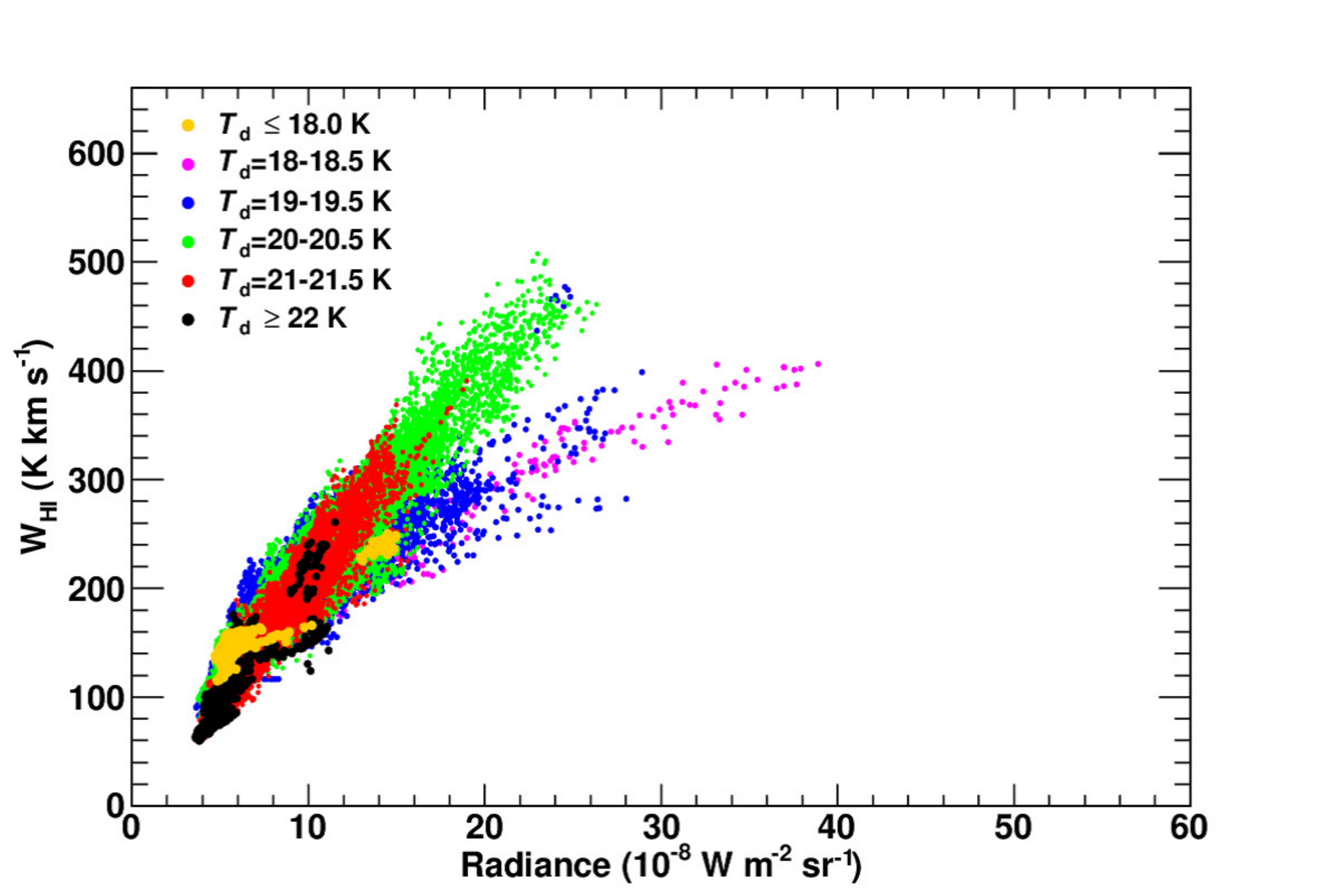}
{0.5\textwidth}{(a)}
\fig{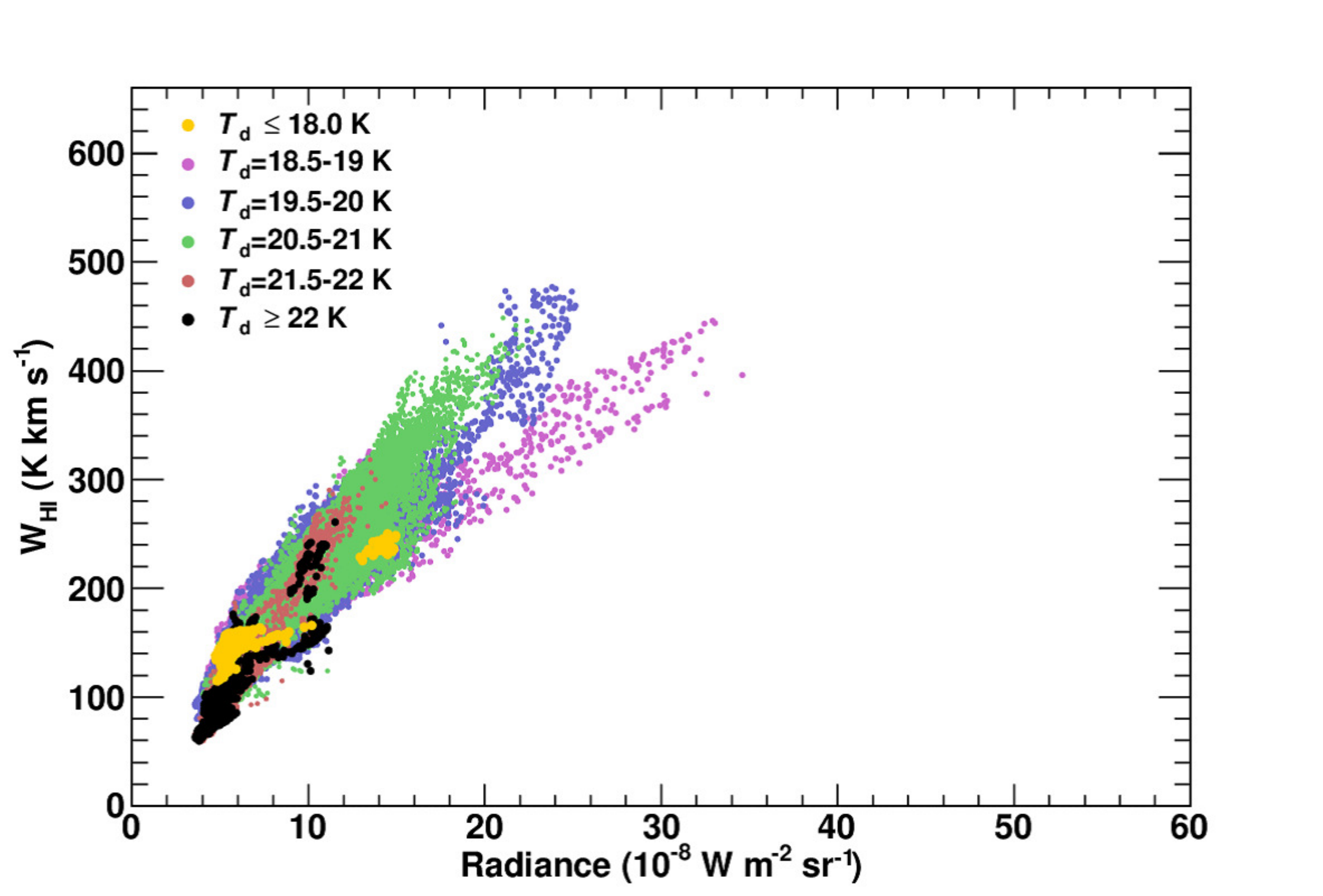}
{0.5\textwidth}{(b)}
}
\gridline{
\fig{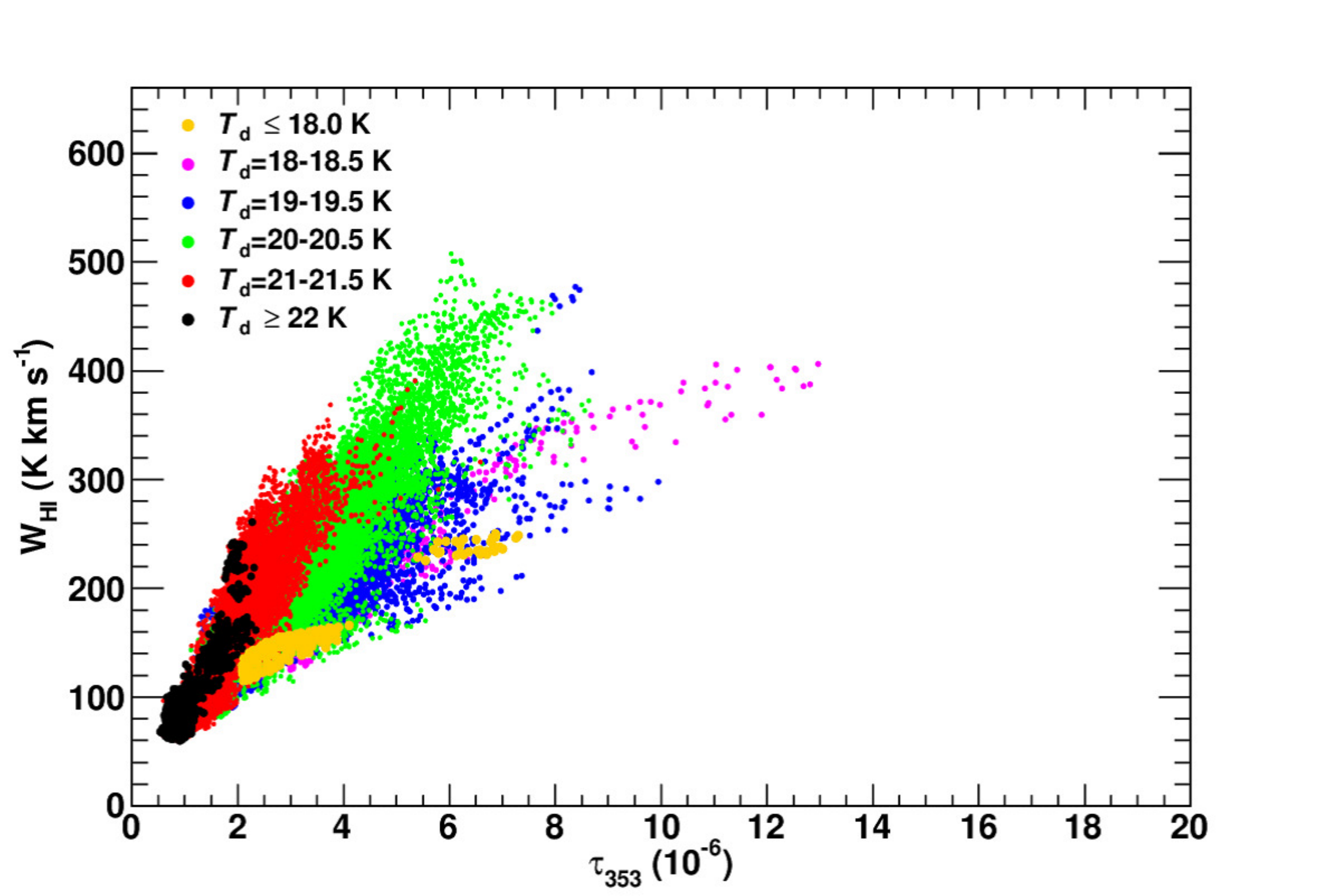}
{0.5\textwidth}{(c)}
\fig{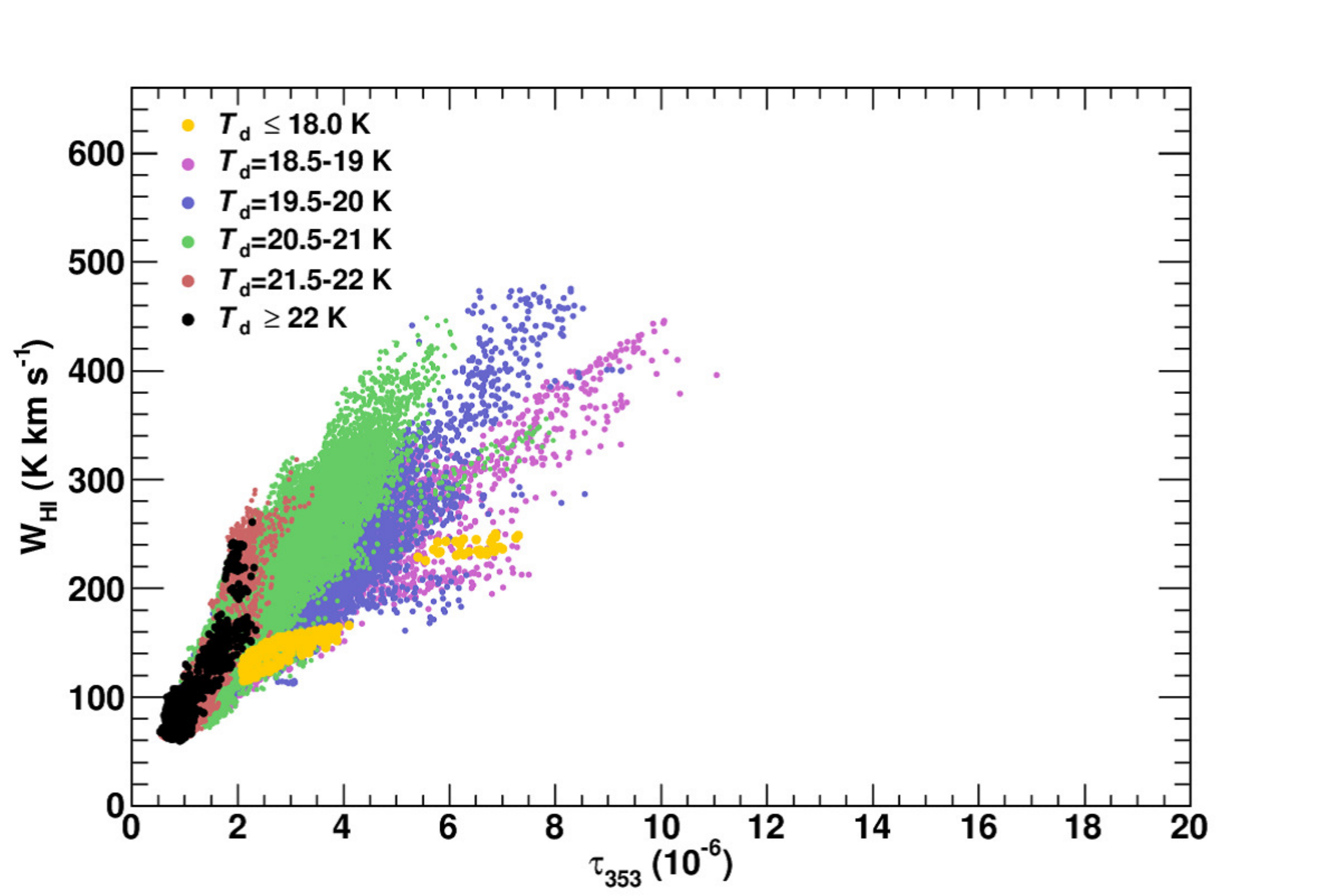}
{0.5\textwidth}{(d)}
}
\caption{
Correlation between $\WHI$ and dust tracers in the northern region. Panels (a) and (b)
show the $\WHI$--$R$ relations and panels (c) and (d) show the
$\WHI$--$\tau_{353}$ relations.
\label{fig:105}
}
\end{figure}

\begin{figure}[ht!]
\figurenum{19}
\gridline{
\fig{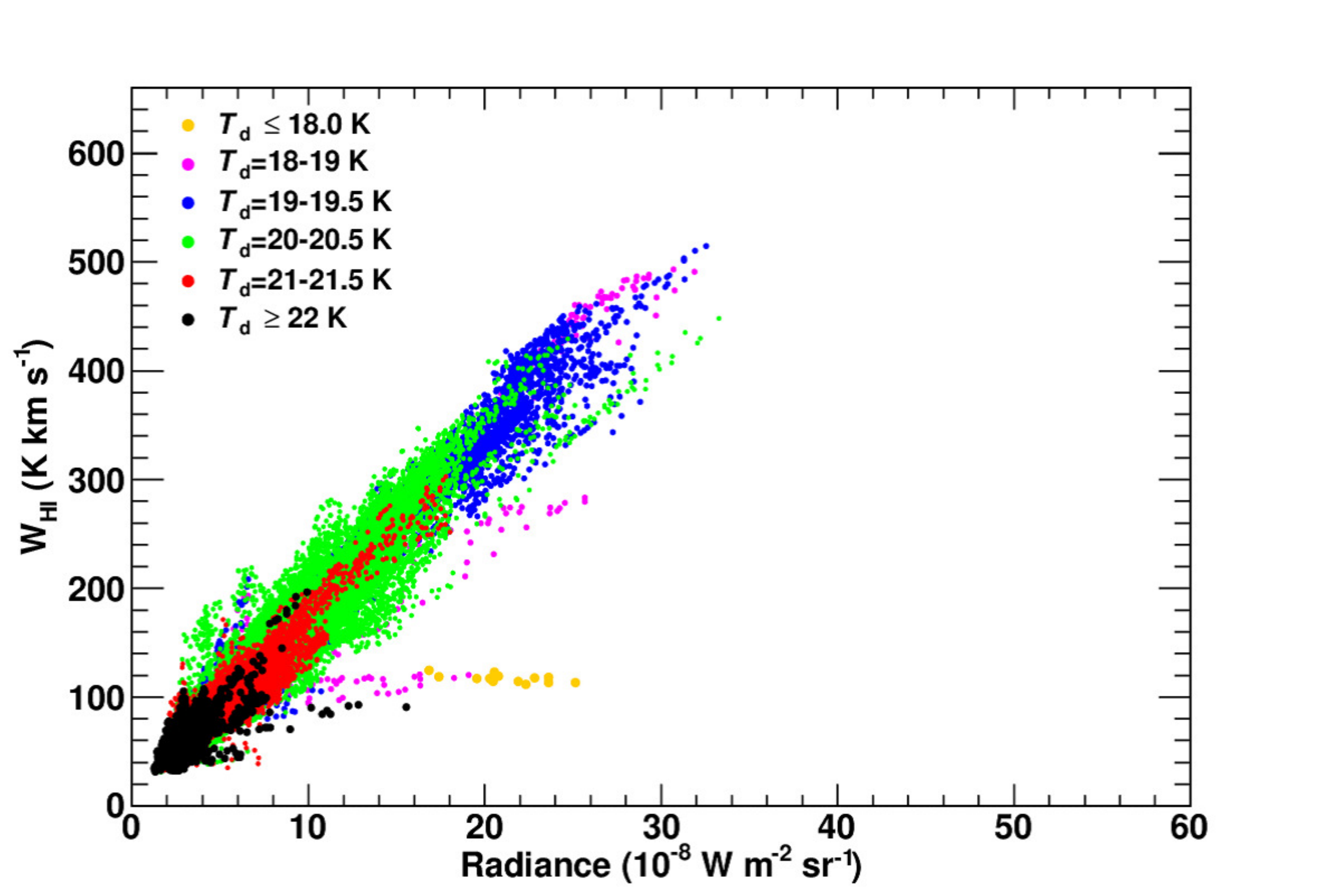}
{0.5\textwidth}{(a)}
\fig{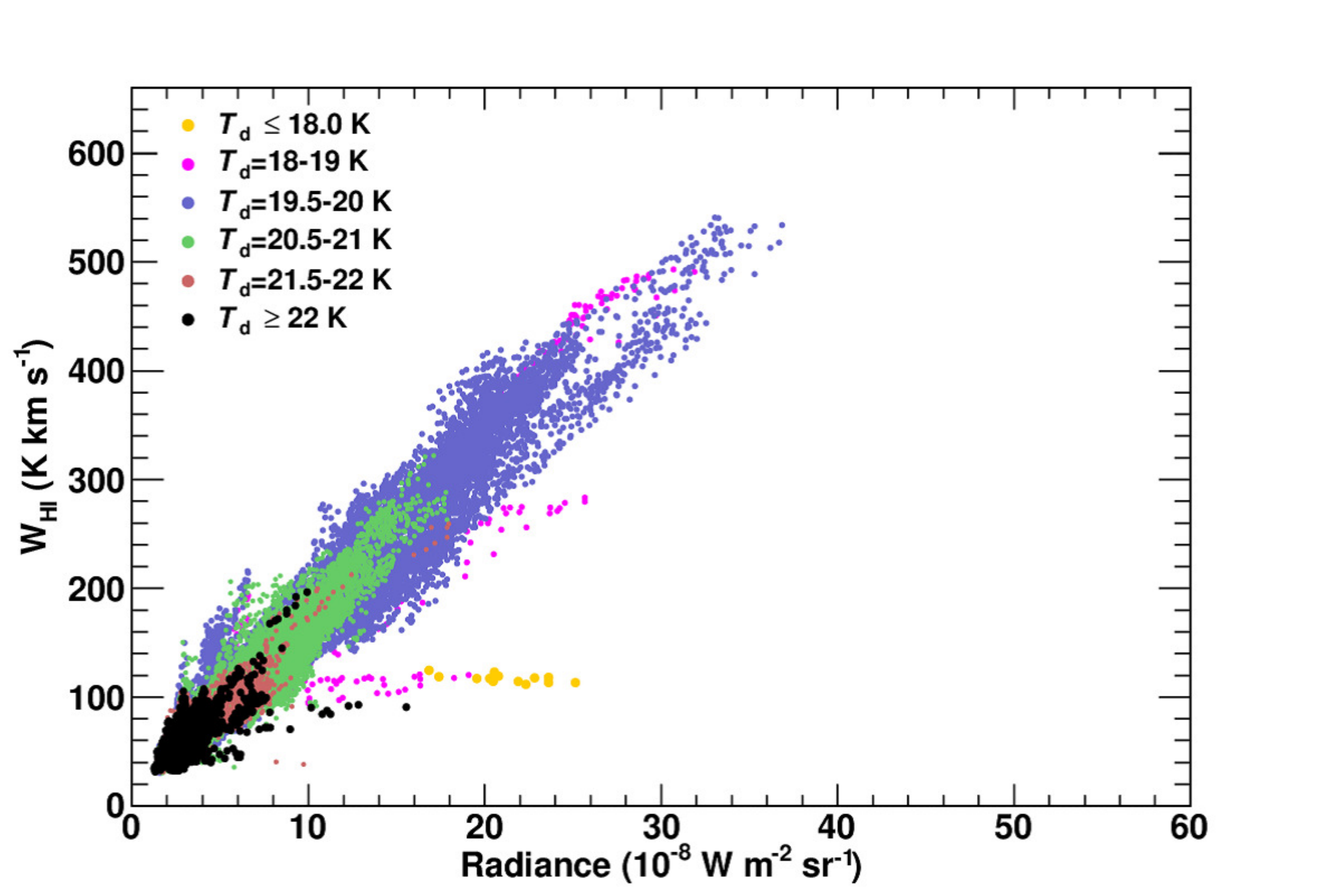}
{0.5\textwidth}{(b)}
}
\gridline{
\fig{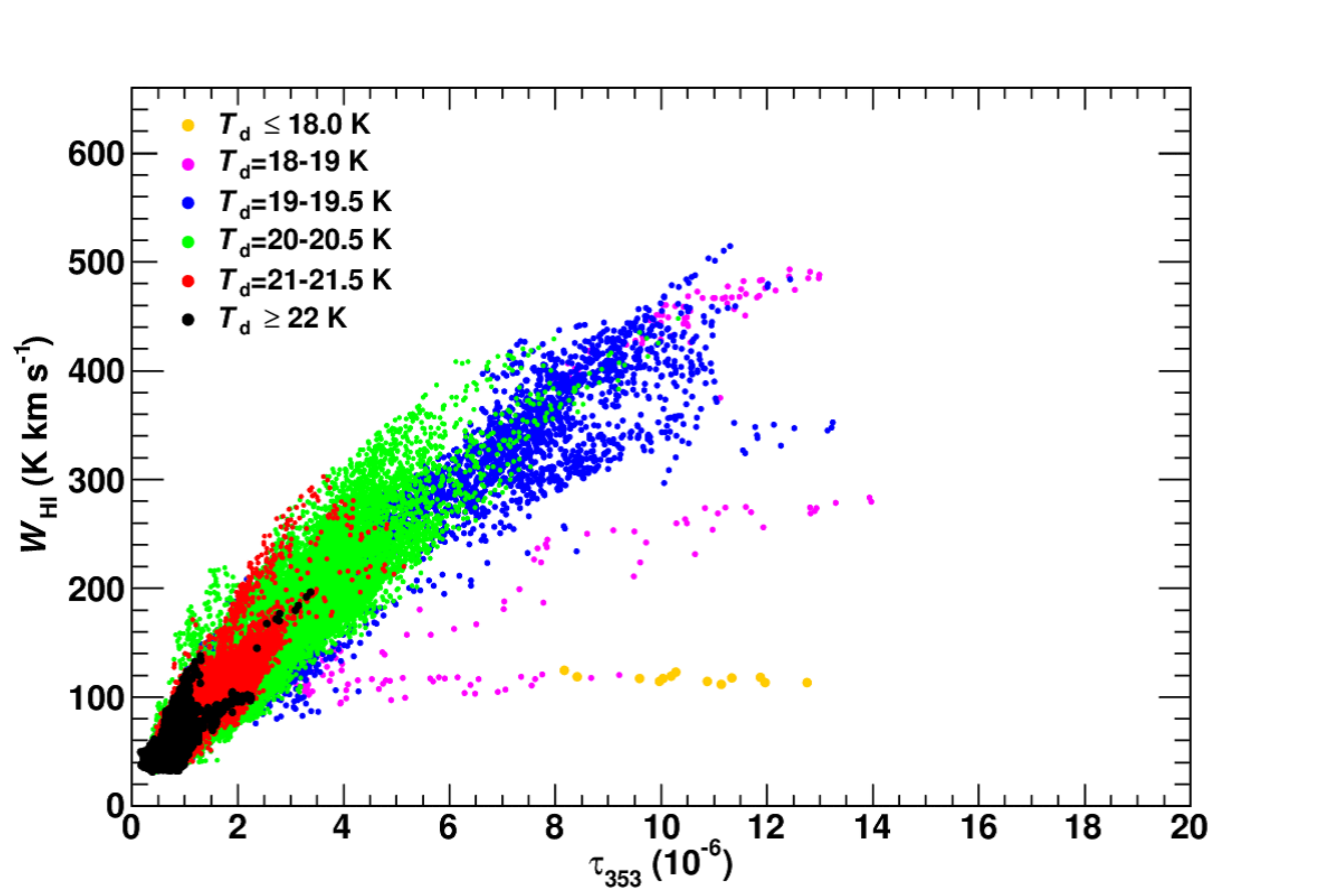}
{0.5\textwidth}{(c)}
\fig{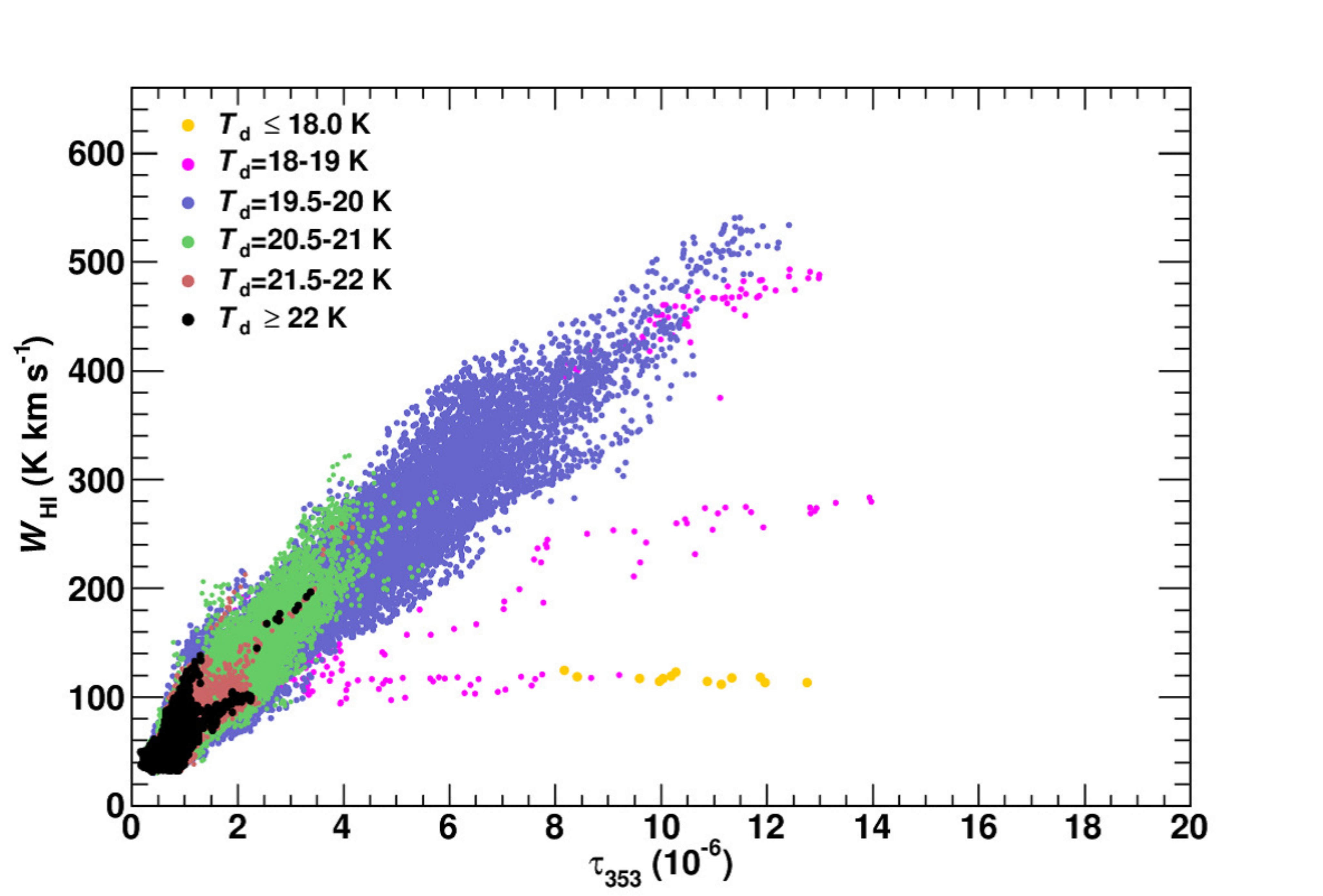}
{0.5\textwidth}{(d)}
}
\caption{
The same as Figure~18 but for the southern region.
\label{fig:106}
}
\end{figure}

\clearpage

\end{document}